\documentclass[10pt,sigconf,letterpaper,nonacm]{acmart}

\usepackage{amsmath}
\usepackage{subfigure}

\usepackage{hyperref}

\usepackage{filecontents}

\usepackage{balance}

\usepackage{tikz}

\usepackage{amssymb}
\usepackage{mathtools}

\usepackage[english]{babel}
\usepackage[utf8]{inputenc}
\usepackage{algorithm}
\usepackage[noend]{algpseudocode}

\DeclareMathOperator*{\argmax}{argmax} 

\newcommand{\sysname}{COLA}

\newenvironment{parafont}{\fontfamily{ptm}\selectfont}{}
\newcommand{\Para}[1]{\vspace{2pt}\noindent\begin{parafont}\textbf{#1}\end{parafont}}

\makeatletter
\newlength{\trianglerightwidth}
\algnewcommand{\BlankComment}[1]{\Statex \hskip\ALG@thistlm \footnotesize\ttfamily{#1}}
\algnewcommand{\LineComment}[1]{\Statex \hskip\ALG@thistlm \footnotesize\ttfamily\textcolor{blue}{$\triangleright$ #1}}
\algnewcommand{\LineCommentCont}[1]{\Statex \hskip\ALG@thistlm%
  \parbox[t]{\dimexpr\linewidth-\ALG@thistlm}{\hangindent=\trianglerightwidth \hangafter=1 \strut$\triangleright$ #1\strut}}
\makeatother

\usepackage{float}

\usepackage{amsfonts}
\usepackage{booktabs}
\usepackage[singlelinecheck=false,font={normalsize}]{caption}

\usepackage{capt-of}
\usepackage{tabularx}
\usepackage{xurl}

\usepackage{array}
\usepackage{arydshln}
\usepackage{filecontents}

\def\compactify{\itemsep=0pt \topsep=0pt \partopsep=0pt \parsep=0pt}
\let\latexusecounter=\usecounter

\newenvironment{CompactEnumerate}
  {\def\usecounter{\compactify\latexusecounter}
   \begin{enumerate}}
  {\end{enumerate}\let\usecounter=\latexusecounter}

\usepackage[english]{babel}
\usepackage{blindtext}
\renewcommand\footnotetextcopyrightpermission[1]{} 
\setcopyright{none}

\acmDOI{}

\acmISBN{}

\acmPrice{}

\begin{document}

\date{}

\title{Collective Autoscaling for Cloud Microservices}
\author{Vighnesh Sachidananda}
\affiliation{%
  \institution{Stanford University}
  \city{Stanford}
  \country{USA}
}
\author{Anirudh Sivaraman}
\affiliation{%
  \institution{New York University}
  \city{New York}
  \country{USA}
}


\begin{abstract}
As cloud applications shift from monoliths to loosely coupled microservices, application developers must decide how many compute resources (e.g., number of replicated containers) to assign to each microservice within an application. This decision affects both (1) the dollar cost to the application developer and (2) the end-to-end latency perceived by the application user. Today, individual microservices are autoscaled independently by adding VMs whenever per-microservice CPU or memory utilization crosses a configurable threshold. However, an application user's end-to-end latency consists of time spent on multiple microservices and each microservice might need a different number of VMs to achieve an overall end-to-end latency.

We present \sysname{}, an autoscaler for microservice-based applications, which {\em collectively} allocates VMs to microservices with a global goal of minimizing dollar cost while keeping end-to-end application latency under a given target. Using 5 open-source applications, we compared \sysname{} to several utilization and machine learning based autoscalers. We evaluate \sysname{} across different compute settings on Google Kubernetes Engine (GKE) in which users manage compute resources, GKE standard, and a new mode of operation in which the cloud provider manages compute infrastructure, GKE Autopilot. \sysname{} meets a desired median or tail latency target on 53 of 63 workloads where it provides a cost reduction of 19.3\%, on average, over the next cheapest autoscaler. \sysname{} is the most cost effective autoscaling policy for 48 of these 53 workloads. The cost savings from managing a cluster with \sysname{} result in \sysname{} paying for its training cost in a few days. On smaller applications, for which we can exhaustively search microservice configurations, we find that \sysname{} is optimal for 90\% of cases and near optimal otherwise.


\end{abstract}

\settopmatter{printacmref=false}
\setcopyright{none}
\renewcommand\footnotetextcopyrightpermission[1]{}
\pagestyle{plain}

\maketitle


\begin{sloppypar}
\section{Introduction}
Cloud applications are increasingly built out of loosely coupled microservices communicating over RPCs\cite{netflix-ms:online, twitter-ms:online, uber-ms:online}. Building applications out of microservices allow teams in large organizations to independently develop code with fewer concerns about programming language choice and code dependencies. This reduces friction in development.

However, this decomposition into microservices complicates autoscaling: the process of automatically allocating compute resources to a cloud application as its workload changes. Today, each microservice within an application is typically autoscaled independent of other microservices using a combination of two mechanisms: (1) horizontal pod autoscaling (HPA), which changes the number of replicas supporting a microservice and (2) cluster autoscaling, which changes the number of VMs to accommodate the change in the number of replicas induced by  HPA.

This independent per-microservice autoscaling design is sub-optimal because it ignores the fact that each microservice operates in the broader context of a Web service application. In particular, while allocating additional VMs to a particular microservice does improve the response latency of RPCs served by this microservice, this improvement may have little impact on end-to-end latency of the application because of a much more sluggish microservice upstream or downstream.

Instead, we \emph{collectively} autoscale all of an application's microservices and decide how much to allocate each microservice,  with a view towards a global objective. Effectively, we replace many distributed and independent autoscalers with one centralized autoscaler with global visibility.

To collectively autoscale a microservice-based application, we frame autoscaling of microservices as constrained optimization. The constraint is meeting an end-to-end mean/tail latency target. The optimization objective is minimizing dollar cost. Ideally, we would solve this optimization problem online as the workload  and the microservices' business logic change. But, finding the right number of VMs for each microservice takes time because it requires iteratively identifying and eliminating bottlenecked microservices until the end-to-end latency target is met. Running this iteration online can seriously disrupt the application's user experience.

Instead, we propose COLA, an offline search process to find the right allocation of VMs to each microservice given some information on the workload that is expected during deployment.\footnote{This information doesn't have to be perfect: COLA will interpolate between known workloads if possible and fall back to default autoscalers if not.} COLA's search process works as follows: for each workload, COLA applies the workload, identifies the most congested microservice (the microservice whose CPU utilization increases the most in response to the workload), determines the right allocation of VMs to this microservice using a bandit problem with a reward capturing the latency-vs.-cost tradeoff, then iterates this procedure with the next most-congested microservice until the latency target is met.  

Effectively, from the perspective of an application developer, COLA performs profile-guided optimization~\cite{lattner2002llvm, homescu2013profile} in the same spirit as a compiler that exploits knowledge of what program paths are more or less common to optimize generated code. Naively, this offline search process needs to explore all possible allocations of VMs to microservices, which blows up quickly. Hence, we develop optimizations to efficiently search this space and save both time and cost.

\begin{table}[!t]
  \begin{tabular}{lll}
    \toprule
    Application& Microservices & Cost Reduction \\
    \midrule
  Simple Web Server \cite{sws:online}  & 1 &  13.95\%\\  Book Info \cite{binfo:online} & 4 &  18.01\% \\
  Online Boutique \cite{ss:online} & 11 &   33.11\%\\
  Sock Shop \cite{ob:online} & 14 &  1.34\% \\
  Train Ticket \cite{zhou2018poster} & 64 &   26.25\%\\
  \bottomrule
\end{tabular}
\caption{\sysname{} compared to next best of 5 baselines -- 2 based on Kubernetes' autoscaler~\cite{kub-cluster-autoscaler:online} and 3 ML-based research autoscalers~\cite{alipourfard2017cherrypick,venkataraman2016ernest, qiu2020firm}}
\label{tab:summary-table}
\vspace{-1.00cm}
\end{table}


We summarize our evaluations of COLA below:
\begin{CompactEnumerate}
\item Across 5 open source applications including the largest available open source application in terms of number of microservices~\cite{zhou2018poster}, on the Google Kubernetes Engine (GKE) platform, COLA outperforms 5 baselines (2 based on Kubernetes' autoscaler~\cite{kub-cluster-autoscaler:online} and 3 ML-based research autoscalers~\cite{alipourfard2017cherrypick,venkataraman2016ernest, qiu2020firm} adapted from the literature) (Table~\ref{tab:summary-table}). Across 53 of 63 workloads where \sysname{} meets a desired latency target, \sysname{} reduces cost of microservice deployment by 19.3\% when compared to the next cheapest autoscaler across our 5 baselines. On the remaining 10 workloads, the lowest cost policy which meets the corresponding latency target costs 69.9\% more than \sysname{}. Further, we evaluate and find that \sysname{} outperforms baselines on a set of different compute environments where we vary node size and cluster type. A full set of tabular results is provided in Appendix Tables \ref{tab:bi-fr-tabular}-\ref{tab:tt-ap-tabular}.
\item We demonstrate that initial training costs for \sysname{} are paid for by the cost savings of deployment in less than one day in some cases and within a week for most cases. Additionally, we find \sysname{} can be incrementally retrained for new latency targets and workloads with small overhead.
\item Empirically, we find that COLA's results are near-optimal: typically, all other allocations of VMs to microservices either require more total VMs than COLA or do not satisfy the latency target.
\item Theoretically, using simplified models of interconnected microservices, we provide a mathematical rationale for COLA's training and inference procedures. We show why utilization is a sensible metric to determine which microservice is most congested and also why linearly interpolating \sysname{}'s policies can generalize to unseen workloads.
\end{CompactEnumerate}

\section{Background}
\subsection{Kubernetes}
\label{subsec:kubernetes}
When developing a microservice application, developers deploy each microservice on one or more containers. Kubernetes \cite{kubernetes:online} is a platform that orchestrates the deployment, scaling, and management of these containers. We review Kubernetes concepts and control plane actions to describe how containers are deployed and are allocated compute resources.

\Para{Kubernetes Concepts.}
A Kubernetes cluster is made up of a set of \textit{nodes}. Each node is a physical or virtual machine. A master node is responsible for hosting an entry point for interacting with the cluster, managing cluster state and scheduling resources. Worker nodes run an agent, known as a kubelet, to communicate with this master node. They also run containers within an application. Within worker nodes, the master node deploys {\em pods}: the atomic element of deployment, replication, and scaling in Kubernetes. A pod is an encapsulation of one or more containerized applications (typically one) along with a shared context of storage and network resources. A pod also has associated CPU and memory {\em limits}, which restrict the amount of the node's resources that the pod is allowed to use.

\begin{figure*}
\captionsetup[subfigure]{justification=centering}
\centering     
\subfigure[Horizontal Pod Autoscaler - Latency \newline vs. Number of VMs for Book Info App]{\label{fig:a}\includegraphics[width=.30\textwidth]{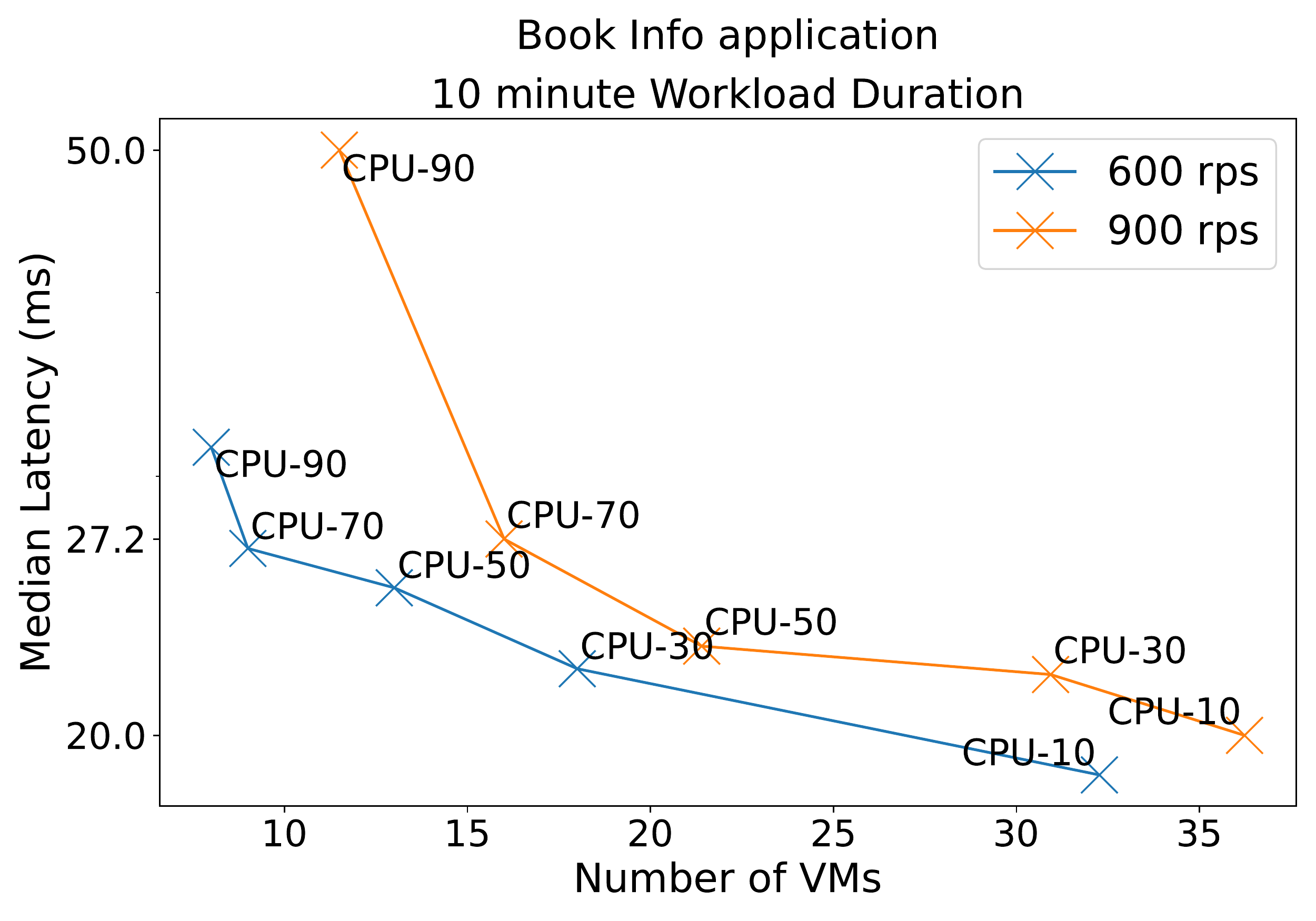}}
\subfigure[Horizontal Pod Autoscaler - Latency \newline vs.  Number of VMs for Online Boutique App]{\label{fig:b}\includegraphics[width=.30\textwidth]{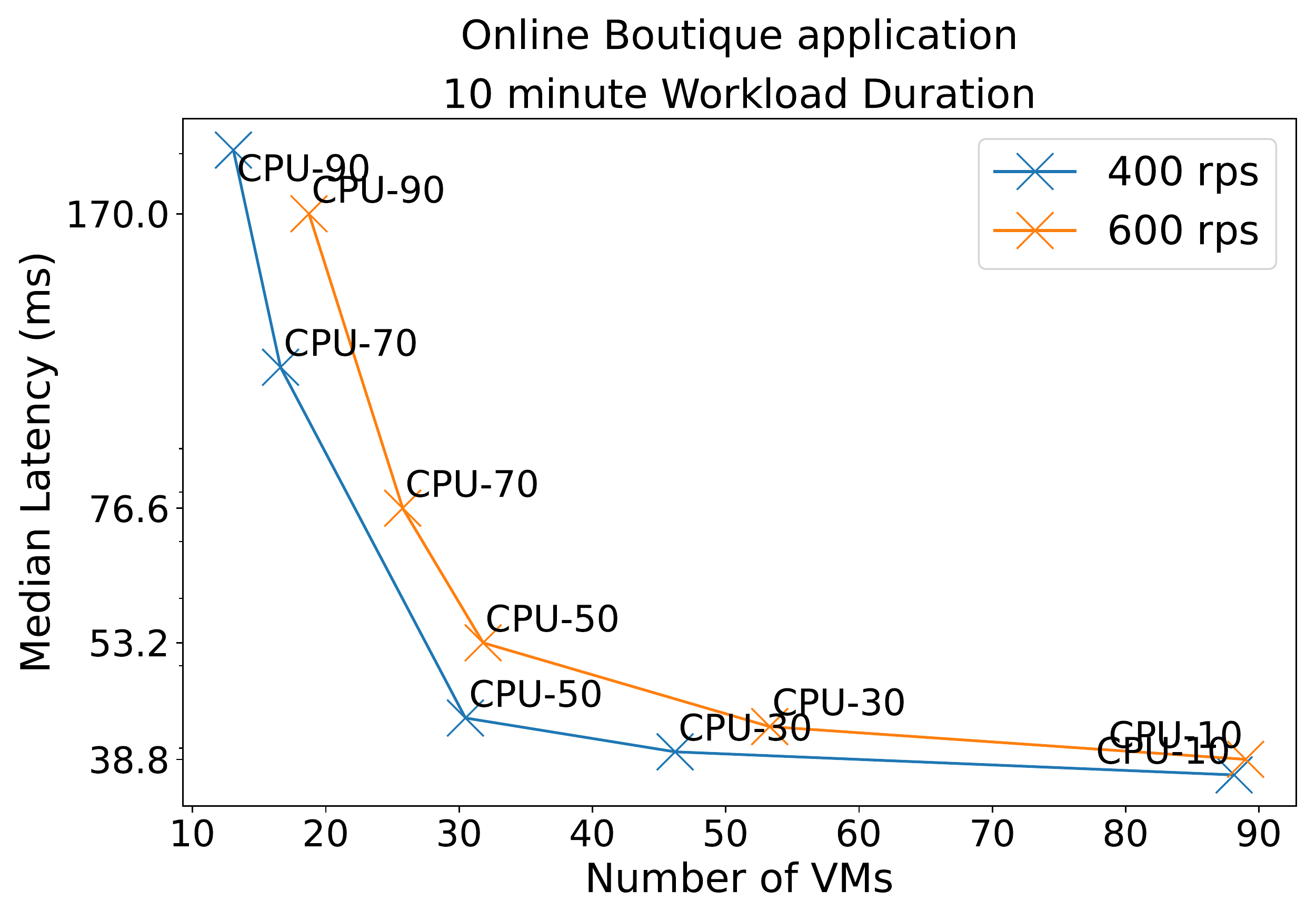}}
\subfigure[Single Span Latency vs. Number of VMs  for Online Boutique App]{\label{fig:c}\includegraphics[width=.30\textwidth]{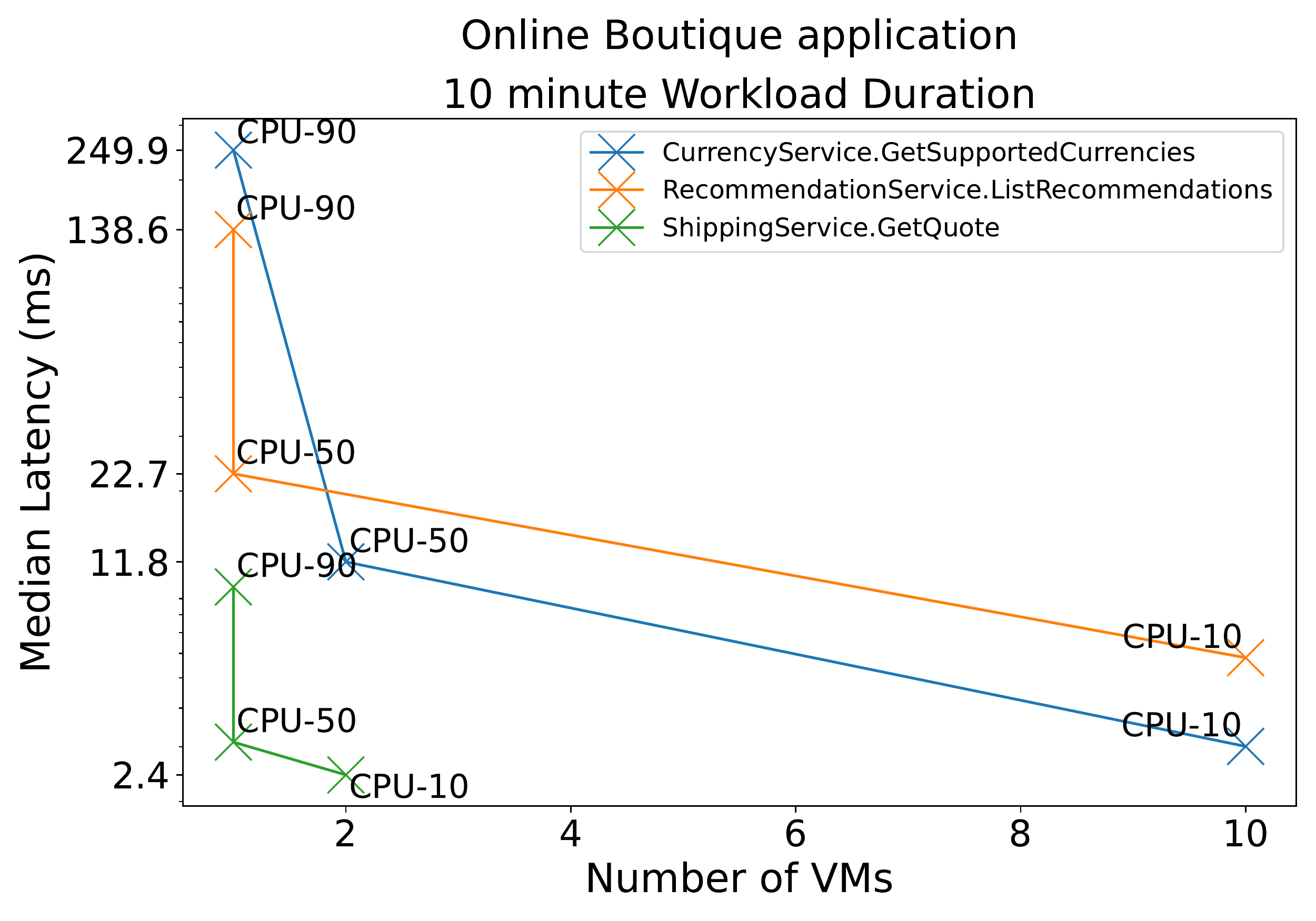}}
\caption{CPU Threshold Horizontal Pod Autoscalers}
\label{fig:hpa-evals}
\end{figure*}

\Para{Autoscaling Mechanisms in Kubernetes.} There are three primary mechanisms to automatically scale microservice application clusters managed by Kubernetes: Horizontal Pod Autoscaling (HPA), Vertical Pod Autoscaling (VPA) and Cluster Autoscaling (CA). HPA refers to automatically increasing or decreasing the number of replicas/pods of a pod deployment (in our case each deployment corresponds to one microservice), in response to changes in application or system metrics (i.e., CPU or memory utilization). VPA automatically increases the CPU or memory limits associated with one or more pods. VPA can increase or decrease either the minimum resource available for a pod (``CPU or Memory Request'') or the maximum resource available for a pod (``CPU or Memory Limit'').  CA adds or removes additional nodes (physical or virtual machines) to a cluster. While HPA and VPA are provided by the Kubernetes system, CA is provided by the underlying infrastructure/cloud provider, e.g., Google Kubernetes Engine (GKE)\cite{GKE:online}, Amazon Elastic Kubernetes Service (EKS)~\cite{EKS:online}, and Microsoft's Azure Kubernetes Service (AKS)~\cite{AKS:online}. CA in the cloud modifies the cluster size and adjusts the amount of resources rented from a cloud provider, directly affecting the dollar cost required to operate a cluster.

\Para{Autoscaling Policies in Kubernetes.} We now describe the Kubernetes autoscaling policies that leverage the autoscaling mechanisms above. We focus on HPA as it is generally more aligned with facilitating cost efficient policies in latency sensitive applications for a few technical reasons. First, VPA reduces cluster cost less generally than HPA. Reducing cost with VPA relies on decreasing the size of underlying VMs which is a procedure not generally available across cloud providers. In the public cloud, it is in a cloud tenant's interest to statically set their pod limits so that the pod can fully utilize the VM since tenants pay full price for a VM regardless of how well utilized it is. Further, when containers are scaled by increasing pod limits with VPA, they are restarted and potentially rescheduled to new VMs which meet their new limits. As a result availability and end-to-end latency can often be negatively impacted more than with a procedure such as HPA. As a result, managing a cluster without disruptions which is scaled with VPA can lead to provisioning compute resources for peak load. Since our aim is to find cost efficient autoscaling for latency sensitive services, we examine \sysname{} in the context of HPA. The technical challenges in using VPA for our goal may contribute to HPA being used much more commonly than VPA; for instance, a recent survey of Datadog's customers \cite{DdogContainerReport:online} (who in aggregate deployed 1.5 billion containers) shows that roughly 40\% of their customers have adopted HPA, while less than 1\% use VPA.

\Para{Kubernetes HPA Autoscaling} For HPA, Kubernetes natively offers utilization-based autoscalers which match the Autopilot \cite{rzadca2020autopilot} implementation. These autoscalers are implemented as a control loop with an update period (default of 15 seconds). At each period, the controller queries the resource utilization for the pods of each microservice as a percentage of the total resources requested by those pods. At any point in time, $t$, the number of desired replicas is given by the following formula where $R_t$ is the current number of replicas, $M_{t}$ is the mean utilization and $T$ is the target utilization value: $R_{t+1} = \lceil R_{t} * ( M_{t} / T ) \rceil$. Each microservice within an application is designated a target utilization value and scaled independently of others based on the above equation. The utilization here could refer to either CPU utilization or memory utilization.

\subsection{Limitations of Kubernetes HPA}
\label{subsec:dda}
Utilization-based HPA is natively available in Kubernetes and simple to deploy. But, it suffers from key issues when trying to optimize an end-to-end latency target. We highlight these using controlled experiments.

\Para{Evaluation Setup.}
We evaluate 2 open source microservice applications on GKE~\cite{GKE:online}: Online Boutique~\cite{ob:online} and Book Info~\cite{binfo:online}. For each application, we run 2 workloads; each workload has a different number of requests per second submitted to the application. For each workload, we examine the effects of using 5 different CPU utilization thresholds (10\%, 30\%, 50\%, 70\% and 90\%) for a CPU utilization threshold HPA. For our evaluations, each replica requests all of the resources of one virtual machine. Consequently, the number of replicas equals the number of VMs a cloud customer would pay for. The results of the evaluation are shown in Figure \ref{fig:hpa-evals}.

\Para{HPA Shortcomings.}  We find that HPA utilization threshold choice dramatically effects the end-to-end latency and dollar cost of an application. For example, the Online Boutique application's median latency ranges from 37 ms to 202 ms with 88 and 13 VMs rented respectively for the 400 request per second workload. The choice of utilization threshold in this case can lead to a 5.5$x$ range in latency or a 6.8$x$ range in cost. We also find that choosing the ``best'' policy, i.e., the utilization threshold which achieves a developer's latency target at the lowest cost, is difficult because:
\begin{CompactEnumerate}
    \item \textbf{The best policy depends on the application.} Given a latency target and/or cost objective there is no easy way to select a utilization threshold which meets the target without experimentation. From Figure \ref{fig:hpa-evals} we can see that maintaining a lower utilization results in lower latency, but no closed-form mapping between these two quantities is available to a user. Further, as seen in Figure \ref{fig:hpa-evals} for a fixed utilization threshold, end-to-end latencies are very different across Online Boutique and Book Info.
    \item \textbf{The best policy depends on the workload.} Depending on the workload, the appropriate policy to meet a latency target changes. In Figure \ref{fig:hpa-evals} we compare two workloads where one workload has 50\% more requests per second than the other. As seen for Online Boutique in the center graph, a 50\% CPU utilization threshold for 400 requests per second is able to achieve a latency target of 50 ms. However for the 600 request per second workload, we need to reduce the CPU utilization threshold to 30\% to meet the target.
\end{CompactEnumerate}
\section{\sysname{}'s Autoscaling Procedure}
\label{sec-design}

In terms of policy, \sysname{} aims to tackle the shortcomings of utilization-based autoscalers. Thus, \sysname{} focuses on satisfying two properties. First, we would like \sysname{} to allocate the number of pods for each microservice such that this allocation meets a developer provided latency objective cost effectively. Secondly, \sysname{} should contextualize this policy: it should alter the number of pods per microservice based on the observed workload and application.

In terms of mechanism, \sysname{} supports user-defined pod limits, for CPU, memory and disk size, per microservice. These limits dictate the extent of a VM which each pod can use. When these limits are not defined, \sysname{} defaults these resource limits to use the full extent of one core of a virtual machine for all microservices. To autoscale, \sysname{} increases or decreases the number pods for each microservice. If the aggregate pod limits of our application surpasses the resources available across VMs, \sysname{} scales up the number of VMs to facilitate allocation of these additional pods. Thus, \sysname{} is both a horizontal and a cluster autoscaler. Given the scope of this paper and based off of the trade-offs discussed in \S~\ref{subsec:dda} we chose to evaluate \sysname{} with Horizontal Pod Autoscaling rather than Vertical Pod Autoscaling. 







\subsection{A Strawman Solution}
\label{ssec:naive}
We first describe a naive approach to training an autoscaling policy for an application and workload. For the microservice autoscaling problem, we assume we have a set of \textit{workloads} which the application may encounter when exposed to users. Formally, a workload is a vector that consists of an aggregate requests per second presented to the application, followed by a probability distribution of the requests over {\em endpoints}. Each endpoint represents a distinct application URL that can be accessed by a user request and corresponds to a sequence of microservices touched by the user request through recursive RPC calls. For each workload, we would like to assign it a \textit{cluster state} which specifies the number of VMs for each application microservice. The state we choose for a given workload is chosen to maximize some \textit{reward} which encapsulates a developer's latency and cost objectives.

Under this setup, we can explore the possible states with a contextual bandit~\cite{langford2007epoch} where the workload vector is the bandit's context---or even more naively by using a multi-armed bandit for each workload. However, our setting presents challenges for such a naive algorithm. For even a small application with 10 microservices and up to 5 VMs per microservice, there are $5^{10}$ possible cluster states. Microservice applications can consist of 10-100s of microservices, which makes such an approach infeasible. Our setting further exacerbates the efficient implementation of a multi-armed bandit because the sampling time (the time to observe the reward for each sample) can be long. Specifically, this sampling time includes generating a workload and observing steady-state latencies of the generated workload. We find that to accurately estimate steady-state latency we need a sample spanning 30-60 seconds of wall-clock time (shown in Appendix Figures \ref{fig:sampling-error}.

\subsection{\sysname{}'s Solution}

\sysname{} iteratively identifies the most congested microservice and scales up this microservice alone, stopping the iteration once the latency target is met. This allows us to reduce a \textit{global} search problem for the right number of VMs for each application microservice to a \textit{local} problem that finds the right number of VMs for the most congested microservice alone. To identify the most congested microservice, we select the microservice whose CPU utilization increases most in response to the workload. Our evaluations show that this is better than other selection strategies (e.g., the microservice with the longest or most spans) because CPU utilization reflects how overworked a microservice is (\S\ref{subsec:deconstruct-cola}).

To find the right number of VMs for the most congested microservice, we employ a \textit{multi-armed bandit} algorithm~\cite{auer2002finite}, with the arms corresponding to the number of VMs. The bandit's reward is to minimize the number of allocated VMs while respecting the latency target. We provide psuedocode for our solution in Appendix \S~\ref{subsec:pseudocode} and describe its main components below.



\Para{Selecting a Congested Microservice.}
\label{subsubsec:select-service}
At each iteration of our algorithm, we must select a microservice to optimize and then select the best number of VMs for that microservice. The heuristic we use is selecting the microservice with highest increase in CPU utilization under the current workload. We implement this selection by taking the difference of each microservice's CPU utilization with and without our workload applied. At the beginning of a training iteration, we first wait for a short period of time (60 seconds) during which we do not issue any requests to the cluster. We then measure the CPU utilization without the workload for each microservice by averaging the utilization of the microservice's consituent pods. Then, we apply the workload we are optimizing for and measure the CPU utilization of each microservice with the workload applied. Intuitively, a higher utilization means that an input queue to a microservice is empty less frequently. Increasing the number of VMs in such a microservice can reduce queuing time as more VMs are available to drain the corresponding input queue. Several works have highlighted the correlation between utilization and median/tail latency in networked systems \cite{li2014tales, vulimiri2013low, patwardhan2004communication}.





\Para{Optimizing the Congested Microservice.}
\label{subsubsec:optimize-service} Given a microservice to optimize, we use a UCB1 multi-armed bandit \cite{auer2002finite} to select the best number of VMs. This choice is motivated by two properties of UCB1: it (1) explores each possible action at least once as long as the number of possible actions is less than the number of trials and (2) heavily biases actions taken in later trials to those that obtain the highest rewards---unlike a Uniform multi-armed-bandit (Chapter 1.2 of ~\cite{slivkins2019introduction}) where each action receives an equal budget of trials.

These properties are important for our us. First, we do not assume any analytic relationship between our reward and the number of VMs and hence should explore all actions to observe their associated rewards. Second, because \sysname{} runs in a cloud environment, we assume the presence of noise in latency measurements. So, it is helpful to take more samples of the optimal action we choose in order to accurately measure its expected latency. Third, we use our latency estimate for the optimal action to determine early stopping, so it is even more important that we get a good estimate for the optimal action, which UCB1's bias towards optimal actions helps with.

\Para{Reward.}
We would like a reward such that higher rewards promote cluster states which \textit{economically} achieve a latency target, $l_{target}$. Our reward is consequently defined as the combination of two weighted objectives. These objectives are to (1) obtain an observed latency, $l_{obs}$, which is less than or equal to the target latency, $l_{target}$, and (2) to use as few virtual machines, $M$, as possible. Given a latency target $l_{target}$, weight parameter $\lambda$ and cluster state $S$ the reward formulation is shown below:

\begin{equation}
\label{equation:reward-function}
\begin{aligned}
R(l_{target}, \lambda, S) \triangleq \lambda \cdot min((l_{target} - l_{obs}), 0) - M(S) \\
\end{aligned}
\end{equation}

As inputs to the reward formulation, developers enter an acceptable end-to-end latency of their choice (e.g. median, average, tail latency), $l_{target}$, in milliseconds. When our observed latency has not met the target, ($l_{obs} > l_{target}$), we are penalized by $\lambda$ for every millisecond we are above the target. For GKE standard, each VM we add to the microservice cluster incurs a cost. We compute the number of VMs for a cluster state $S$ based on the state's aggregate pod requests for CPU and memory. When deploying on GKE Autopilot \cite{GKE:online} we are charged for aggregate pod requests rather than VMs. In our case and by default when CPU and memory requests are unassigned, Autopilot automatically assigns uniform limits across microservices. As a result, for the Autopilot reward, $M(S)$ calculates the number of pods in a cluster state rather than number of VMs. The reward formulation is similar to a Lagrange multiplier optimization problem, where we seek to minimize cluster cost while constraining latency. Another way to interpret our reward is: an application developer is willing to allocate one more VM as long as doing so reduces the observed end-to-end latency by $1/\lambda$ milliseconds (if we are above our target). The reward also penalizes cluster states that needlessly allocate more VMs to microservices beyond what is needed to just meet the latency target. Lastly, the smaller the value of $\lambda$, the fewer VMs we are willing to allocate to pursue our latency target. So, \sysname{} runs Figure~\ref{fig:training-diagram}'s loop with an initially small value of $\lambda$. If the loop does not meet the latency target after a fixed number of iterations, \sysname{} retries it with an increased $\lambda$.




\Para{Workload Selection.}
\label{subsubsec:context-discretization} The set of possible workloads for an application can grow substantially for an application depending on the range of expected requests per second (RPS) and the number of external endpoints. Training an autoscaler for every possible workload that the application might encounter is impractical. Instead, application developers provide 2 inputs to \sysname{}: (1) a set of probability distributions over endpoints exposed by their microservice application, which we refer to as {\em request distributions}, and (2) an upper and lower bound for the RPS the application expects and a step size. For each request distribution, we sample the RPS range uniformly by the step size and train only for this subset of workloads in the request range (see Chapter 8.2 in \cite{slivkins2019introduction}). For example, with a lower bound of 100 RPS, an upper bound of 1000 RPS, and step size of 100 RPS, \sysname{} will train over RPS values [100, 200, ..., 1000]. As a result, developers using \sysname{} need not know how many RPS their application will handle during deployment but should be able to produce an upper and lower bound. When the observed RPS falls outside of the RPS range provided by the application developer, \sysname{} falls back gracefully to the Kubernetes HPA (\S\ref{subsec:model-robustness}).

\Para{Warm Starting.}
\label{subsubsec:warm-starting} When training, we optimize each workload with the same request distribution from our set in increasing order of number of RPS. Let us denote this set as $C_1,...C_G$ and the optimal number of VMs for $C_i$ as $S_{C_i}$. Given the knowledge that $C_i$ performed well under $S_{C_i}$ VMs, we start our search for the best cluster state for $C_{i+1}$ from $S_{C_i}$. This procedure of warm starting the training of our \sysname{} with another policy is also applied when we perform model retraining (\S\ref{subsec:model-retrain}). Further, developers may warm start with an existing autoscaler such as the Kubernetes HPA.

\section{Deploying \sysname{}'s Policies}
\label{sec-implementation}

After training, \sysname{} acts as an online cluster controller. The online controller uses \sysname's learned policies to map a request workload (i.e., requests per second concatenated with a probability distribution over endpoints) to a cluster state (i.e., number of pods/VMs for each microservice). The controller has 3 components:
\begin{CompactEnumerate}
    \item Metrics agent: Queries the cluster for the observed request workload as logged by an ingress load balancer.
    \item Horizontal pod autoscaler: Updates the number of pods by applying the learned policy to the observed request workload.
    \item Cluster autoscaler: Updates the number of VMs to reflect the updated number of pods.
\end{CompactEnumerate}

\subsection{Metrics Agent}
\Para{Querying Metrics.} We run a metrics agent which pulls aggregate request counts along with their associated endpoint names from microservice monitoring tools every minute. This gives us a minute-level RPS and request distribution. We concatenate the total RPS and a probability distribution of requests by endpoint to create our current request workload, $C_{obs}$. During deployment, we may run into an aggregate RPS that is larger than the largest trained value. In this case \sysname{} fails over to the Kubernetes HPA. An evaluation of this scenario is shown in \S\ref{subsec:model-robustness}.



\subsection{Horizontal Pod Autoscaler}
Given a learned policy from \sysname{}, we use the following procedure to interpolate VM allocations for unseen request rates and request distributions within our training range. In Appendix \S~\ref{subsec:q-theory} Proposition \ref{prop:inference-prop} we discuss why this procedure is theoretically reasonable in simple queuing networks.

\Para{Interpolating Between Trained RPS Values.} During training time, given a request distribution we learn a cluster state for a  sampled set of RPS values. At deployment, we are likely to be presented with a request per second rate we have not trained for. When this occurs, we interpolate the cluster states associated with nearby trained request per second rates as follows. Our metrics agent computes the current requests per second, $R_{obs}$. We bracket $R_{obs}$, between the 2 nearest trained RPS values, $R_{upper}$ and $R_{lower}$. Then, we weigh how close $R_{obs}$ is to these 2 training values. For example, the distance to the upper workload (same request distribution with higher RPS value) is: $d_{upper} = | R_{obs} - R_{upper} |$. We take a weighted average of the number of VMs in the associated cluster states which were learned during training, $C_{upper}$ and $C_{lower}$, using $d_{upper}$ and $d_{lower}$ as the weights.


\Para{Interpolating Between Trained Request Distributions.} We follow a similar interpolation procedure for an unseen request distribution with two modifications: (1) the distances are no longer the absolute value of difference of RPS, but rather the Euclidean distance between the unseen request distribution vector and the trained request distribution vector and (2) we perform a weighted average over all trained request distribution vectors rather than just the 2 nearest RPS values.

\subsection{Cluster Autoscaler}
\label{subsec:cluster-autoscaler}

We compute the total number of required VMs by summing over the VMs needed for each microservice. After computing the total VMs, we scale up/down as needed. When scaling up, we trigger the cluster autoscaler, then horizontal autoscaler. First the cluster autoscaler issues a request to a cloud provider to add more VMs to our cluster. Then after our request for new VMs is completed, we use the horizontal autoscaler to scale the number of pods within the cluster's microservices according to the cluster state. For scaling down, we trigger the horizontal autoscaler, then cluster autoscaler. We first scale down the pods in our cluster to the new cluster state. We then determine which VMs are currently unused by our cluster, meaning these VMs are serving no microservice deployments within our application. We cordon these VMs, drain any non-application containers (e.g. monitoring, proxies) off of the node, and delete them from our cluster.


\begin{table*}[t]
  \begin{tabular}{lllllr}
    \toprule
    Application& Replica Range  & Distinct Microservices & Autoscaled Microservices &  Operations & Source \\
    \midrule
   Simple Web Server     &  1-30      &  1   &  1  &   1  & Istio \cite{sws:online}                \\
  Book Info        &  4-60      &  4   &  4   &   1  & Istio \cite{binfo:online}                \\
  Sock Shop        &  14-100  & 14    &  9   &   5  & Weaveworks \cite{ss:online}          \\
  Online Boutique  &  11-130    & 11   &  11 &   6 & Google \cite{ob:online}              \\
  Train Ticket     &  74-700  & 64    &  63  &  10 & Fudan SE \cite{zhou2018poster}             \\
  \bottomrule
\end{tabular}
\caption{Benchmark Microservice Applications}
\label{tab:bench-applications}
\vspace{-0.75cm}
\end{table*}
\section{Evaluation}
\label{sec-evaluation}
We answer the following questions:
\begin{CompactEnumerate}
    \item Across microservice applications and workloads, how does \sysname{} compare to other autoscalers (\S\ref{subsec:comparing-policies})?
    \item Do the \sysname{}'s policies generalize well to workloads outside our trained workloads(\S\ref{subsec:comparing-policies})?
    \item Why does \sysname{} work (\S\ref{subsec:deconstruct-cola})?
    \item Can we train quickly and with acceptable cost(\S\ref{subsec:training-cost})?
\end{CompactEnumerate}

\subsection{Methodology} 
\Para{Cluster.}
For evaluations we use managed Kubernetes clusters from Google Kubernetes Engine. All microservice replicas request 600 millicpu and 2400 MB of memory. All VMs hosting these microservices belong to one node pool. The number of VMs in this node pool autoscales based on either \sysname{} or a baseline horizontal autoscaler coupled with GKE's cloud autoscaler.

\Para{Load Generator.}
\label{subsec:load-generator}
We use a VM located in the same datacenter within Google Cloud but outside of our Kubernetes cluster to issue a workload to our application using Locust ~\cite{locust:online}. For workload generation, we create a connection pool where each connection emulates an application user. Every 2 seconds, we draw an action at random from a fixed distribution of actions for each connection. This action corresponds to issuing one or more requests. The expected aggregate requests per second is proportional to the number of connections (or emulated users) within the load generator. We timeout requests on the client side after 2 seconds, which upper bounds the maximum end-to-end latency. In our evaluations, timeouts occur rarely (<.1\% of requests for most workloads) for all autoscalers other than the memory-utilization-threshold autoscalers.

\Para{Microservice Applications.}\label{subsubsec:microservices} We evaluate autoscaling policies on 5 open source applications of varying size, listed in Table \ref{tab:bench-applications}. The applications chosen vary in terms of number of microservices and endpoints and complexity of application logic. The Train Ticket microservice application ~\cite{zhou2018poster} on which we evaluate is the largest open source microservice application in terms of distinct microservices we could find. We make a few modifications to these open source microservice applications which we list in Appendix \S~\ref{subsec:app-mods}.

We considered using the popular DeathStarBench open-source benchmark suite~\cite{gan2019open}. While DeathStarBench has been used for vertical autoscaling evaluations~\cite{zhang2021sinan, sinan-gcp:online,qiu2020firm}, where the fraction of utilized CPU is scaled up/down, we find it is ill-suited to the horizontal+cluster autoscaling at the heart of \sysname{}. First, DeathStarBench applications fix both the number of deployed VMs and the assignment of each microservice to these deployed VMs. In such a setting, the number of VMs and hence the dollar cost does not change, making it impossible to show cost improvements from \sysname{} or other autoscalers. Second, the benchmark applications require a startup script to be invoked on each VM to add data, code, and libraries to the VM. Most managed Kubernetes systems (e.g., GKE, EKS, AKS) do not support custom startup scripts; they assume that applications are self contained within a container image for each microservice. Both shortcomings could be fixed with sufficient engineering, but we found it easier to use other applications instead. We note that our largest application, Train Ticket, has more microservices than the largest DeathStarBench application (64 vs. 41).






\Para{Workloads.}\label{subsubsec:workload} We evaluate on 4 workloads:
\begin{CompactEnumerate}
    \item \textbf{Constant Rate:} Requests are issued for a set of constant rates and the request distribution is identical across timesteps.
    \item \textbf{Diurnal Workload:} A predetermined schedule of requests per timestep are issued. The number of requests increases then decreases.
    \item \textbf{Unseen Request Distribution:} Requests are issued at a sequence of constant rates. The distribution of requests across endpoints is unseen in training.
    \item \textbf{Alternating Constant Rate:} Requests rates alternate between randomly sampled ``high'' and ``low'' rates.
\end{CompactEnumerate}
We were unable to find a realistic production trace to generate workloads. To address this, for all workloads, each request rate is run for only 10--15 minutes depending on the workload. For all workloads except the diurnal workload, we run a constant rate for 10 minutes. The diurnal workload is a schedule of four different rates, each run for 15 minutes, for a total of 1 hour of evaluation. Our evaluation setup stresses autoscalers in a worst-case sense by adversarially changing request rates more frequently and drastically than they would change in practice. Hence, we believe our workloads are more challenging than trace-based workloads, where autoscaling triggers much less frequently than once in 10-15 minutes~\cite{rzadca2020autopilot}.



\subsection{Autoscaling Baselines}

\Para{Kubernetes CPU-Threshold Autoscaling.} We evaluate the performance of CPU threshold autoscaling (\S\ref{subsec:kubernetes}) for a few different thresholds. For evaluations "CPU-x" denotes a CPU based autoscaler with x\% of the requested CPU as the target CPU utilization. 

\Para{Kubernetes Memory-Threshold Autoscaling.} We evaluate the performance of memory threshold autoscaling (\S\ref{subsec:kubernetes}) for a few different thresholds. For evaluations "MEM-x" denotes a Memory based autoscaler with a x\% target on the requested memory usage. 





\Para{Linear Regression.} Inspired by Ernest ~\cite{venkataraman2016ernest}, we implement a ordinary least squares (OLS) regression autoscaler. We train a OLS regression model which predicts reward (as defined for \sysname{} in Equation \ref{equation:reward-function}) given the number of replicas for the microservice, the ratio of the workload's total requests per second divided by the number of microservice replicas, and the total requests per second. We train on randomly sampled cluster states. For inference, we randomly sample 20,000 possible cluster states and use the trained linear regression model to predict the cluster state with the highest reward for the current workload. Ties for highest reward are broken by choosing the lowest cost state among tied candidates.

\Para{Bayesian Optimization.} Inspired by Cherry Pick~\cite{alipourfard2017cherrypick}, we train a Gaussian Process Regressor based on an implementation from the scikit-learn ~\cite{scikit-learn} package. For this regression, we construct a feature vector consisting of the number of replicas for each microservice concatenated with the aggregate requests per second for the cluster. The predicted variable is the reward as defined for \sysname{}, shown in Equation \ref{equation:reward-function}. We perform inference similarly to the Linear Regression model: by picking the highest reward candidate for the given input RPS among 20,000 random samples.


\Para{Deep Q-Network (DQN).} Inspired by FIRM~\cite{qiu2020firm}, we implement a Deep Q-Network algorithm, Deep Deterministic Policy Gradient~\cite{lillicrap2015continuous}, which aims to make decisions on cluster state to maximize \sysname{}'s reward in Equation \ref{equation:reward-function}. The input to the DQN is a vector of features including the requests per second as well as per-microservice CPU utilization, memory utilization and number of replicas. The output of the DQN is a vector whose size is the number of microservices in our cluster. Each cell in this vector is a value in $[-1,1]$ which we map linearly to the minimum and maximum replica range for each service. During inference time, we provide a workload vector including the requests per second and per-microservice CPU, memory and replicas; we then allocate cluster resources based on the DQN's output.

\Para{\sysname{} Policies.}
In our evaluations we consider two \sysname{} policies: "\sysname{}-50" and "\sysname{}-tail-100". The "\sysname{}-50" policy is trained to scale microservice clusters with a median latency objective of 50 ms. The "\sysname{}-tail-100" policy targets 90\%ile latency below 100ms. We include a full list of training hyperparameters and associated values for these policies in Table \ref{tab:hyperparams}.

\Para{Terminology.} We define the {\em closest autoscaling policy} as the policy whose latency is closest in absolute value to \sysname{}. Secondly, we define the {\em closest objective matching autoscaling policy} as the most economical baseline  policy which meets the median/tail latency objective. We use {\em in sample} to refer to request contexts that \sysname{} and machine learned autoscalers have explicitly trained for and {\em out of sample} to refer to request contexts that these models have not observed in training, thus requiring \sysname{} to interpolate. If \sysname{} meets its latency target we compare directly with the closest objective matching autoscaling policy. If \sysname{} does not meets its target we compare with the closest autoscaling policy.


\begin{figure*}%

\subfigure{%
  \label{fig:ob-fr-500}%
  \includegraphics[width=0.33\textwidth]{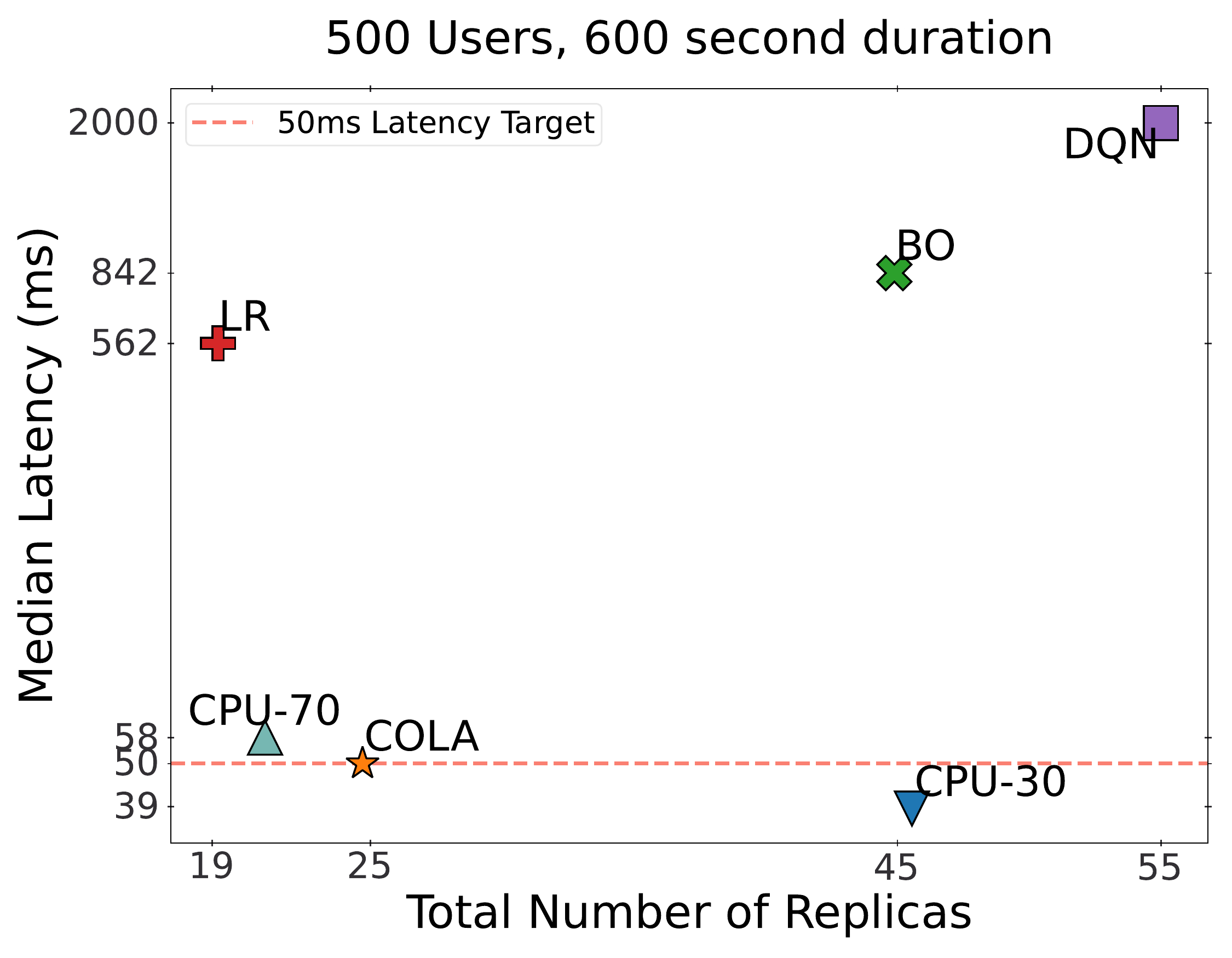}%
}%
\hspace*{\fill}
\subfigure{
  \label{fig:ob-fr-800}%
  \includegraphics[width=0.33\textwidth]{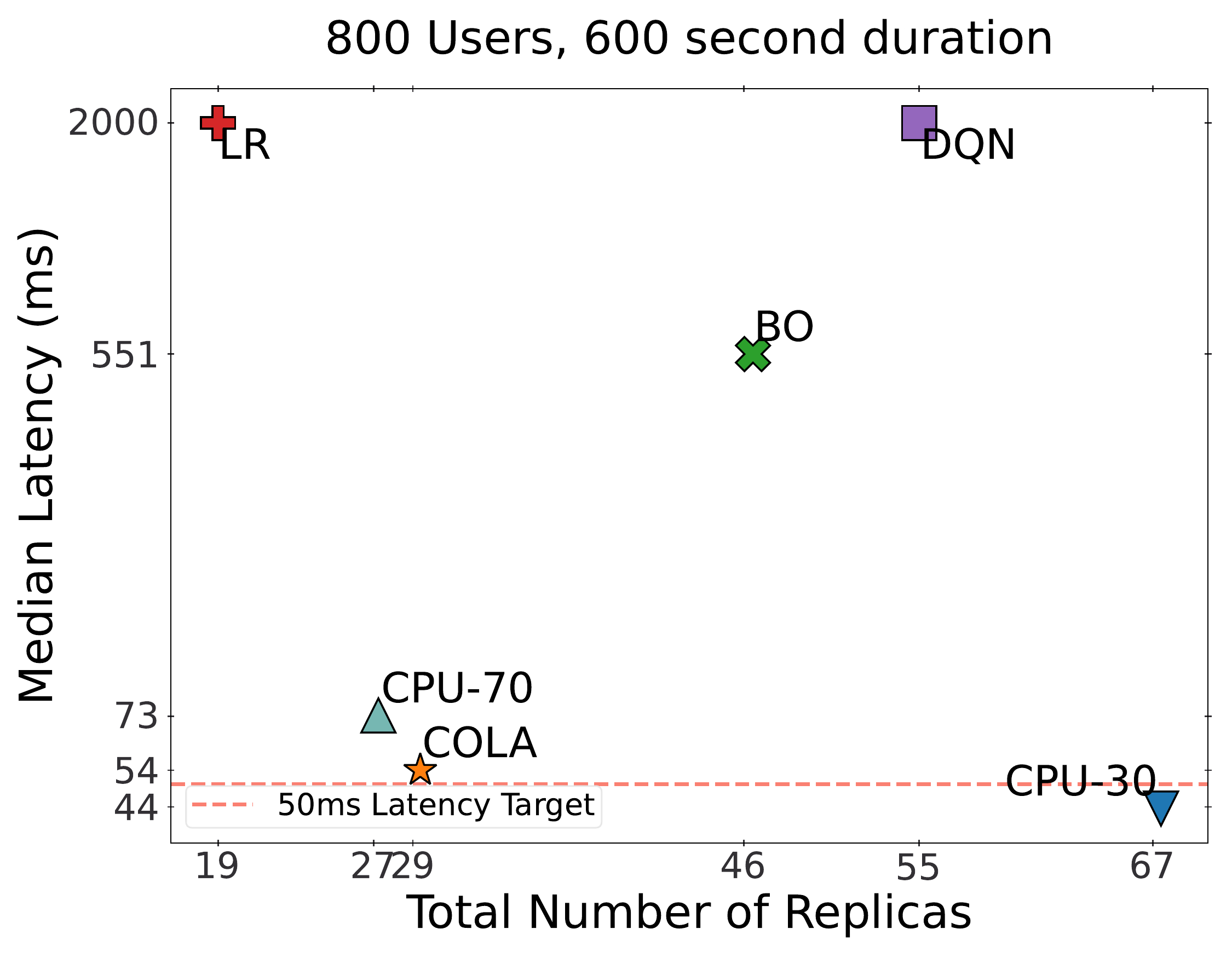}%
}%
\hspace*{\fill}
\subfigure{
  \label{fig:ob-fr-tail-600}%
  \includegraphics[width=0.33\textwidth]{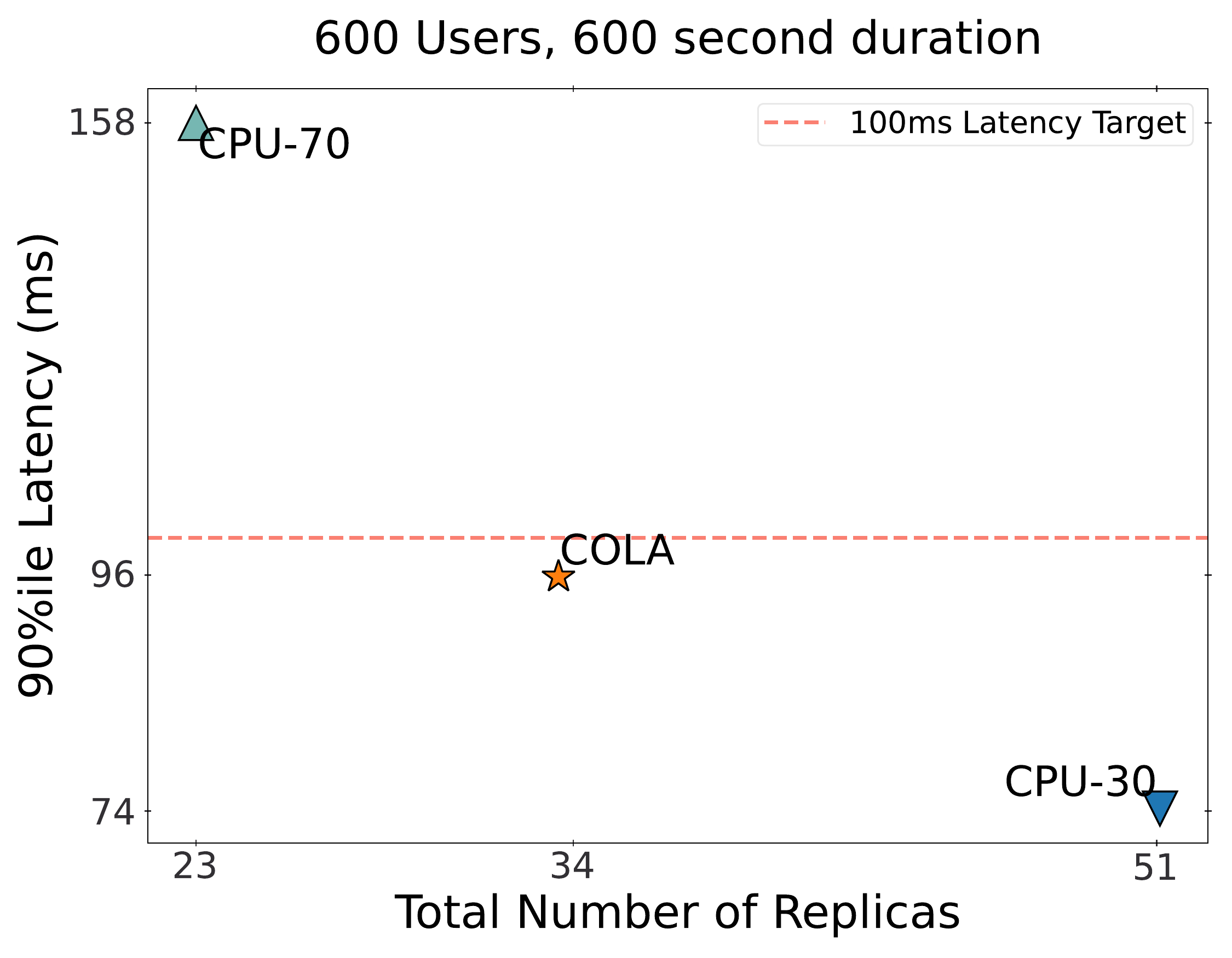}%
}
\captionsetup{justification=centering}
\vspace{-0.5cm}
\caption{Online Boutique, Constant Rate Evaluation on Median Latency (Left, Middle) and 90\%ile Latency}\label{fig:ob-fr}

\subfigure{%
  \label{fig:tt-fr-250}%
  \includegraphics[width=0.33\textwidth]{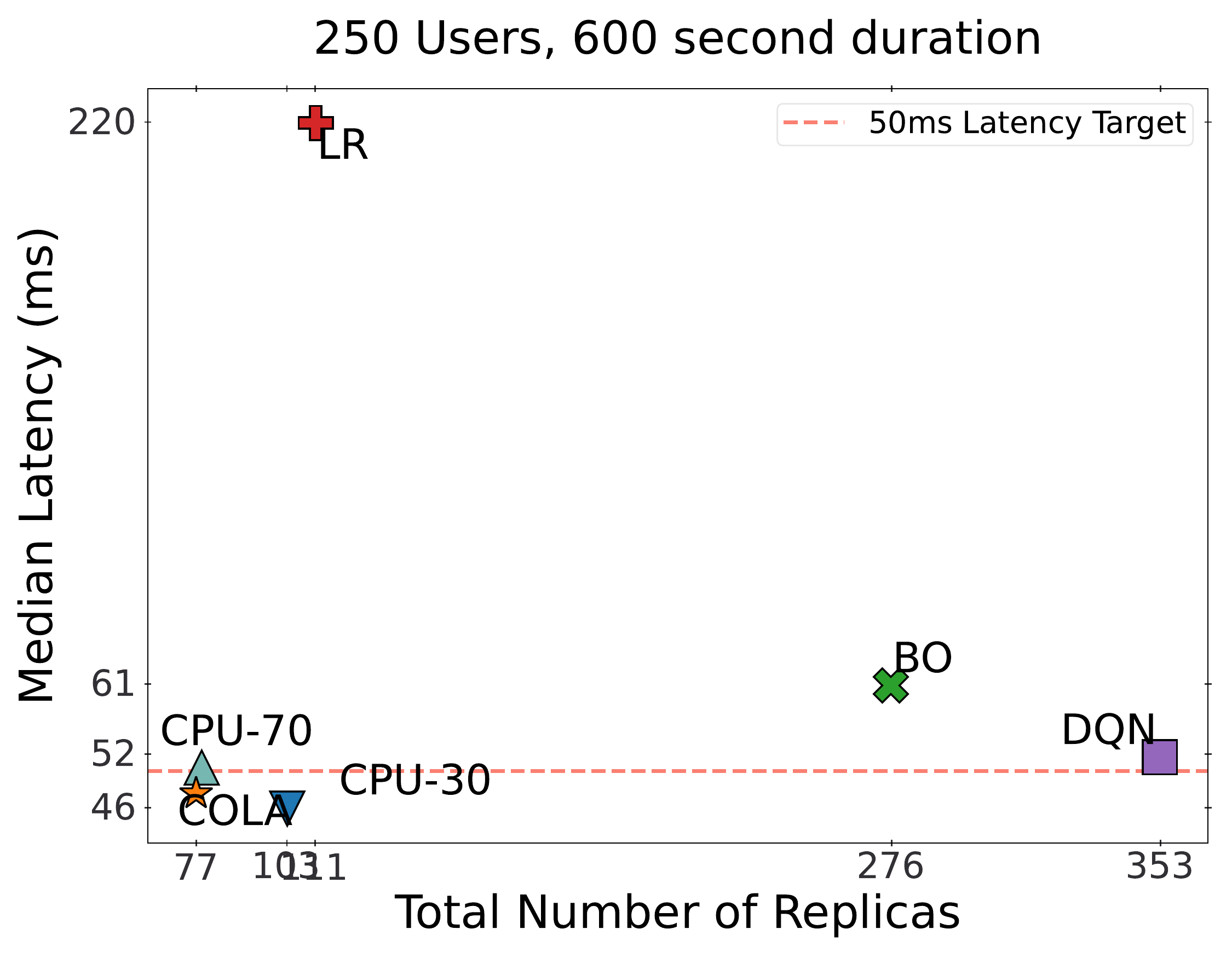}%
}%
\hspace*{\fill}
\subfigure{
  \label{fig:tt-fr-600}%
  \includegraphics[width=0.33\textwidth]{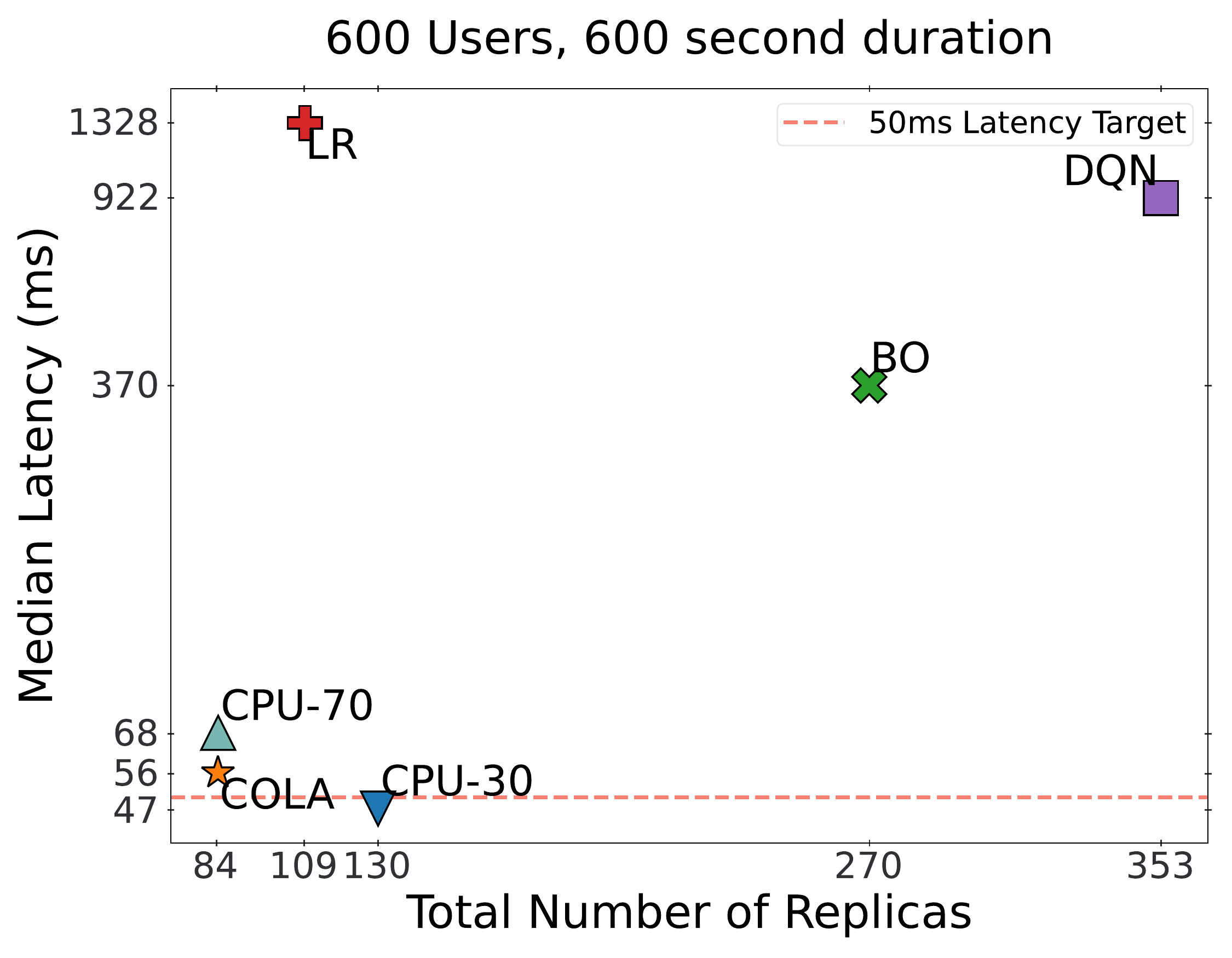}%
}%
\hspace*{\fill}
\subfigure{
  \label{fig:tt-fr-tail-250}%
  \includegraphics[width=0.33\textwidth]{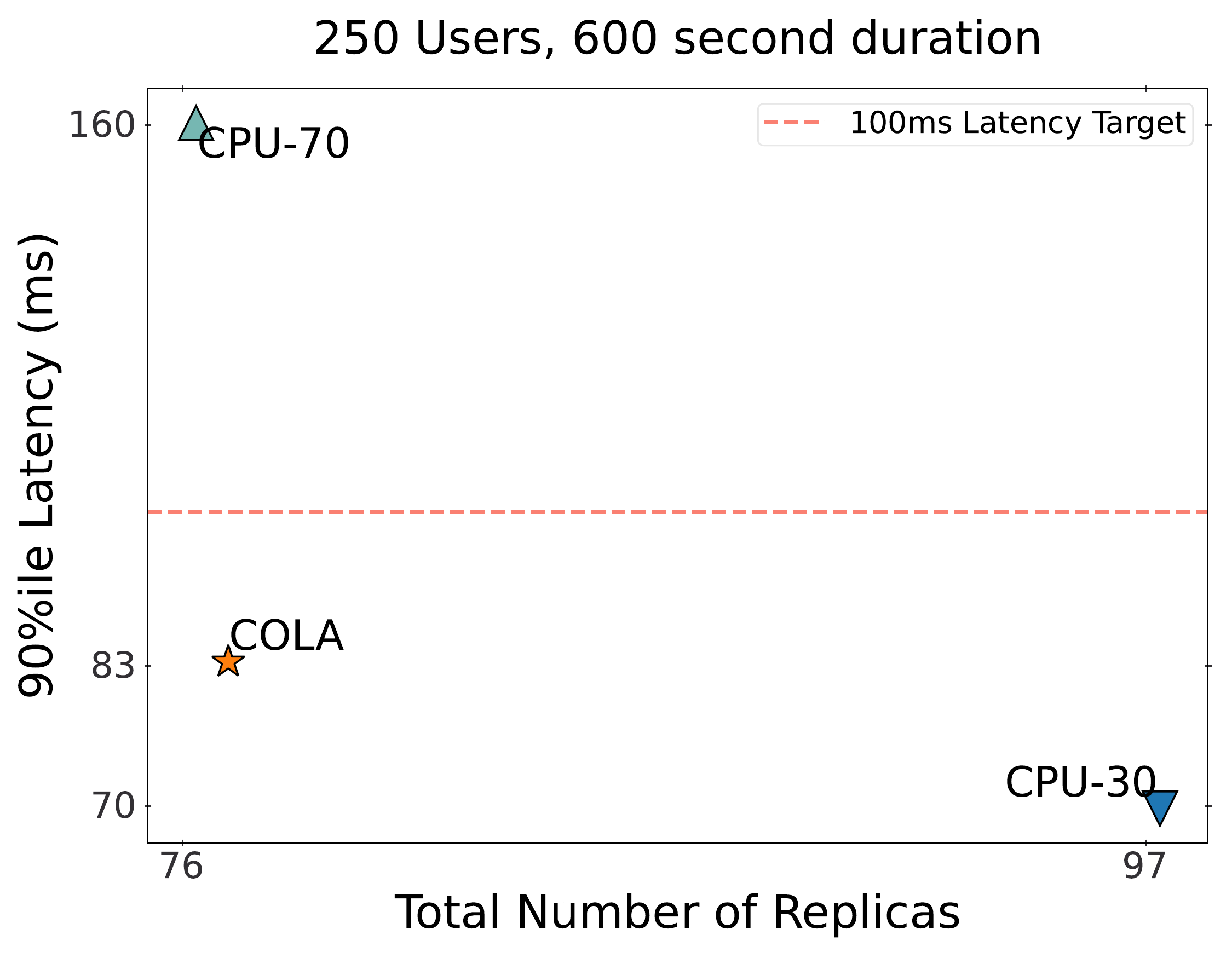}%
}

\captionsetup{justification=centering}
\vspace{-0.5cm}
\caption{Train Ticket, Constant Rate Evaluation on Median Latency (Left, Middle) and 90\%ile Latency (Right)}\label{fig:tt-fr}

\subfigure{%
  \label{fig:bi-fr-ramp-oos}%
  \includegraphics[width=0.33\textwidth]{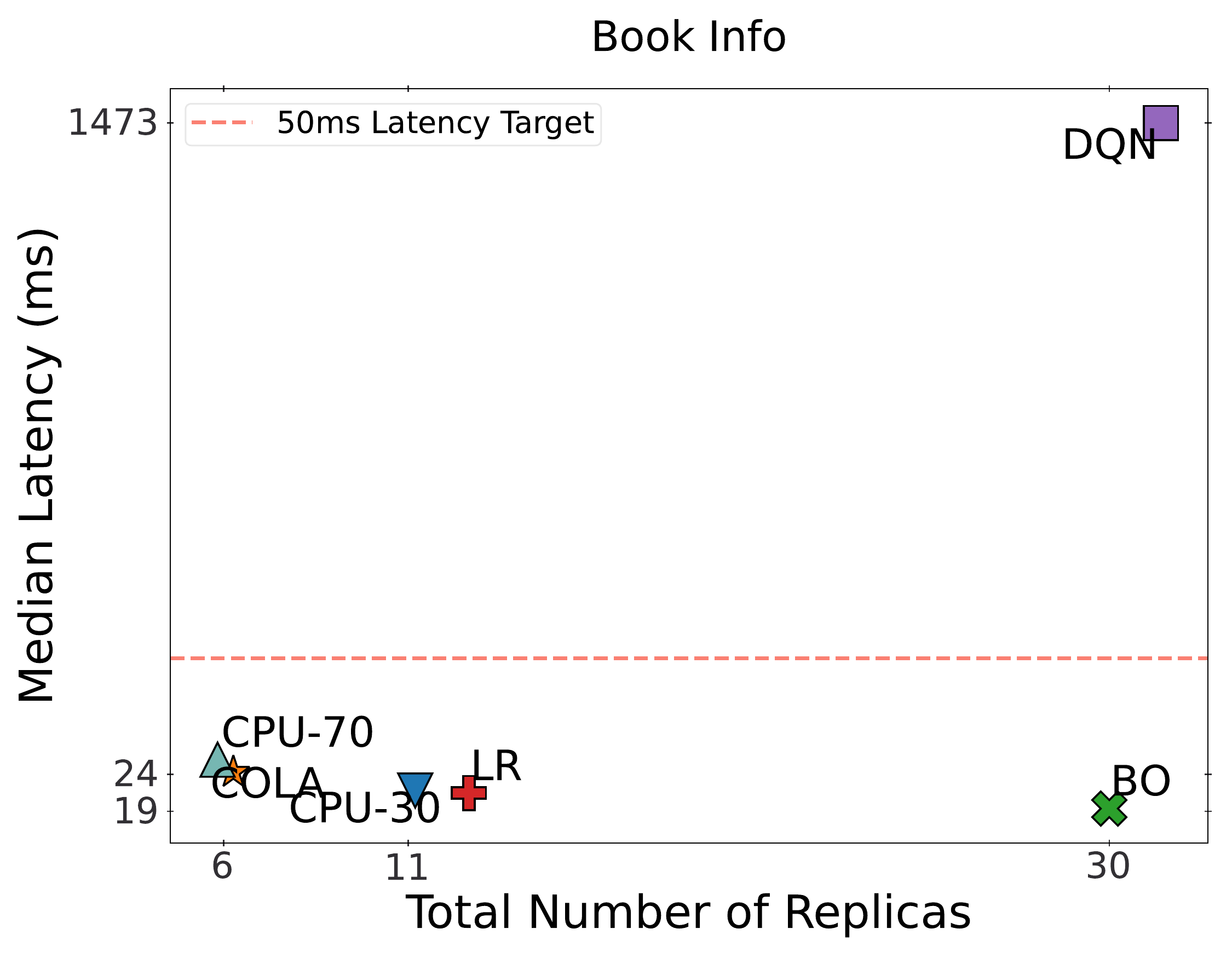}%
}%
\hspace*{\fill}
\subfigure{
  \label{fig:ob-fr-ramp-oos}%
  \includegraphics[width=0.33\textwidth]{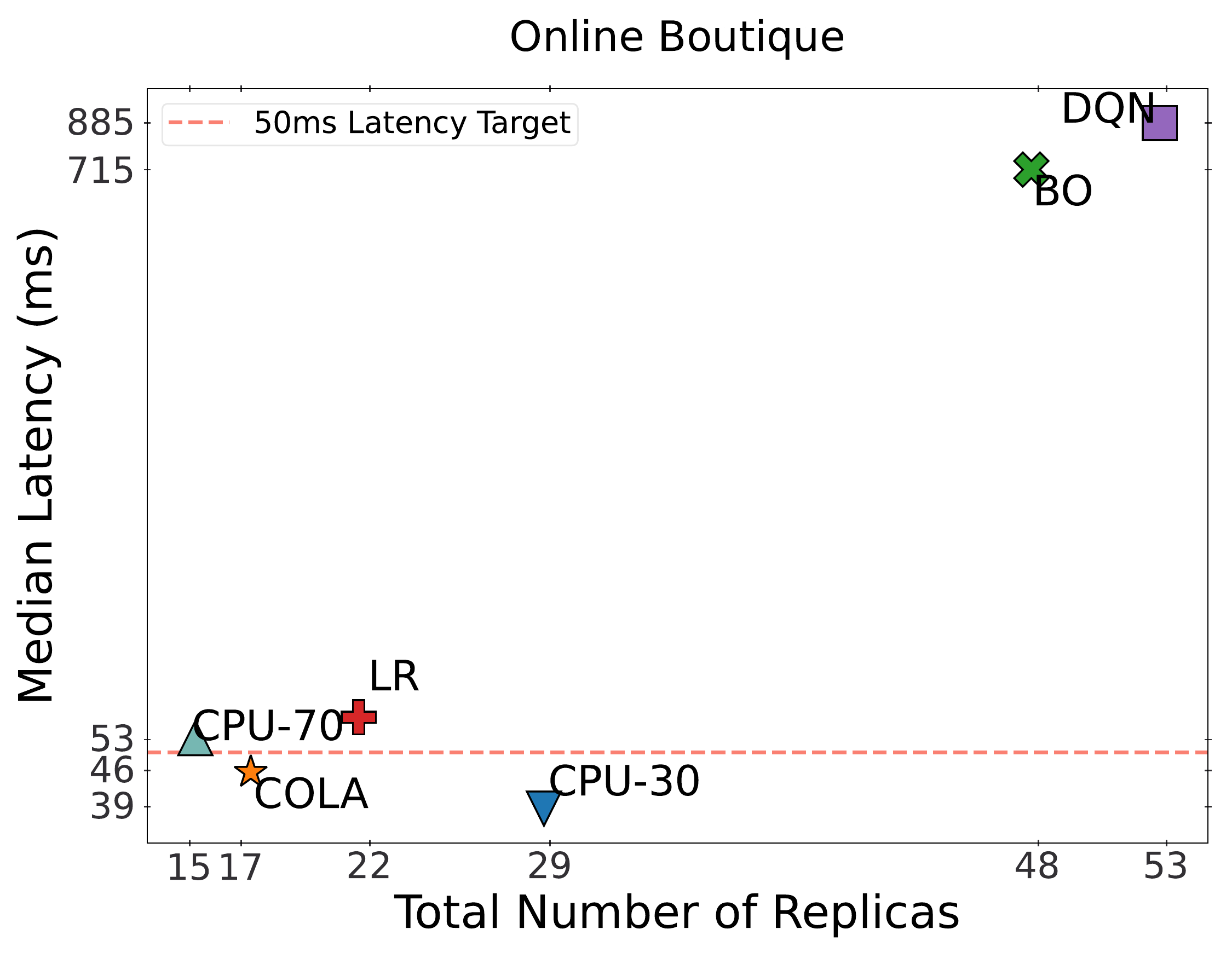}%
}%
\hspace*{\fill}
\subfigure{
  \label{fig:tt-fr-ramp-oos}%
  \includegraphics[width=0.33\textwidth]{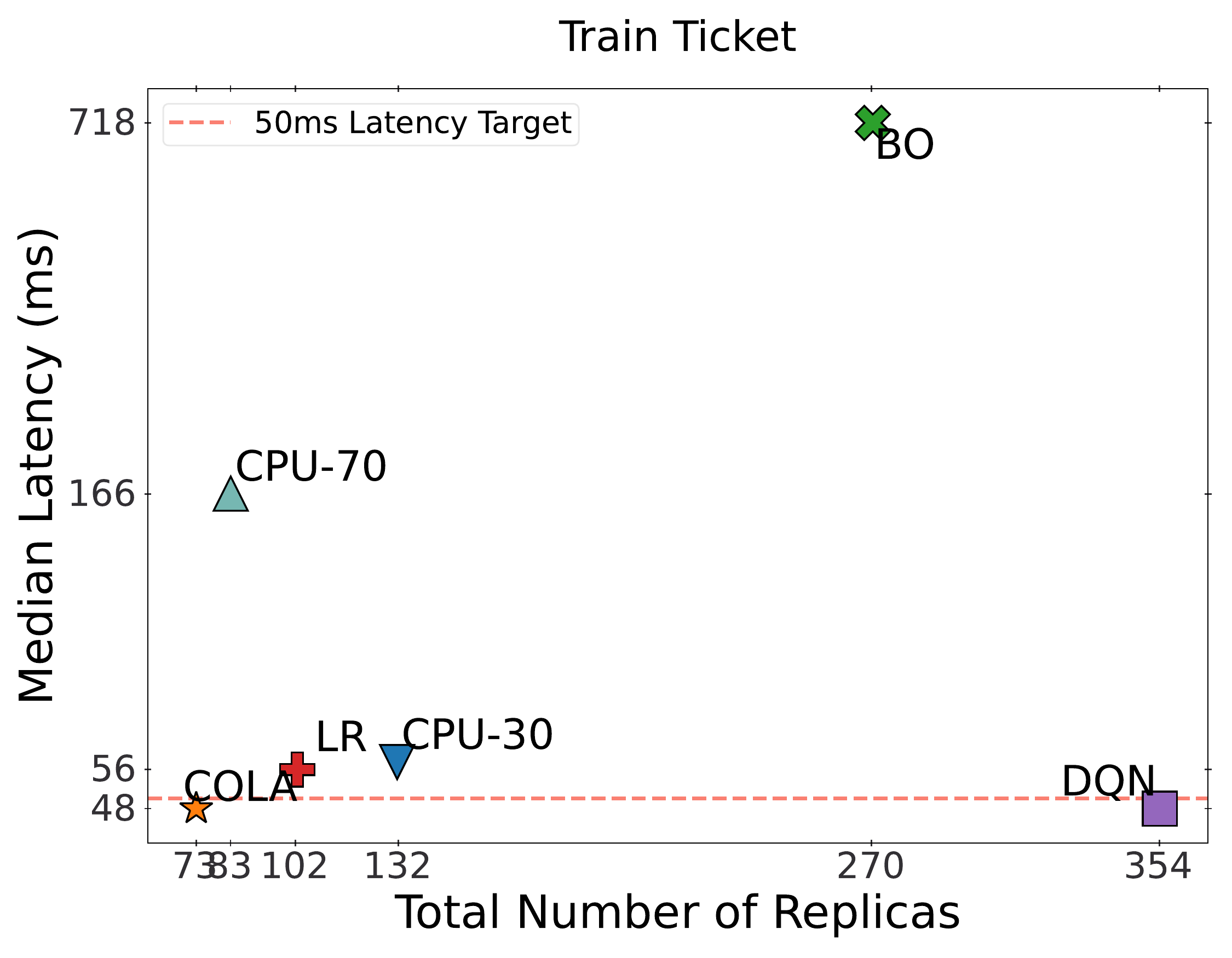}%
}
\captionsetup{justification=centering}
\vspace{-0.5cm}
\caption{Out of Sample Diurnal Workload for Book Info (Left), Online Boutique (Middle), Train Ticket (Right)}\label{fig:oos-diurnal}

\subfigure{%
  \label{fig:ob-req-dist-200}%
  \includegraphics[width=0.31\textwidth]{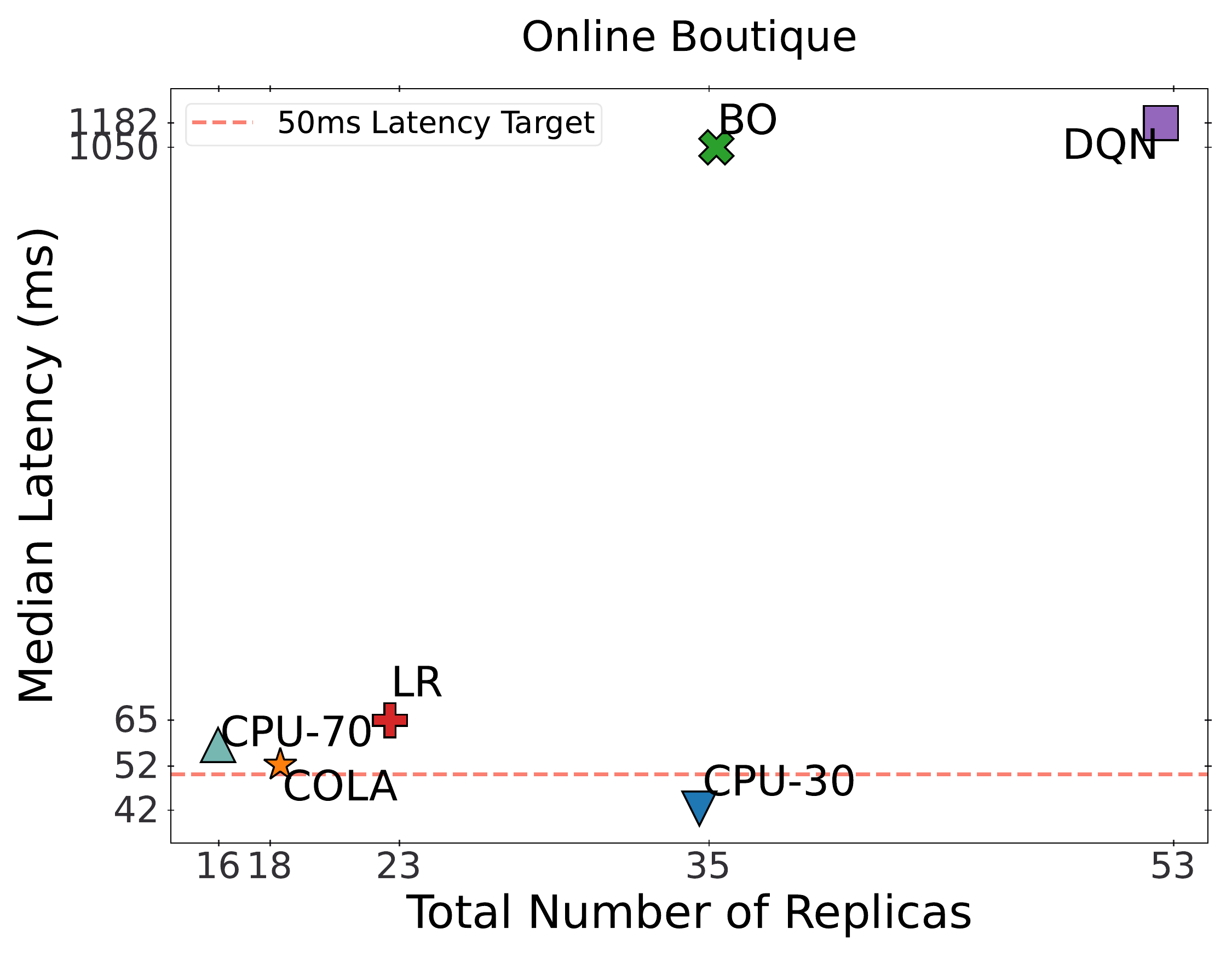}%
}%
\hspace*{\fill}
\subfigure{
  \label{fig:ob-req-dist-300}%
  \includegraphics[width=0.31\textwidth]{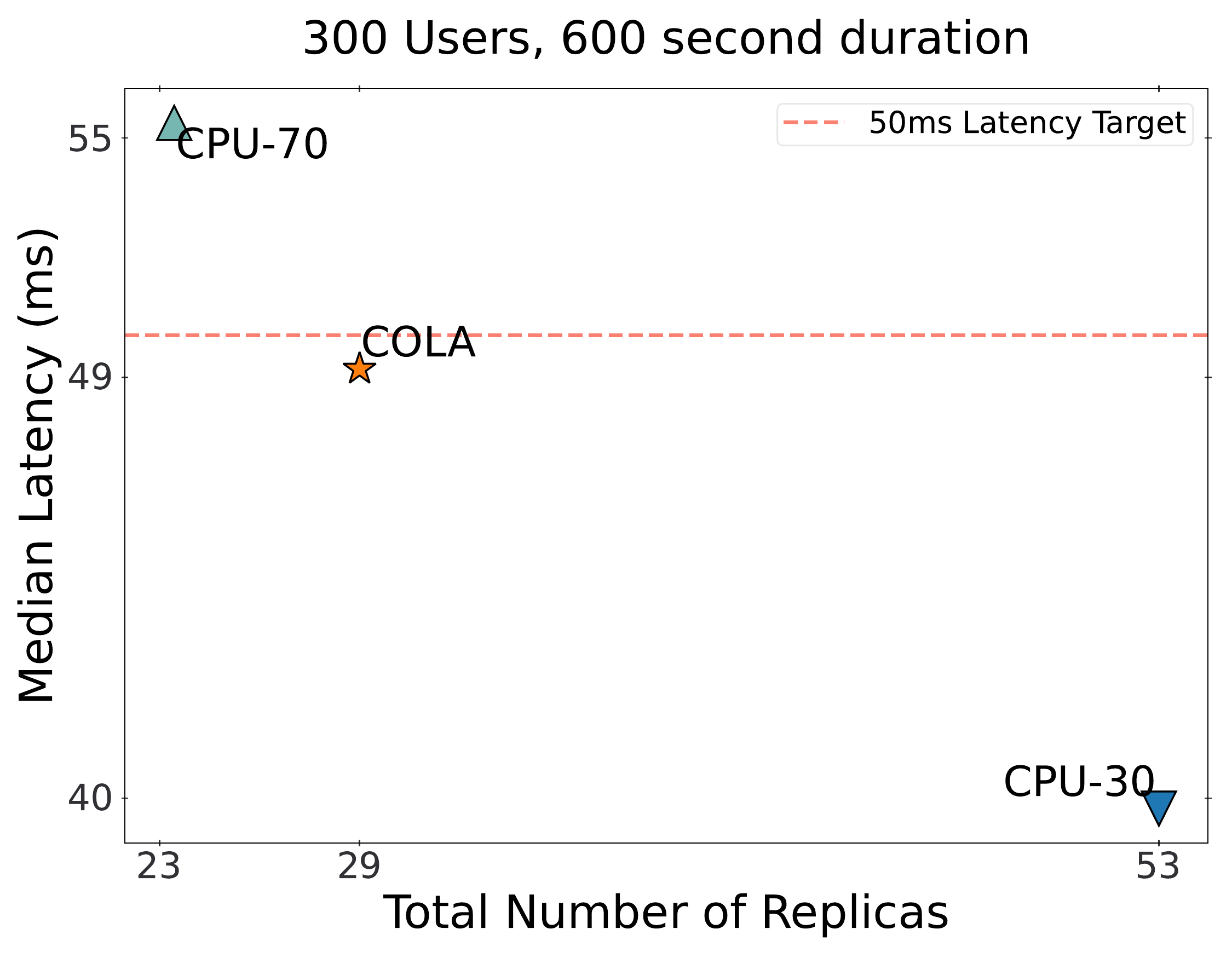}%
}%
\hspace*{\fill}
\subfigure{
  \label{fig:ss-ramp-hilo}%
  \includegraphics[width=0.31\textwidth]{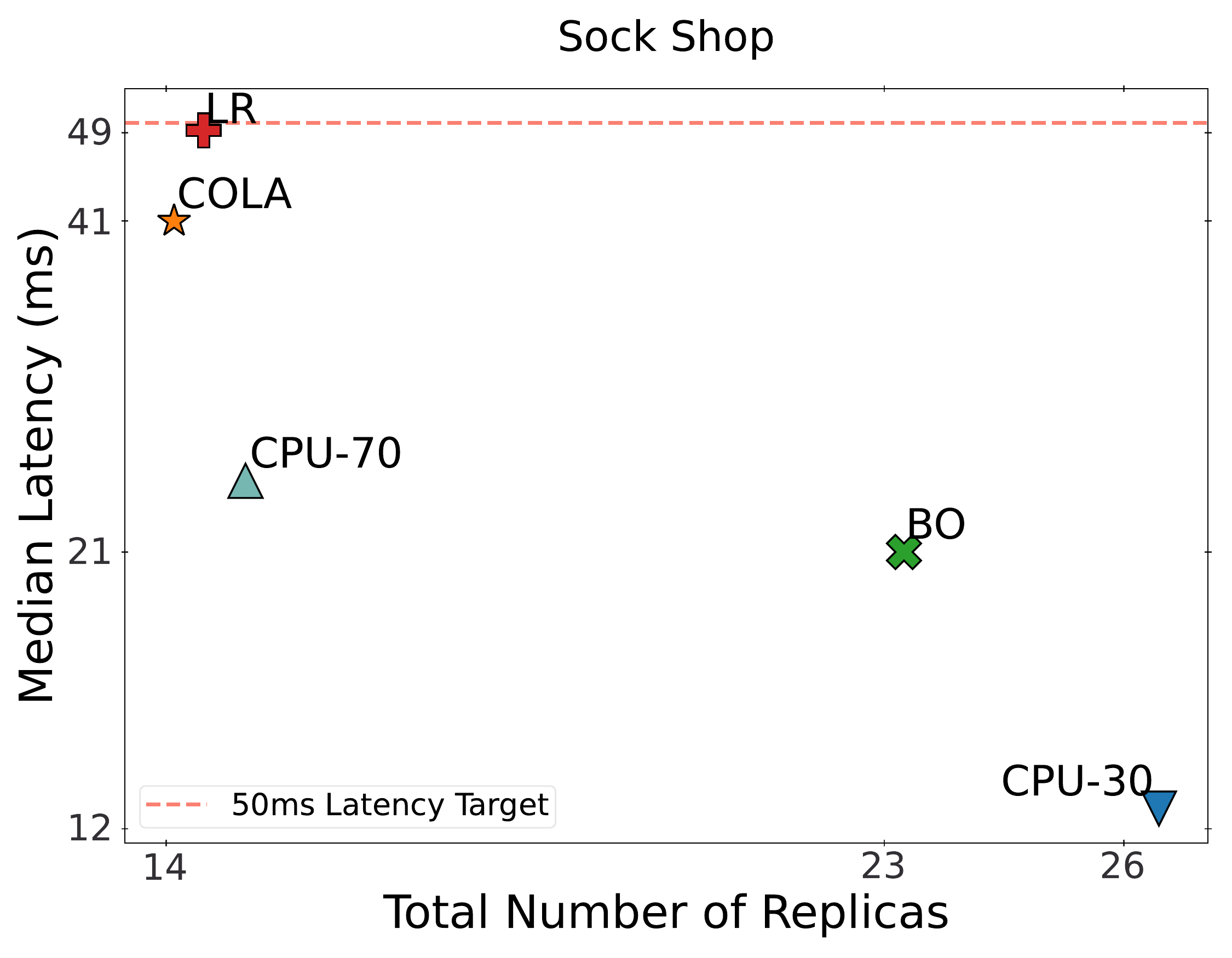}%
}
\captionsetup{justification=centering}
\vspace{-0.5cm}
\caption{In Sample Diurnal Workload for Online Boutique (Left), Online Boutique Unseen Request Distribution (Left, Middle) and Sock Shop Alternating Constant Rate (Right)}\label{fig:req-dist-acr}

\end{figure*}

\begin{figure*}[!t]

\subfigure{%
  \includegraphics[width=0.31\textwidth]{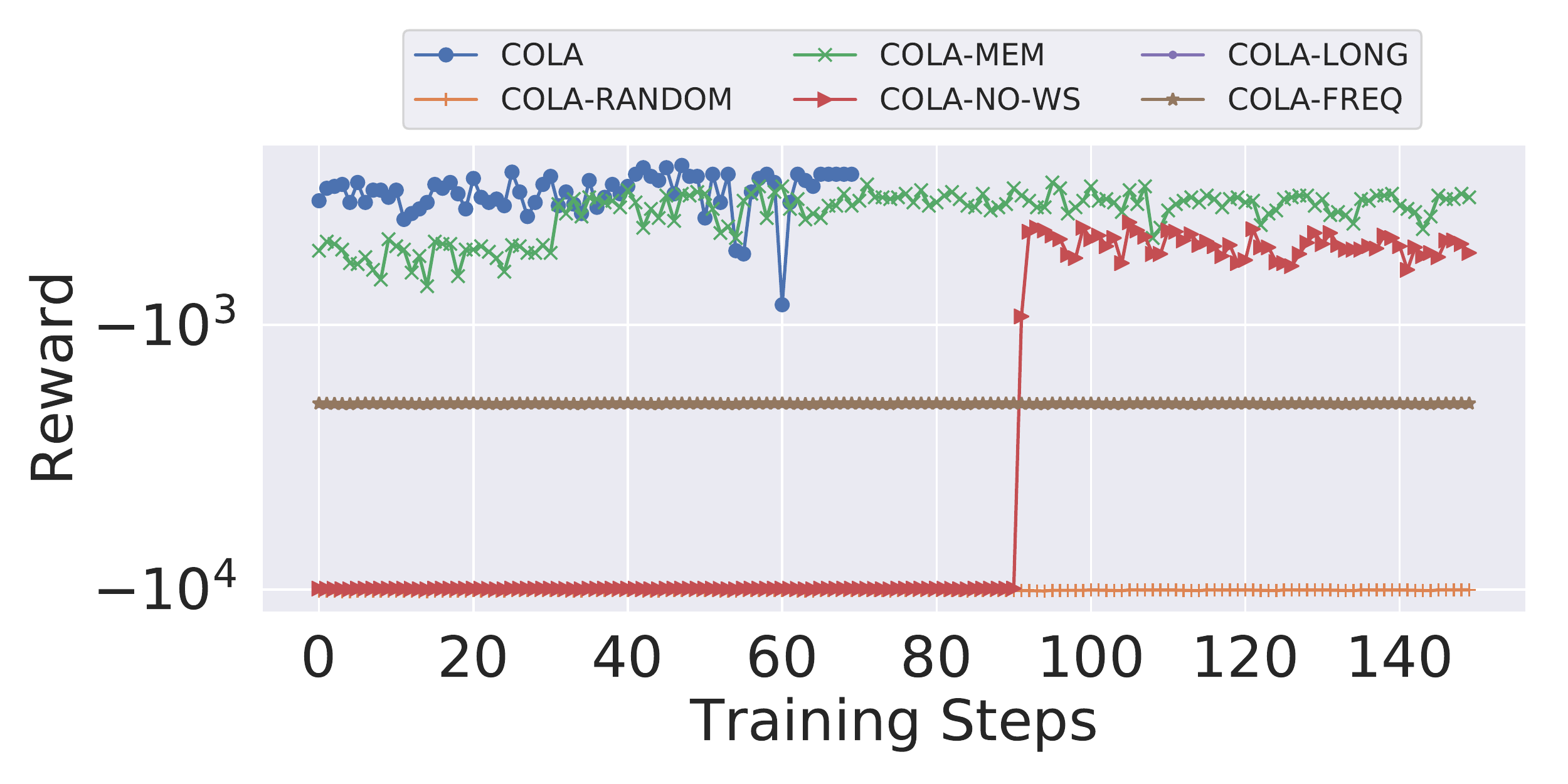}%
}  %
\hspace*{\fill}
\subfigure{
  \includegraphics[width=0.31\textwidth]{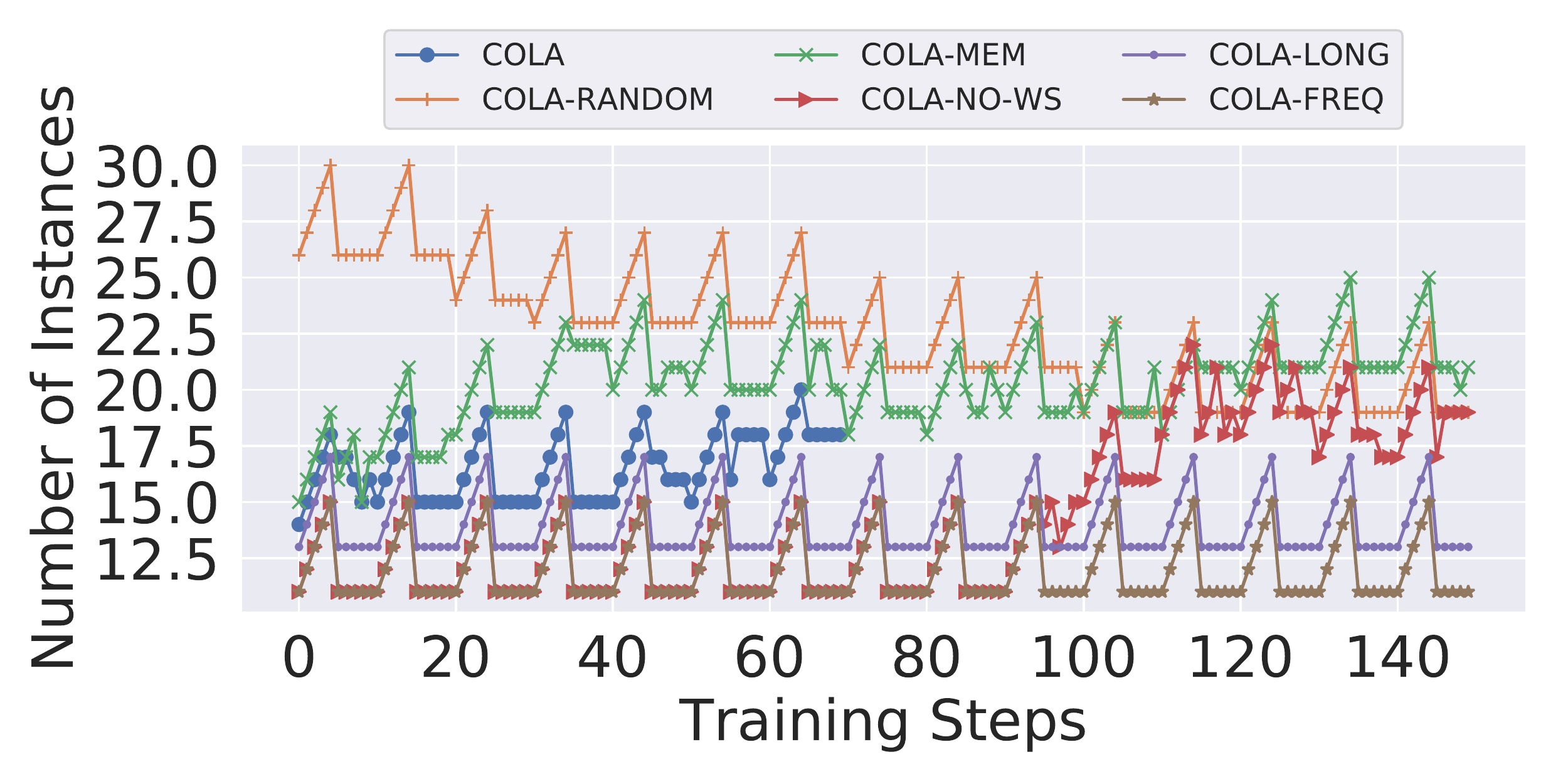}%
}  %
\hspace*{\fill}
\subfigure{
  \includegraphics[width=0.31\textwidth]{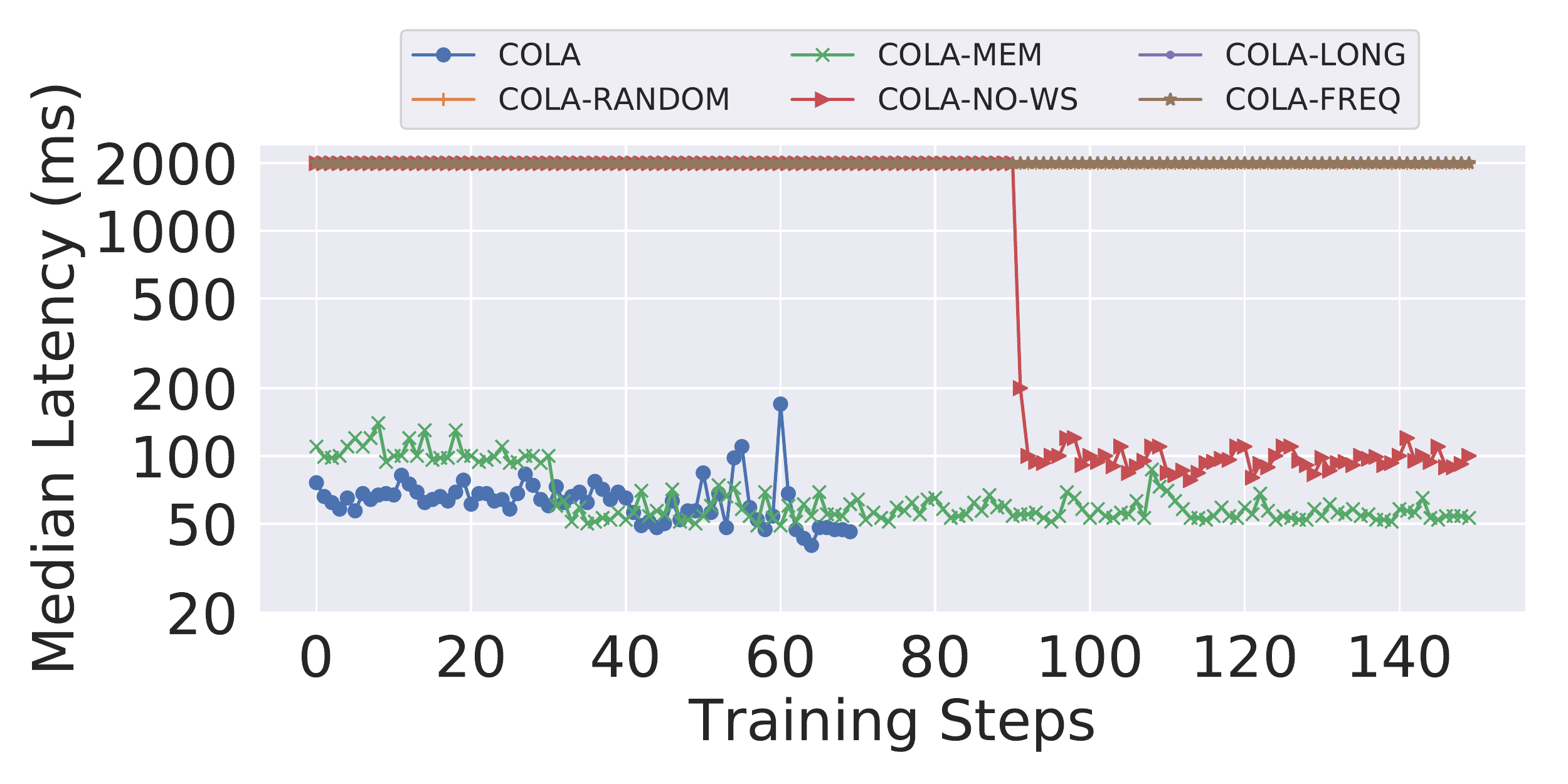}%
}
\captionsetup{justification=centering}
\caption{Online Boutique Training Trajectory with different Service Selection Strategies, 400 Users}\label{fig:ob-training-service-selection}

\end{figure*}

\subsection{Comparing Autoscaling Policies}\label{subsec:comparing-policies}
We examine the performance of \sysname{} compared with baselines across the 5 microservices introduced in \S\ref{subsubsec:microservices} and 4 workloads introduced in \S\ref{subsubsec:workload}. For all applications we report evaluation numbers where half of the evaluated request per second rates are in-sample and half are out-of-sample. We summarize results and include full tabular results for all applications in Appendix tables~\ref{tab:bi-fr-tabular}--\ref{tab:tt-ap-tabular}.

\Para{Constant Rate Workload.} On all applications, \sysname{} provides the lowest dollar cost autoscaling policy to meet our latency targets. We train and evaluate autoscalers on two latency targets, for 50 ms median latency and 100 ms tail latency. For the 50 ms median latency target, \sysname{} costs \textbf{22.1\%} less than the next cheapest policy. When evaluated on a 100 ms tail latency target, \sysname{} costs \textbf{20.8\%} less than the next cheapest policy. Across applications, workloads, and latency targets the closest autoscaling policy to \sysname{} varies. This uncertainty in terms of which policy works best for a given setting indicates a key benefit of training an autoscaling policy. We show a few of these evaluations for the Online Boutique and Train Ticket applications in Figures \ref{fig:ob-fr}-\ref{fig:tt-fr} and detail a full set of tabular results in Appendix Tables \ref{tab:bi-fr-tabular}--\ref{tab:ob-tail-fr-tabular}.

\Para{Diurnal Workload.} We evaluate autoscalers under a diurnal workload which consists of 5 different request per second rates run for 10 minutes each (a total of 3000 seconds). Both an in-sample and out-of-sample diurnal workload are run for each application. \sysname{} produces the cheapest autoscaling policy for this diurnal workload for 3 out of 4 applications. On average \sysname{} reduces costs by \textbf{20.1\%} compared to the next cheapest policy. On the Book Info application \sysname{} is outperformed by the CPU-70 policy, costing \textbf{5\%} more than this policy. We show a few diurnal workloads in Figures \ref{fig:oos-diurnal}-\ref{fig:ob-req-dist-200} and include full results in Appendix Tables \ref{tab:ob-tail-diurnal-tabular}--\ref{tab:ss-diurnal-tabular}.

\Para{Unseen Request Distribution and Alternating Constant Rate Workloads.} We evaluate \sysname{} and CPU autoscalers on two additional workloads: (1) Unseen Request Distribution where the distribution of requests over operations is different from that seen at training time and (2) Alternating Constant Rate where the requests per second jumps immediately between a high and low request rate. For the Online Boutique application we train \sysname{} on two different distributions of request workloads, one with a low frequency of purchasing an item and one with a frequency of purchases 3x higher, and evaluate on a third distribution with 2x the frequency of purchased orders. Results for this evaluation are shown in Figure \ref{fig:ob-req-dist-300}. We find that \sysname{} is able to reduce the cost of our cluster by \textbf{45.2\%} when compared to the CPU-30 autoscaler, the next cheapest policy, for this unseen request distribution. On the Sock Shop application we evaluate an alternating workload which switches between high and low request rates shown in Figure \ref{fig:ss-ramp-hilo}. We find \sysname{} is able to reduce the cost of our cluster by \textbf{2.6\%} compared to the next cheapest policy, the CPU-70 autoscaler.

\Para{Performance across Cluster Architectures} We evaluate \sysname{} across a variety of cluster architectures. First, we alter the underlying VM sizes which host containers in our cluster. We find that \sysname{} reduces cluster cost over the next best CPU or ML baseline by \textbf{21.6\%} for single core VMs, \textbf{12.4\%} for two core VMs, and \textbf{10.3\%} for four core VMs. As VM machine size increases granularity in resource allocation decisions play less of an effect on cost (e.g. in the extreme case where 1 large VM can host all containers, there is no way to reduce the cost a cloud tenant pays) and \sysname{} provides less benefit over other policies. Secondly, we evaluate \sysname{} on the newly introduced GKE Autopilot clusters, where users do not manage VMs but instead pay for only the aggregate pod requests made by launched containers. When compared with CPU baselines, we find that \sysname{} reduces cost by \textbf{45.4\%} on Autopilot (Tables \ref{tab:bi-ap-tabular}-\ref{tab:tt-ap-tabular}). Overall, we find that \sysname{} consistently reduces costs across a set of different cluster configurations and offers the most benefits when deployed on Autopilot clusters.



\begin{figure}[!t]
	\centering
	\includegraphics[width=.5\textwidth]{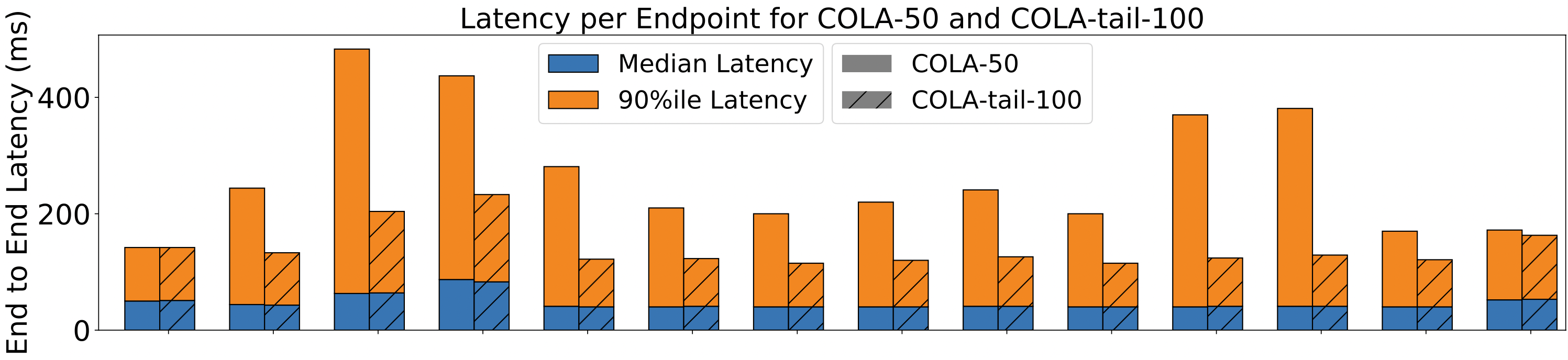}
	\caption{\sysname{} Endpoint Response Times for Median and Tail Latency Policies. Each set of bars is a distinct URL the application exposes.}
	\label{fig:median-latency-endpoint-dist}
	\vspace{-0.5cm}
\end{figure}

\subsection{Deconstructing \sysname{}'s Gains}\label{subsec:deconstruct-cola}

\Para{Service Selection.} We find that selecting microservices for scaling by CPU utilization is crucial. In Figure \ref{fig:ob-training-service-selection} we compare replacing this selection heuristic with several other heuristics during training for the Online Boutique application. These heuristics are selecting services by memory utilization (COLA-MEM), randomly (COLA-RANDOM), without the warm starting optimization (COLA-WS), by the microservice with longest average span duration (COLA-LONG) and most frequently accessed microservice (COLA-FREQ). We find that these other heuristics take 1.58x--2.06x the amount of samples to train Online Boutique and 1.6x--12x the amount of samples to train Book Info (shown separately in Appendix Figure \ref{fig:bi-training-trajectory-ablation}). Also, in most cases, replacing CPU utilization based service selection leads to policies which converge to a median latency above our target. This illustrates the importance of using domain knowledge in designing the right heuristic to train systems such as \sysname{}. By contrast, applying machine learning to the autoscaling problem without the right heuristics, as our ML-based baselines do, yields poorer results. This can be seen in comparisons between \sysname{} and ML-based baselines in Appendix Figures \ref{fig:sws-training-trajectory}-\ref{fig:ob-training-trajectory} where ML-based baselines often hit a wall in terms of reward or utilize drastically more resources to find policies which match our latency constraint. We take a deeper look at the gap between \sysname{} and ML-based baselines in \S~\ref{subsec:la-shortcomings}.



\begin{figure*}[!t]
	\centering
	\includegraphics[width=.82\textwidth]{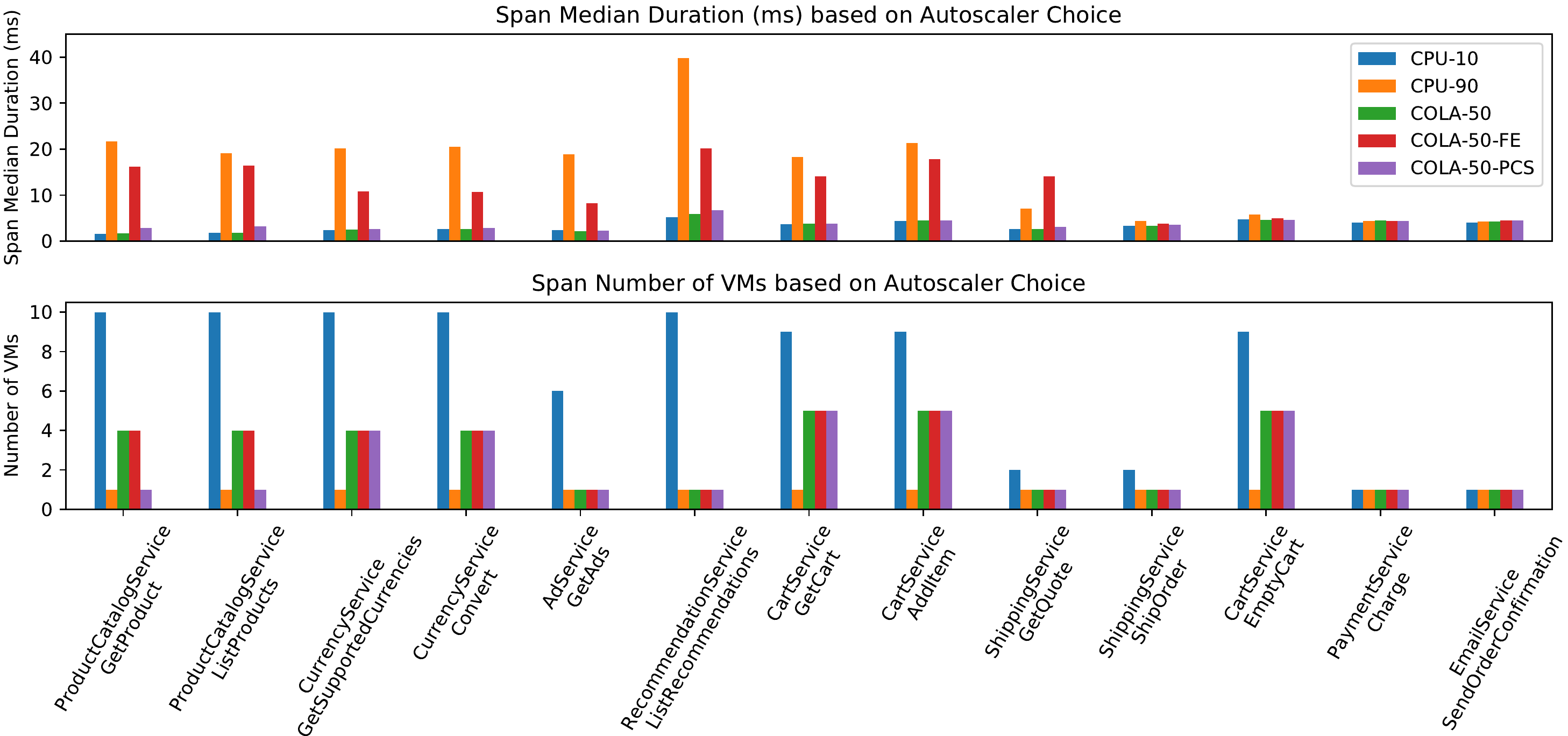}
    \caption{Microservice Span Durations and Number of VMs by Autoscaler Choice.}
	\label{fig:span-duration-by-autoscaler}
	\vspace{-0.45cm}
\end{figure*}
\Para{Scaling up Congested Microservices.} We observe that \sysname{} allocates more VMs to congested microservices by inspecting span-level microservice tracing data. We run an experiment in which we simulate 400 users for the Online Boutique application and compare \sysname{} with underutilized (CPU-10) and overutilized (CPU-90) autoscaling policies. We collect tracing data from 4000 requests for each setting. In Figure \ref{fig:span-duration-by-autoscaler} we sort the microservice spans from left to right in terms of decreasing \textit{congestion}, defined as the \textit{percentage increase in span duration} from the underutilized to overutilized setting. Microservices whose constituent spans have higher congestion, for example \texttt{productcatalogservice}, are the most benefited by \sysname{}'s allocation of instances. As seen in Figure \ref{fig:span-duration-by-autoscaler}, \sysname{} (shown in green) allocates more VMs to the microservices which are running these congested spans. As a result, \sysname{}'s span durations are on average only 2.3\% longer than CPU-10 with 65.8\% fewer VMs. 

From our evaluations, we find that \sysname{} outperforms machine learned autoscalers (Linear Regression, Bayesian Optimization and the Deep Q-Network) on all applications. We take a detailed look at the Online Boutique and find these approaches are not as successful as \sysname{} in discerning congested microservices (discussed further in \S~\ref{subsec:la-shortcomings}). As a result these machine learned autoscalers allocate too many or too few resources to microservices which most heavily contribute to end to end latency (seen in Figures \ref{fig:ml-baseline-allocation} and \ref{fig:frontend-time-series}).


\Para{Tailoring a Policy to a Latency Target.} We compare the policies \sysname{} learns when trained to optimize median vs. 90\%ile latency. For both of these latency targets, shown in Figure \ref{fig:median-latency-endpoint-dist}, we find that \sysname{} has comparable median latency (47 and 46 ms respectively) across the different URL endpoints the Online Boutique application exposes. However we see 90\%ile latency is reduced by close to 50\% when \sysname{} is trained to explicitly optimize 90\%ile latency.

The largest difference between the two policies in terms of VM allocations is that \sysname{}-tail-100 scales up the \texttt{cartservice} microservice with 5$x$ as many instances as \sysname{}-50. From Figure \ref{fig:median-latency-endpoint-dist} we confirm that the two highest tail latency request URLs for the application relate to checking out and viewing the cart in Online Boutique. We find that \sysname{} is able to tailor its policy to not only an application and workload but to a specified latency target itself as well.

\Para{How close is \sysname{} to optimal?} We evaluate \sysname{} for empirical optimality -- finding the best possible cluster state for our given reward. For all applications and all workloads, we find that \sysname{} is able to find a cluster state which meets our latency target. Since our latency target is met, we know that the only possible way to obtain a higher reward than \sysname{} is to use fewer instances than \sysname{} to meet the latency target. We exhaustively search all cluster states for Simple Web Server (which only has 30 possible states) and all cluster which use fewer instances than \sysname{} for Book Info (which has 320,000 possible cluster states). We find that \sysname{} finds the optimal cluster state in 9 of 10 total workload-application pairs and is the second best configuration when it was not optimal. Overall, \sysname{} is .9\% more expensive on average than the optimal configuration across these 10 evaluations. These results are shown in Appendix Figure \ref{fig:optimality-sws-bi}.


\subsection{Training Cost}\label{subsec:training-cost}

To place the efficiency gains we've seen from using \sysname{} in perspective, we review the cost to train our learned autoscaling policies. In Tables \ref{tab:training-cost}-\ref{tab:training-cost-dqn} we list these costs in terms of time, instances, and dollars needed to train a 50 ms median latency autoscaling policy for each application. These costs consider VMs in our application node pool, a second node pool for the ingress gateway and load balancer specific pods, and lastly the load generator itself. Virtual machines in the application node pool are \texttt{n1-standard-1} machines which cost \$.047 per hour, those in the load balancer node pool are \texttt{e2-highmem-8} and cost \$.361 per hour and our load generator operates on a custom defined 20 core, 52 GB mem machine which costs \$.836 per hour.

\begin{table}[!t]
  \begin{tabular}{llll}
    \toprule
    Application&Time (hrs) & Instance hrs & Cost (\$)\\
    \midrule
   Simple Web Server&  0.61&  44.95&  \$2.03 \\
  Book Info &  0.65&  58.04&  \$2.64 \\
  Online Boutique&   3.36&  350.08&  \$16.06 \\
  Sock Shop &  1.34&  193.21&  \$8.95 \\
  Train Ticket&  19.56&  4772.43&  \$223.39\\
  \bottomrule
\end{tabular}
\caption{Training Cost (\sysname{} Median Latency)}
\label{tab:training-cost}
\vspace{-0.75cm}
\end{table}


In order to pay for the cost of \sysname{}'s training, the system must decrease the instance hours used during deployment by at least as many instance hours as were used for training. For BookInfo, we would need to save roughly 58 instance hours to amortize cost and for Online Boutique we would need roughly 350 instance hours saved. For a 700 and 800 constant rate User/s workload in BookInfo, training cost is paid off in 19.3 and 8.6 hours respectively. In Online Boutique the 700 and 800 User/s workloads pay off training cost in 41.2 hours. For the Train Ticket application, training cost is paid off in 183.0 and 126.7 hours respectively for a 250 and 500 User/s workload respectively. Once training cost is paid, \sysname{} results in a reduction in the cost of operating a microservice application. We include details and evaluations of model retraining in Appendix \S~\ref{subsec:model-retrain}.

\section{Related Work}
\Para{Scheduling.} Several schedulers have been developed to manage how jobs are assigned to specific cluster resources such as  Mesos~\cite{hindman2011mesos}, Tarcil~\cite{delimitrou2015tarcil}, Omega~\cite{schwarzkopf2013omega}, Sparrow~\cite{ousterhout2013sparrow}, and Apollo~\cite{boutin2014apollo}. These schedulers are designed to schedule jobs on clusters where the number of resources are fixed, not scale the cluster size dynamically as \sysname{} does.

\Para{Automatically Selecting Number and Type of Cloud VMs.} Various methods autoscale the {\em number} of physical or virtual machines based on current or forecasted resource demand such as Flexera~\cite{ITManage40:online}, CloudScale~\cite{shen2011cloudscale}, PRESS~\cite{gong2010press}, and AGILE~\cite{nguyen2013agile}. These methods focus on the simpler setting of monoliths. Other projects optimize the {\em type} of VM instance based on application-specific knowledge and predictive algorithms; examples include Ernest~\cite{venkataraman2016ernest}, CherryPick~\cite{alipourfard2017cherrypick} and PARIS~\cite{yadwadkar2017selecting}. While \sysname{} is built to specifically handle the challenges of microservice autoscaling, we adapt approaches inspired by Ernest~\cite{venkataraman2016ernest} and CherryPick~\cite{alipourfard2017cherrypick} for the microservice setting and compare to these adaptations.

\Para{Reward-Driven Autoscaling.} Other works establish a Quality of Experience or Service metric that describes performance in a manner similar to \sysname{}'s reward but do so for a general VM workload, not the microservice setting we target here. For instance, in~\cite{ilyushkin2017experimental}, the authors analyze tradeoffs between a variety of horizontal autoscaling policies. In ~\cite{evangelidis2018performance}, Evangelidis et. al constructs probabilistic rule based models for horizontal autoscaling in the public cloud. 

\Para{Horizontal Autoscaling.} In horizontal autoscaling, container-based cluster management tools such as Kubernetes~\cite{kubernetes:online} allow users to scale the number of container pods based on utilization metrics.  We compare \sysname{} against the combination of the Kubernetes horizontal autoscaler and the GKE cluster autoscaler. 


\Para{Vertical Autoscaling.} In vertical autoscaling~\cite{communit72:online}, a controller maintains estimates of CPU and memory consumption of pods, scaling pods' CPU and memory limits when they surpass a specified threshold. RUBAS~\cite{rattihalli2019exploring} extends this approach using a method which performs vertical autoscaling using the median and standard deviation of utilization metrics. Sinan~\cite{zhang2021sinan} is a recently proposed microservice vertical autoscaler which uses machine learning techniques to scale clusters by granularly allocating CPU cores. These procedures differ from \sysname{} as their objective is to improve resource utilization of already allocated VMs---rather than reducing the dollar cost of adding VMs to microservices to meet an end-to-end latency target.

\Para{Autopilot.} Autopilot~\cite{rzadca2020autopilot} uses recommendation systems and past job utilization metrics to both predict and dynamically adjust pod limits. This autoscaling procedure differs from \sysname{} in that the objective of Autopilot is to reduce \textit{slack} (i.e., improve utilization) in the resources needed for a workload rather than navigating a latency-vs.-dollars tradeoff. The former objective is more aligned with the perspective of a cloud provider while the latter that of a cloud tenant. The authors of Autopilot also explicitly mention that the work does not optimize ``serving jobs' end user response latency''---a primary objective of \sysname{}. To meet this objective at Google, the authors mention that serving workloads can be tuned in a manual and application-specific manner. We view \sysname{} as an automated solution for such serving workloads. 

Since the release of the Autopilot paper~\cite{rzadca2020autopilot}, Google has released a commercial offering called Autopilot, a "node-less" version of Google Kubernetes Engine, on which developers can run a microservice application without having to rent Virtual Machines. In this setting, developers pay only for the aggregate pod requests in their application. We evaluate \sysname{} and the Kubernetes HPA, which incorporates techniques presented in Autopilot, on the Autopilot platform. We show that \sysname{} outperforms the Kubernetes HPA on all evaluations in Tables~\ref{tab:bi-ap-tabular}-~\ref{tab:tt-ap-tabular}.

\Para{Other Systems.} Powerchief~\cite{yang2017powerchief} uses queueing models to optimize the CPU frequencies of machines based on learned workload characteristics. This method's procedure is primarily focused on reducing the energy cost of a cloud tenant for a desired performance rather than reducing the number of VMs. In Autotune~\cite{chang2021autotune}, the authors propose a way to tune microservice cluster allocations. Autotune is complementary to autoscaling. It is meant for situations where the application is overprovisioned to begin with and can then be compacted to save cloud resources without losing performance. Lastly, FIRM~\cite{qiu2020firm} learns to autoscale microservices using deep reinforcement learning. The approach hinges on providing this autoscaler with fine grained memory bandwidth and memory allocation information which requires hardware support (Intel Cache Allocation and Memory Bandwidth Allocation), unavailable in \sysname{}'s cloud setting. Inspired by FIRM, we adapt a Deep Q-Network to operate with information available in our setting and compare \sysname{} to it.
\section{Conclusion}
\label{sec:conclusion}

This paper presents \sysname{}, a system for autoscaling microservice-based applications. While existing autoscalers scale each microservice within an application independently, \sysname{} collectively decides how many VMs to allocate each microservice with the goal of achieving an end-to-end user latency target while minimizing dollar cost.  Such a centralized approach to autoscaling gives COLA greater visibility into which microservices really matter for end-to-end latency and allows it to perform better than other approaches to microservice scaling that we compare against.


\end{sloppypar}

\balance

\clearpage
\bibliographystyle{ACM-Reference-Format}
\bibliography{cola}

 \clearpage
\section{Appendix}
\label{sec:appendix}
\subsection{Queueing Theory Analysis}
\label{subsec:q-theory}

\newtheorem{prop}{Proposition}

We model each microservice as an M/M/c queue. In this case, we assume arrivals form a single queue and are governed by a Poisson process with rate $\lambda$, that there are c servers, and job service times are exponentially distributed with parameter $\mu$. The response time can be composed into two pieces: a queueing time and service time. The service time is fixed, $\frac{1}{\mu}$, and consequently latency is increased/decreased by increasing/decreasing the queueing time. This queueing time is shown below and is based on Erlang's C formula \cite{barbeau2007principles} which governs the number of customers in an M/M/c queue given a number of servers and utilization, $\rho=\frac{\lambda}{c \mu}$.

\begin{equation}
    W_q(c, \rho) = \frac{\frac{(c\rho)^c}{c!}\frac{1}{1-\rho}}{\sum_{k=0}^{c-1} \frac{(c\rho)^k}{k!} + \frac{(c\rho)^c}{c!}\frac{1}{1-\rho}} \frac{1}{c\mu - \lambda}
\end{equation}

In addition to the formula above, for an M/M/c queue, we have the queuing length $L_q(c,\rho) = W_q(c, \rho) \lambda$, which is a result of Little's Law \cite{little2008little}. Lastly, $B$ is the probability that all servers are busy. The expected number of busy servers is $c\rho$ and $B$ is no larger than $\rho$ \cite{grassmann1981stochastic}. 

We refer readers to \cite{barbeau2007principles} for a more detailed treatment of the behavior of queuing theory.

\noindent\begin{prop}\label{prop:util-proof}
Consider a set of $n$ M/M/c queues with utilization $\rho_1 > \rho_2 > ... > \rho_n$. Further, let us say that M/M/c queues have the same $c$. Increasing the number of servers for the highest utilized M/M/c queue, the queue associated with $\rho_1$, has the greatest upper bound in terms of queuing length reduction.
\end{prop}

We rely on the convexity of  M/M/c queuing length with respect to $\rho$. This property is well known and shown in a previous paper by Grassmann \cite{grassmann1983convexity}. As a direct result of this convexity, we know the following is true:

\begin{equation}
L_q(c_1, \rho_1) \geq L_q(c_2, \rho_2) + \nabla L_q(c_2, \rho_2)^T(\rho_1-\rho_2)
\end{equation}

Let us consider the case in which we add a server (increasing $c$ to $c+1$) to our M/M/c queue with $\lambda$ and $\mu$ fixed. We can let $\rho_1$ be the case with $c + 1$ servers and $\rho$ be the case with $c$ servers.

\begingroup
\allowdisplaybreaks
\begin{align*}
    L_q(c+1, \rho_1) - L_q(c, \rho) &\geq \nabla L_q(c, \rho)^T \left(\frac{\lambda}{(c+1)(\mu)} - \frac{\lambda}{c\mu}\right) \\
    &= B \left[ c + \frac{1-\rho-B\rho}{(1-\rho)^2} \right]\left(\frac{-\rho}{c+1} \right) \\
    L_q(c, \rho) - L_q(c+1, \rho) &\leq B \left[ c + \frac{1-\rho-B\rho}{(1-\rho)^2} \right]\left(\frac{\rho}{c+1} \right) \\
    \mathbb{E}[ L_q(c, \rho) - L_q(c+1, \rho)] &\leq \left[\frac{\rho^2 c}{c+1} + \frac{1+\rho -\rho^2}{(1-\rho)^2} \right]\left(\frac{\rho}{c+1} \right) \\
    &\hspace{-4em}= \left[\frac{\rho^2 c}{c+1} + \frac{1}{(1-\rho)^2} + \frac{\rho(1-\rho)}{(1-\rho)^2} \right]\left(\frac{\rho}{c+1} \right)
\end{align*}
\endgroup

All terms above are increasing in $\rho$. Consequently, choosing to add a server to the most utilized M/M/c queue offers the greatest upper bound in terms of queue length reduction: $L_q(\rho_1) - L_q(\rho)$. Since queuing time, $W_q$, simply equals $L_q / \lambda $, adding a server to the most utilized M/M/c queue also offers the greatest upper bound in terms of queuing time reduction with $\lambda$ equal across our $n$ queues.

\noindent\begin{prop}\label{prop:inference-prop}
Consider a linearly interpolating policy between $\rho_{l}$ and $\rho_{u}$ corresponding to input rates $\lambda_{l}$ and $\lambda_{u}$ where $\lambda_{u} > \lambda_{l}$ . We assume that under both $\rho_{l}$ and $\rho_{u}$, the queuing delay $W_q$ satisfies a latency target. We show that in some cases this latency from an interpolated policy of these two policies maintains or reduces queuing delay and consequently achieves the latency target we desire. At most the linear interpolation policy increases queuing delay by $ \frac{\lambda_{u}}{\lambda_{l}}$, a ratio which may be controlled by altering the spacing of the considered policies.
\end{prop}

Let us consider the two possible cases under which we perform linear interpolation between $\rho_{l}$ and $\rho_{u}$. These cases are: (1) $\rho_{l} \leq \rho_{u}$ and (2) $\rho_{l} > \rho_{u}$. We denote the interpolated utilization as $\tilde{\rho}$.

In the first case where $\rho_{l} \leq \rho_{u}$, denote $\Delta \rho = \rho_{u} - \rho_{l} \geq 0$. We know that $\tilde{\rho} \leq \rho_{u}$ since:

\begin{align*}
    \tilde{\rho} &= x\rho_{u} + (1-x)\rho_{l} \\
    &=  \rho_{l} + x(\rho_{u} - \rho_{l})\\
    &=  \rho_{u} - (1-x)(\Delta \rho)\\
    &<=  \rho_{u}
\end{align*}

The last line is a result of $0 \leq x \leq1$ which is true since we are interpolating between two endpoints $\rho_l$ and $\rho_u$.

We begin with a result from Proposition \ref{prop:util-proof} to show our claim that $W_q(\tilde{c}, \tilde{\rho}) \leq W_q(c_u, \rho_u)$. Note that with $\tilde{c}$ and $\tilde{\rho}$ positive, we have $\nabla L_q(\tilde{c}, \tilde{\rho})$ positive.

\begin{align*}
L_q(c_u, \rho_u) &\geq L_q(\tilde{c}, \tilde{\rho}) + \nabla L_q(\tilde{c}, \tilde{\rho})^T(\rho_u-\tilde{\rho}) \\
L_q(\tilde{c}, \tilde{\rho}) - L_q(c_u, \rho_u) &\leq \nabla L_q(\tilde{c}, \tilde{\rho})^T(\tilde{\rho}-\rho_u) \\
L_q(\tilde{c}, \tilde{\rho}) - L_q(c_u, \rho_u) &\leq 0 \\
\end{align*}

With $L=\lambda W$, we have $W_q(\tilde{c}, \tilde{\rho}) \tilde{\lambda} = L_q(\tilde{c}, \tilde{\rho})$ and that $W_q(c_u, \rho_u) \lambda_u = L_q(c_u, \rho_u)$. Let $y\tilde{\lambda} = \lambda_u$ where $y >= 1$.

\begin{align*}
\frac{W_q(\tilde{c}, \tilde{\rho})}{\tilde{\lambda}} - \frac{W_q(c_u, \rho_u)}{y \tilde{\lambda}} &\leq 0 \\
y W_q(\tilde{c}, \tilde{\rho}) &\leq W_q(c_u, \rho_u) \\
\end{align*}

Since $y \geq 1$, we have obtained the desired result: $W_q(\tilde{c}, \tilde{\rho}) \leq W_q(c_u, \rho_u)$ when $\rho_{l} \leq \rho_{u}$.

For the second case we consider when $\rho_{l} > \rho_{u}$. By the same argument as for our first case, we know that $\tilde{\rho} \leq \rho_l$ within our interpolating range.

\begin{align*}
L_q(c_l, \rho_l) &\geq L_q(\tilde{c}, \tilde{\rho}) + \nabla L_q(\tilde{c}, \tilde{\rho})^T(\rho_l-\tilde{\rho}) \\
L_q(\tilde{c}, \tilde{\rho}) - L_q(c_l, \rho_l) &\leq \nabla L_q(\tilde{c}, \tilde{\rho})^T(\tilde{\rho}-\rho_l) \\
L_q(\tilde{c}, \tilde{\rho}) - L_q(c_l, \rho_l) &\leq 0 \\
\end{align*}

Let $\tilde{\lambda} = z \lambda_l$ where $z >= 1$.

\begin{align*}
\frac{W_q(\tilde{c}, \tilde{\rho})}{z \lambda_l} - \frac{W_q(c_l, \rho_l)}{\lambda_l} &\leq 0 \\
W_q(\tilde{c}, \tilde{\rho}) &\leq z W_q(c_l, \rho_l) \\
\end{align*}

Since $\lambda_l \leq \tilde{\lambda} \leq  \lambda_u$ we know $1 \leq z \leq \frac{\lambda_u}{\lambda_l}$. Consequently, we know that $W_q(\tilde{c}, \tilde{\rho})$ is at most $\frac{\lambda_u}{\lambda_l}$ larger than $W_q(c_l, \rho_l)$ as well as our target queuing delay. 

\subsection{Further Evaluations}
\label{subsec:further-evals}

\begin{figure*}
\subfigure{%
  \label{fig:bi-fr-300}%
  \includegraphics[width=0.31\textwidth]{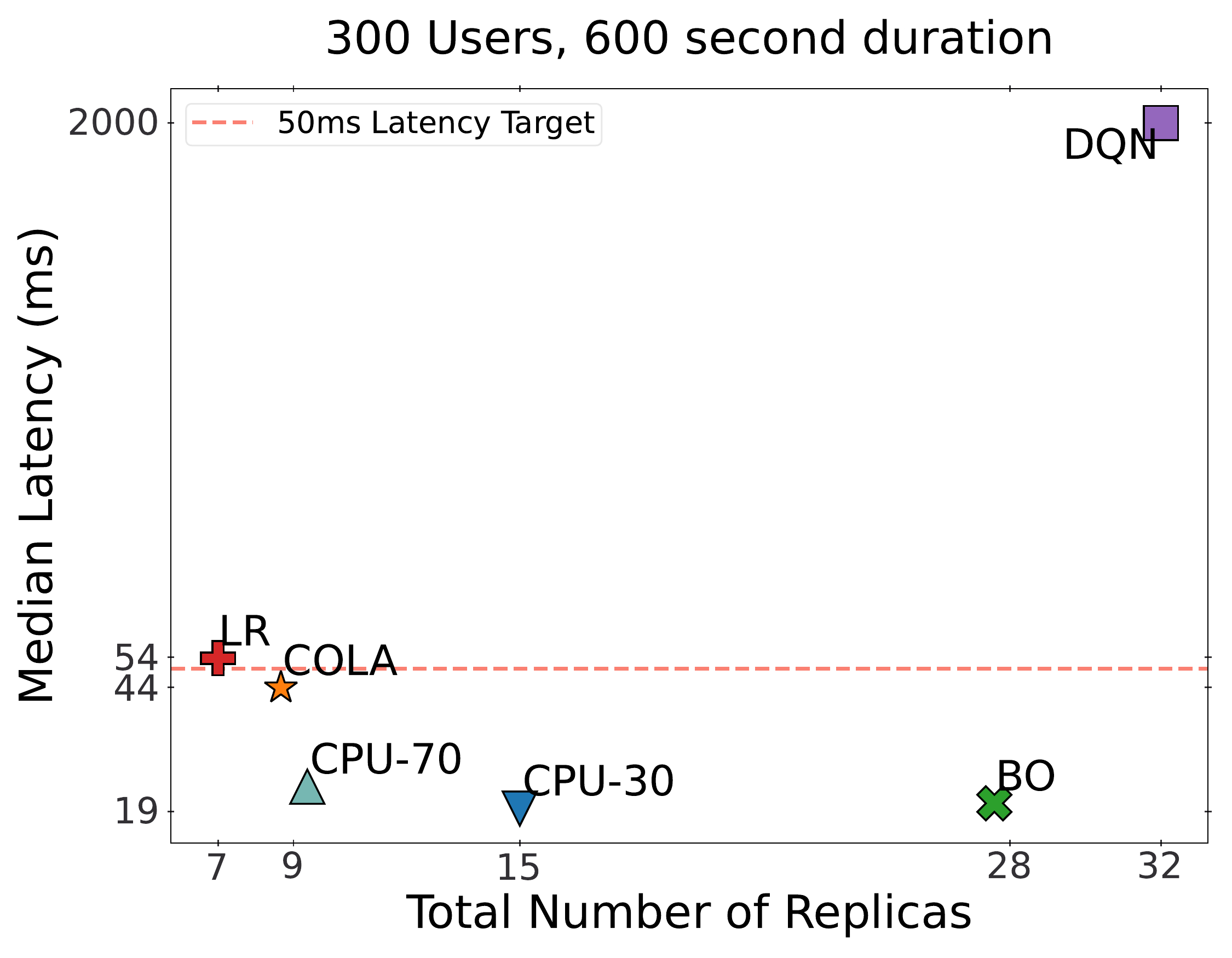}%
}%
\hspace*{\fill}
\subfigure{
  \label{fig:bi-fr-400}%
  \includegraphics[width=0.31\textwidth]{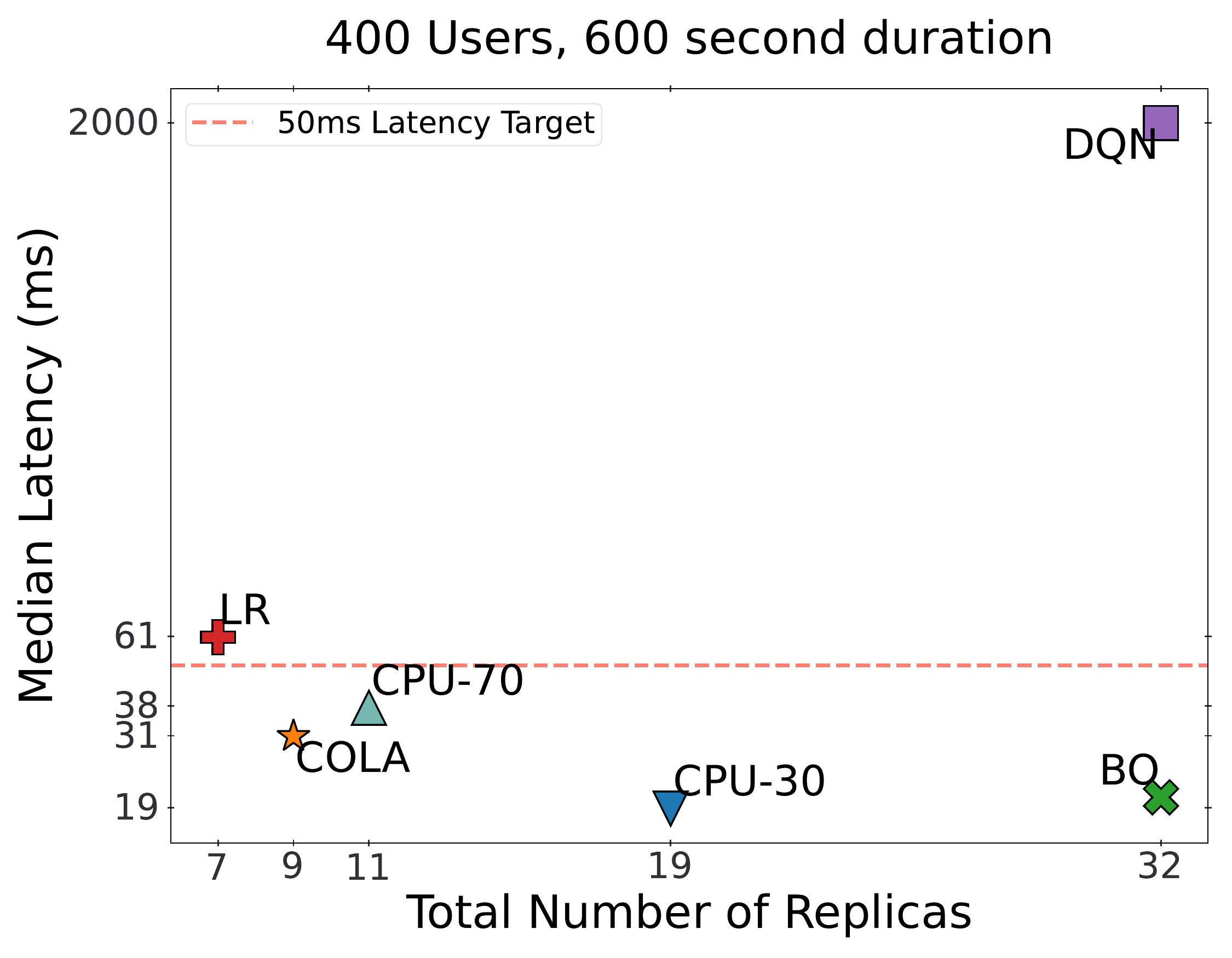}%
}%
\hspace*{\fill}
\subfigure{
  \label{fig:bi-fr-700}%
  \includegraphics[width=0.31\textwidth]{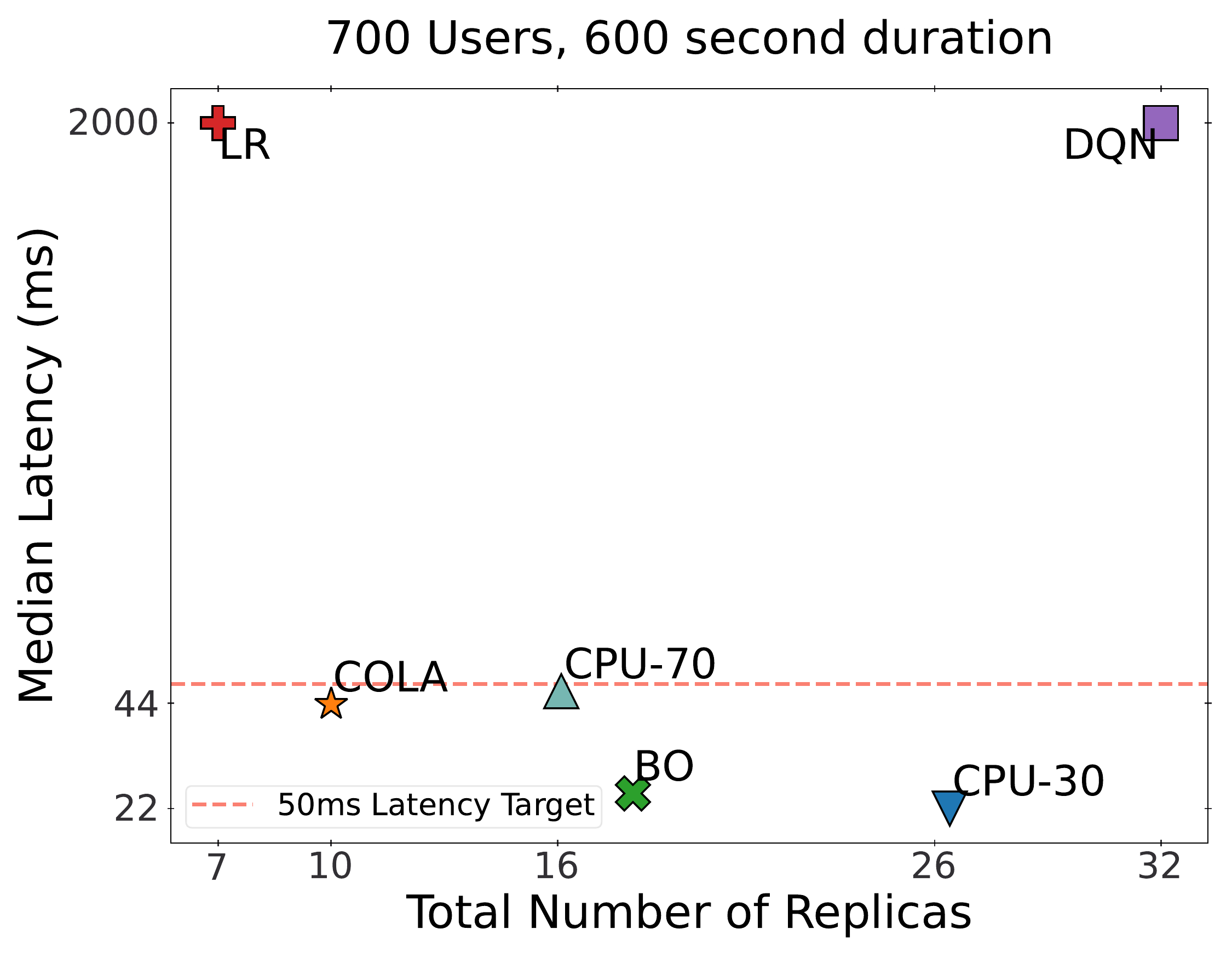}%
}
\captionsetup{justification=centering}
\caption{Book Info Constant Rate Workload}\label{fig:bi-fr}

\subfigure{%
  \label{fig:ss-fr-200}%
  \includegraphics[width=0.31\textwidth]{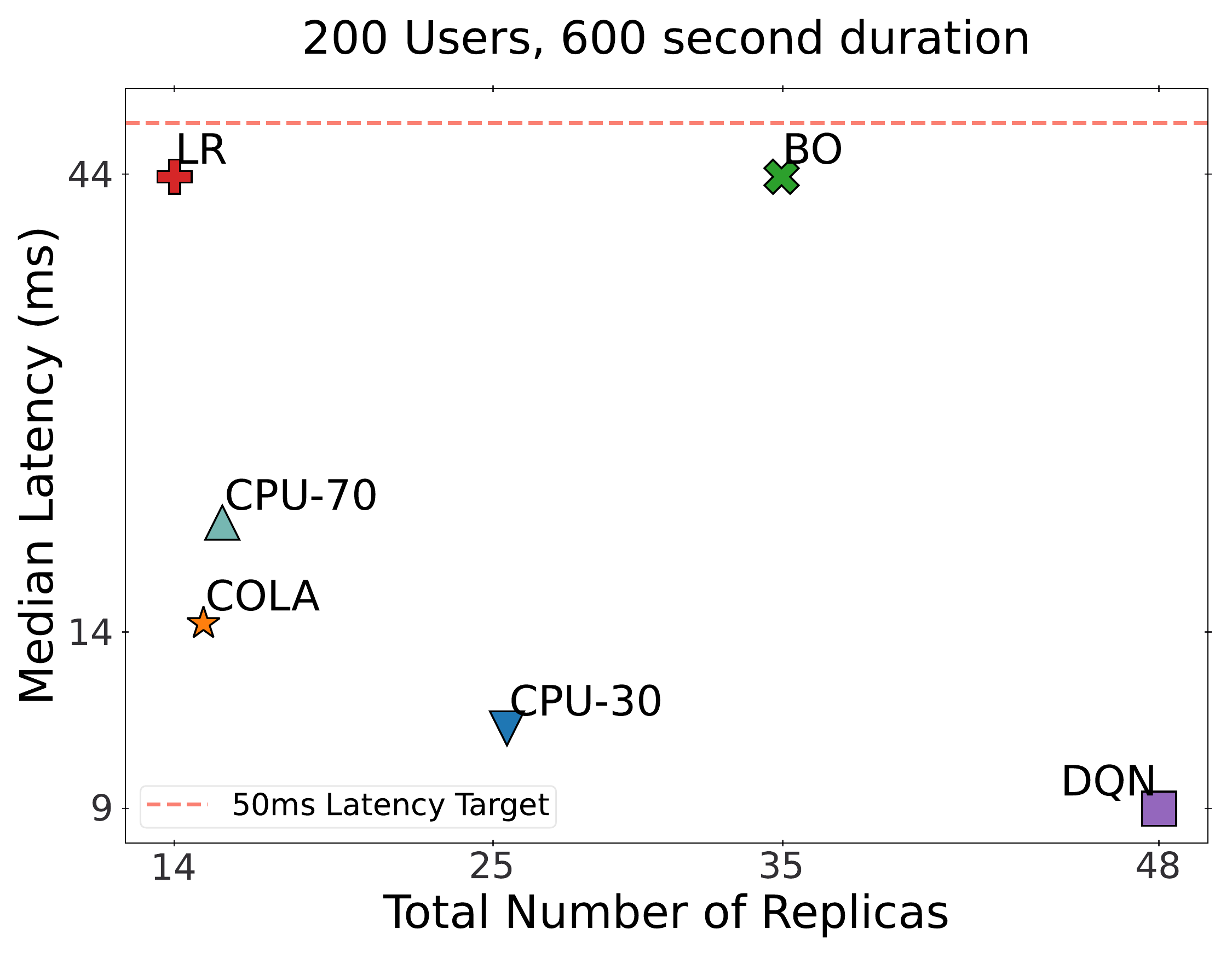}%
}%
\hspace*{\fill}
\subfigure{
  \label{fig:ss-fr-300}%
  \includegraphics[width=0.31\textwidth]{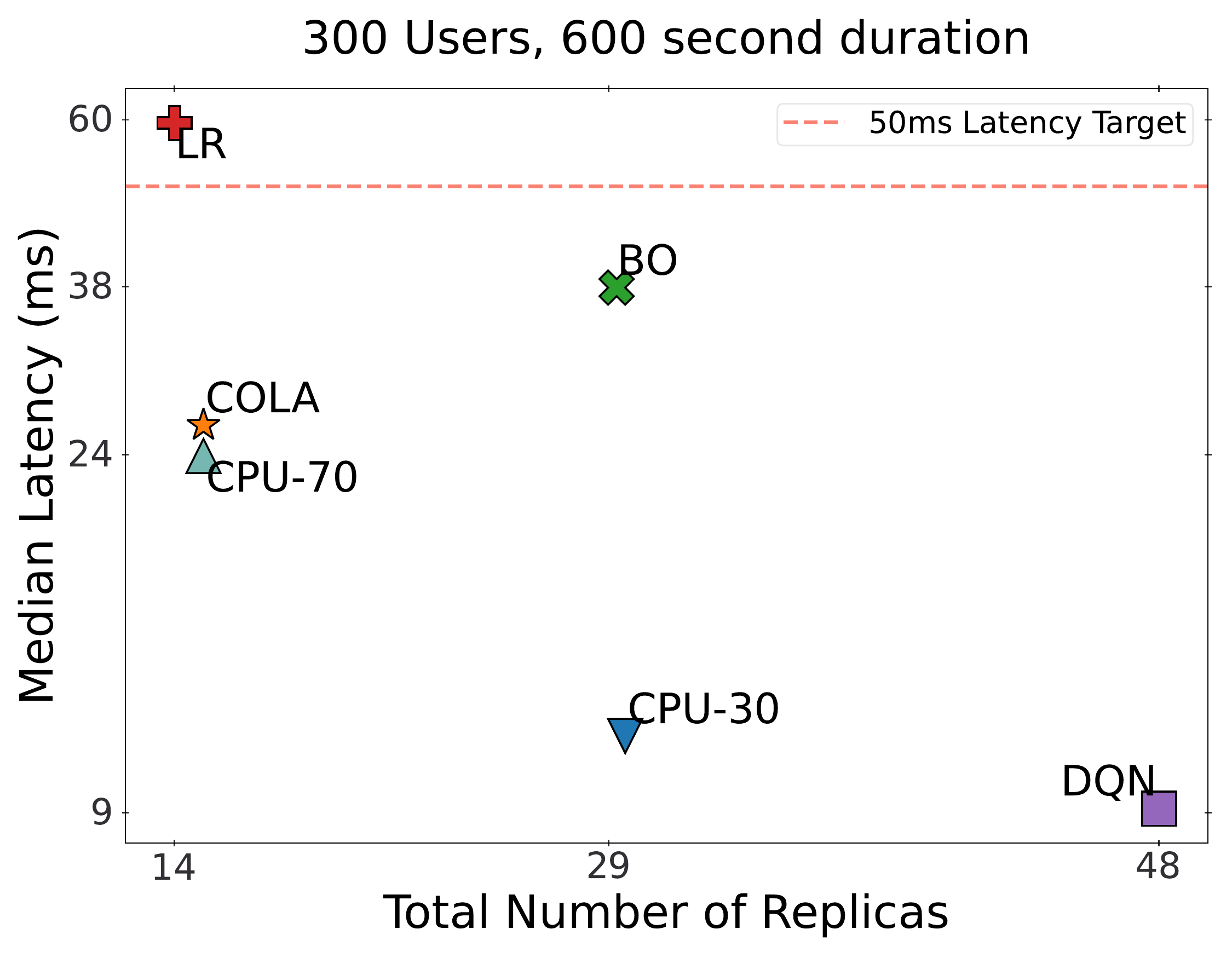}%
}%
\hspace*{\fill}
\subfigure{
  \label{fig:ss-fr-500}%
  \includegraphics[width=0.31\textwidth]{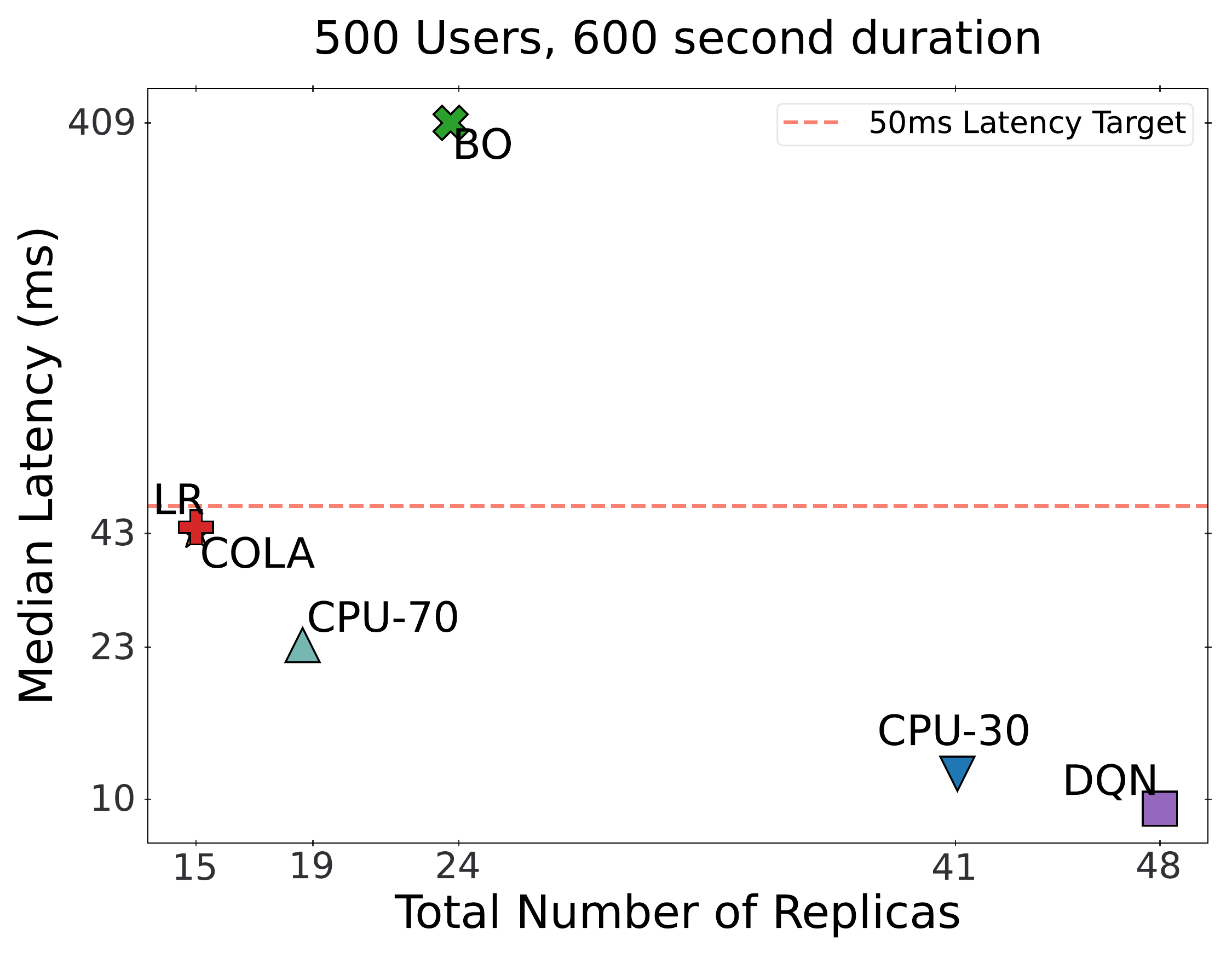}%
}
\captionsetup{justification=centering}
\caption{Sock Shop Constant Rate Workload}\label{fig:ss-constant-rate}

\subfigure{%
  \label{fig:bi-fr-ramp}%
  \includegraphics[width=0.31\textwidth]{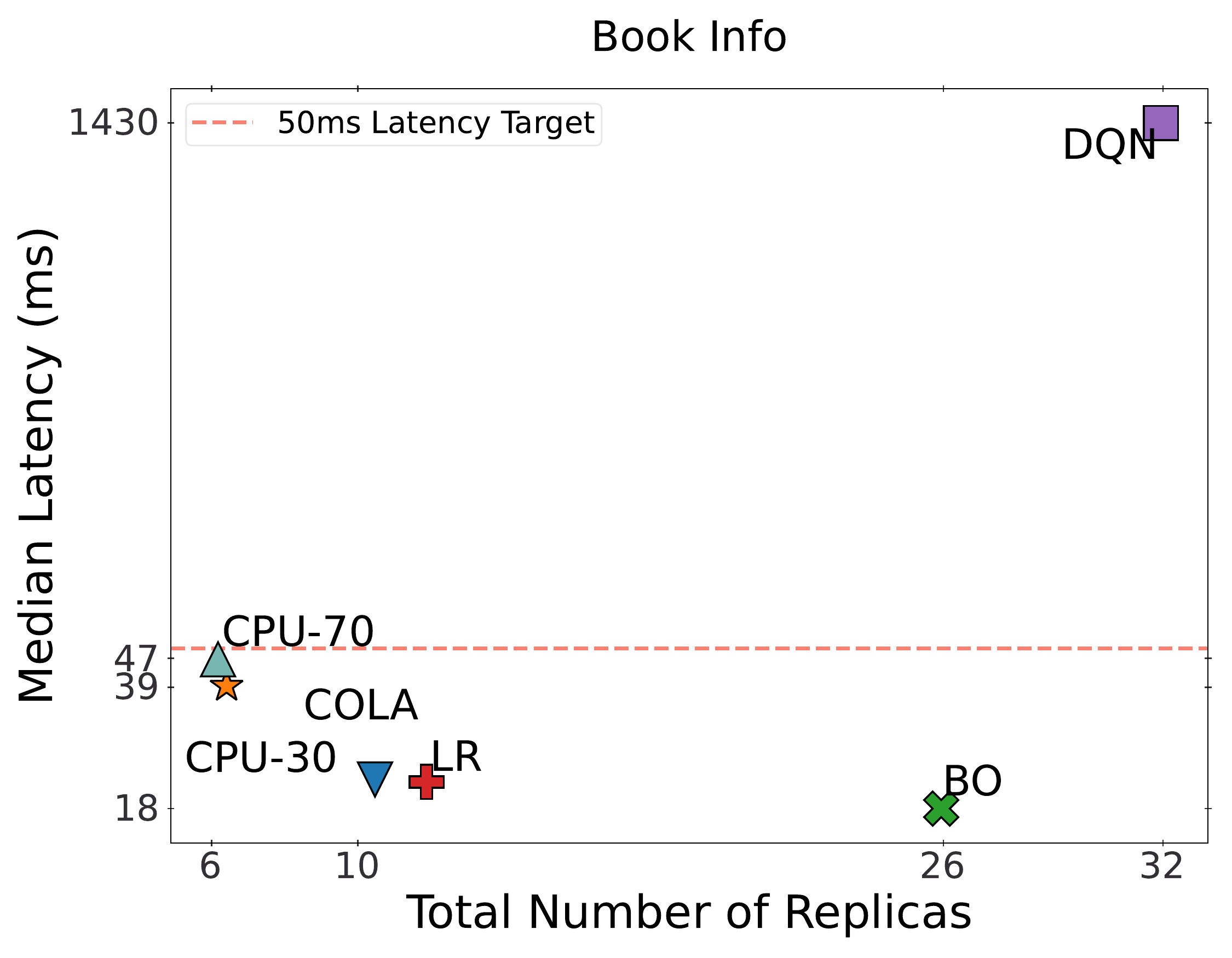}%
}%
\hspace*{\fill}
\subfigure{
  \label{fig:ob-fr-ramp}%
  \includegraphics[width=0.31\textwidth]{figs/updated/onlineboutique/online_boutique_fr_ramp.pdf}%
}%
\hspace*{\fill}
\subfigure{
  \label{fig:tt-fr-ramp}%
  \includegraphics[width=0.31\textwidth]{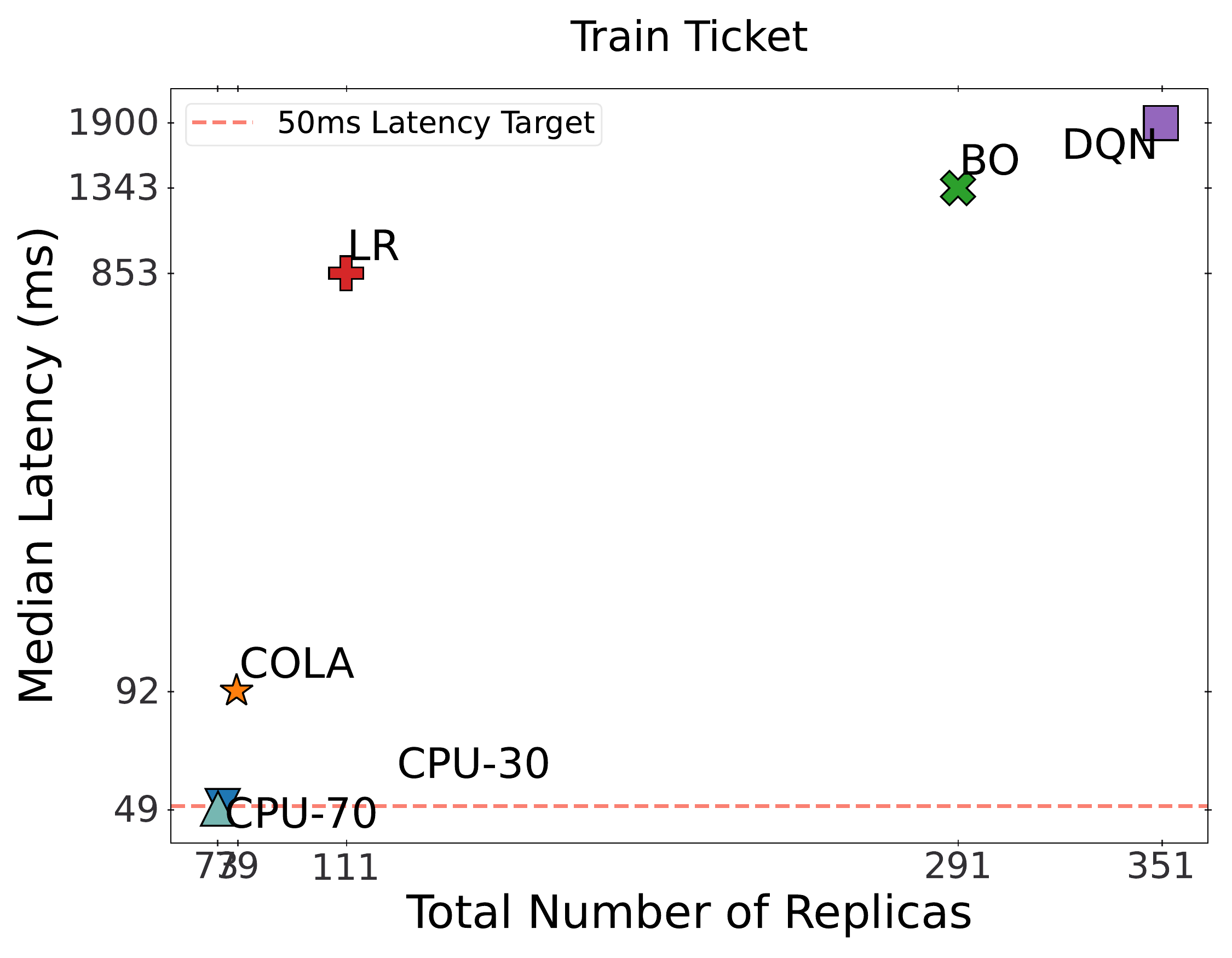}%
}
\captionsetup{justification=centering}
\caption{In Sample Diurnal Workload for Book Info (Left), Online Boutique (Middle) and Train Ticket (Right)}\label{fig:is-diurnal}

\subfigure{%
  \label{fig:bi-small}%
  \includegraphics[width=0.31\textwidth]{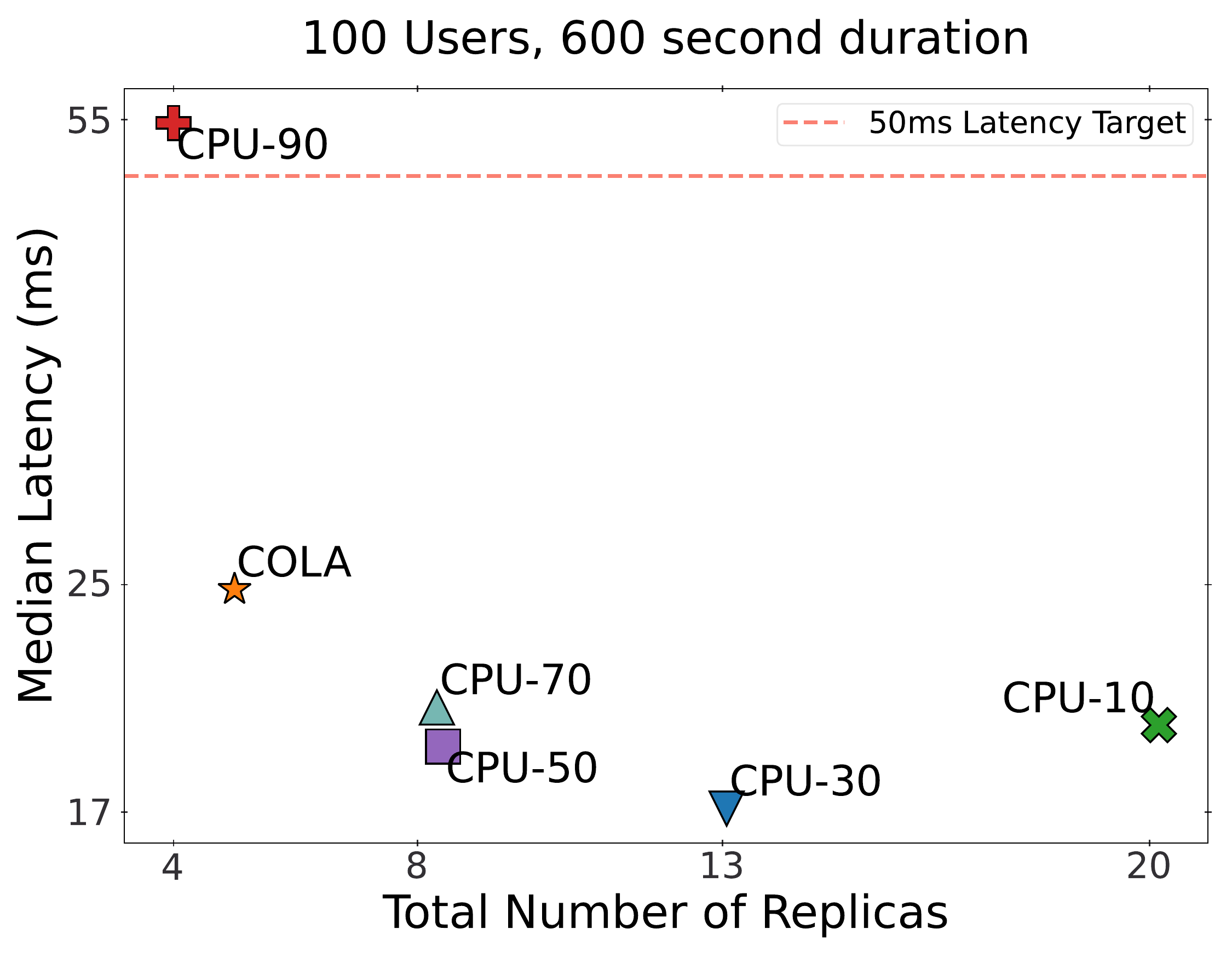}%
}%
\hspace*{\fill}
\subfigure{
  \label{fig:ob-fr-medium}%
  \includegraphics[width=0.31\textwidth]{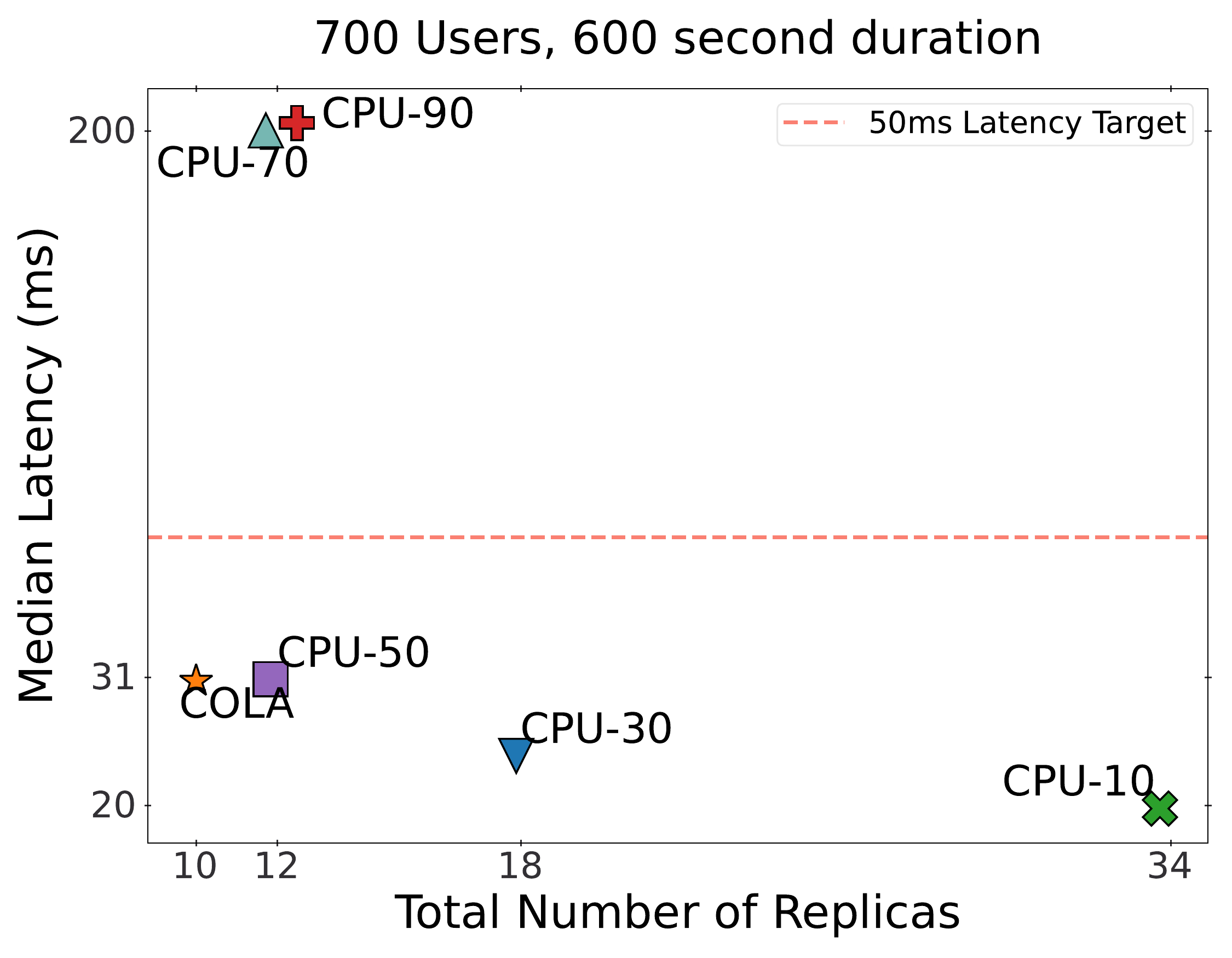}%
}%
\hspace*{\fill}
\subfigure{
  \label{fig:bi-ldr-large}%
  \includegraphics[width=0.31\textwidth]{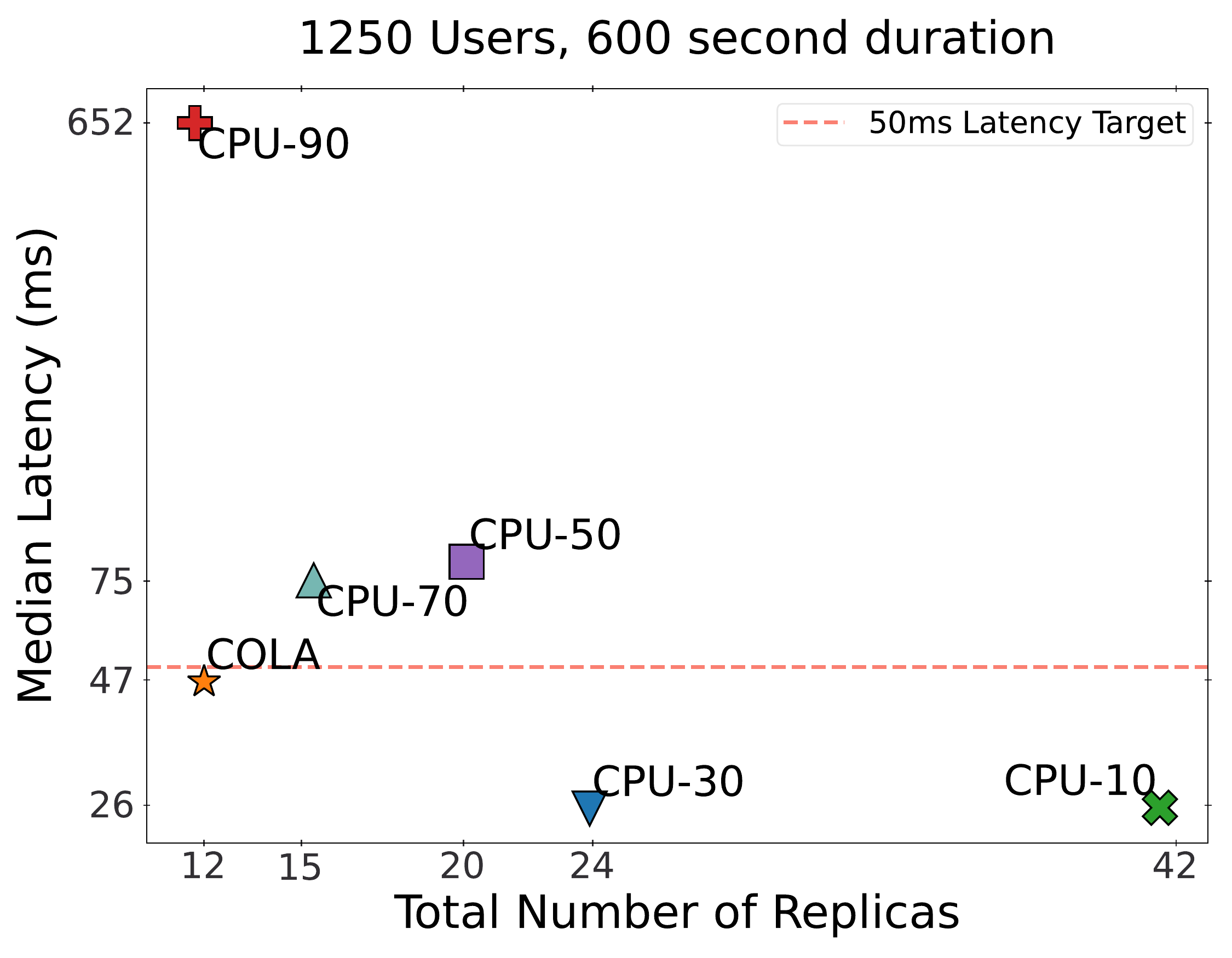}%
}
\captionsetup{justification=centering}
\caption{Large Dynamic Request Range Workload for Bookinfo}\label{fig:bi-ldr}

\end{figure*}

\Para{Constant Rate Workloads} In Figures \ref{fig:bi-fr} and \ref{fig:ss-constant-rate} we show constant rate evaluations for the Book Info and Sock Shop applications. For the BookInfo application, we train COLA on 200, 400, 600 and 800 requests per second with a target of 50ms median latency. It is asked to autoscale for fixed rate workloads both in and out of our training samples; shown in Figure 11. COLA provides the most cost effective policy across all candidate autoscalers with the next cheapest policy (CPU-70) costing 28.5\% more. For the Sock Shop application we train COLA on 200, 300, 400, and 500 requests per second for a median latency of 50ms. On a set of in sample workloads we find that the most cost effective autoscaling policy depends on the workload (Figure 12). Overall, COLA is the cheapest policy to meet the latency target across all workloads. Tables \ref{tab:bi-fr-tabular} and \ref{tab:ss-fr-tabular} detail the full results of these experiments. 

\Para{In Sample Diurnal Workloads} Figure \ref{fig:is-diurnal} shows In-Sample diurnal workloads for the Book Info, Online Boutique and Train Ticket applications. On the Book Info application, COLA meets its latency target but is outperformed by the CPU-70 policy with a cost reduction of -3\%.  For the Online Boutique application, \sysname{} reduces cost over the next cheapest policy which meets the target, CPU-30, by 46\%. For the Train Ticket application, COLA performs worse on the In Sample diurnal workload compared to the CPU-30 and CPU-70 policies. It incurs a cost reduction of -7.4\%. Tables \ref{tab:bi-tail-fr-tabular}, \ref{tab:ob-tail-diurnal-tabular}, and \ref{tab:tt-diurnal-tabular} show the numerical results of these experiments.

\Para{Large Dynamic Request Range} For our evaluations we have trained and evaluated autoscalers on a dynamic request range -- the ratio between the largest and smallest number of requests -- of 4 to 5 depending on the application. Realistically, this dynamic request range could be much higher if the use of an application fluctuates significantly (e.g. based on the time of day). We run evaluations where \sysname{} is trained and then manages a cluster across a dynamic request range of 40, with the smallest number of requests being 25 and the largest being 1000. Results for this setting are shown below in Figure \ref{fig:bi-ldr} and compared to a variety of CPU threshold autoscalers. Across six different request rates, \sysname{} reduces the cost to meet our 50ms latency objective by 24.2\%. Tabular results for this evaluation are listed in Table \ref{tab:bi-ldr-tabular}.

\Para{ Memory Threshold Autoscaling Policies} We include Constant Rate evaluations which compare \sysname{} with a variety of memory threshold autoscalers. For the applications on which we evaluate, we find that memory threshold autoscalers offer worse and/or inconsistent performance compared to CPU threshold autoscalers and consequently present them here in the Appendix.

For the Simple Web Server application, memory autoscaler only perform well on the 500 request per second case as only 1 node is needed to satisfy the load. In all other workloads, the memory autoscalers fail to scale up to the incoming load.

In the Book Info application, only the 10\% memory autoscaler scales up microservice replicas and performs well for the 300 and 400 request per second. However, for 700 and 800 requests per second the memory autoscaler suffers high median and tail latencies for all thresholds.


For the Online Boutique Application, memory autoscalers fail to scale up to the workloads and suffer poor median and tail latencies.

\subsection{Model Retraining}
\label{subsec:model-retrain} Over time, production workloads may drift outside our training request per second range and/or request distribution. We show training cost associated with adding a new workload and modifying the latency target for \sysname{}. Generally, we note that model retraining for a new specification will depend significantly on the application and workload. Several triggers can be applied to retrain \sysname{}, for example observing latency higher than the target, number of times \sysname{} falls back to another autoscaler, or percent of requests which return errors. For context, Autopilot's authors~\cite{rzadca2020autopilot} note that a few anecdotal applications at Google are reconfigured roughly 10 times a month.

\begin{table}[!htb]
  \begin{tabular}{lrrrr}
    \toprule
    \multicolumn{1}{c}{} & \multicolumn{2}{c}{Add Workload} & \multicolumn{2}{c}{Update Latency}\\
    \cline{2-5}
      Application& Time &Cost (\$)&Time &Cost (\$)\\
      \midrule
          Simple Web Server &  0.18&   \$0.59 & 0.31&   \$1.02   \\
          Book Info         &  0.19&   \$0.77 & 0.27&   \$1.09    \\
          Online Boutique   &  0.28&   \$2.03 & 0.31&   \$1.49     \\
          Train Ticket      &  0.51&   \$5.83 & 0.54&   \$6.12     \\
      \bottomrule
    \end{tabular}
\caption{Model Retraining Time (hours) and Cost (\$).}
\label{tab:retraining-cost}
\end{table}

\begin{figure*}
\subfigure{%
  \label{fig:fr-sws-mem}%
	\includegraphics[width=8.5cm]{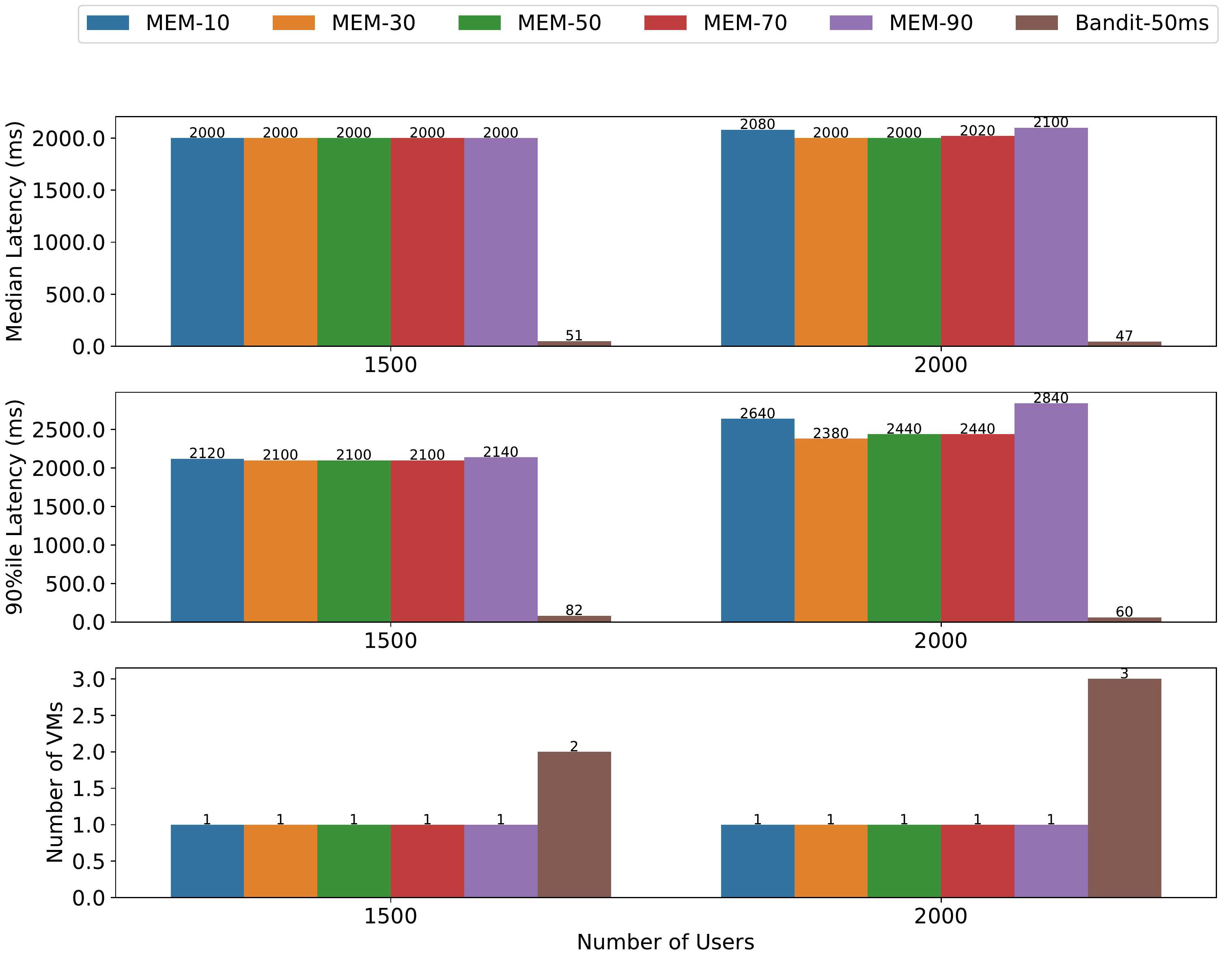}
}%
\hspace*{\fill}
\subfigure{
  \label{fig:fr-bi-mem}%
	\includegraphics[width=8.5cm]{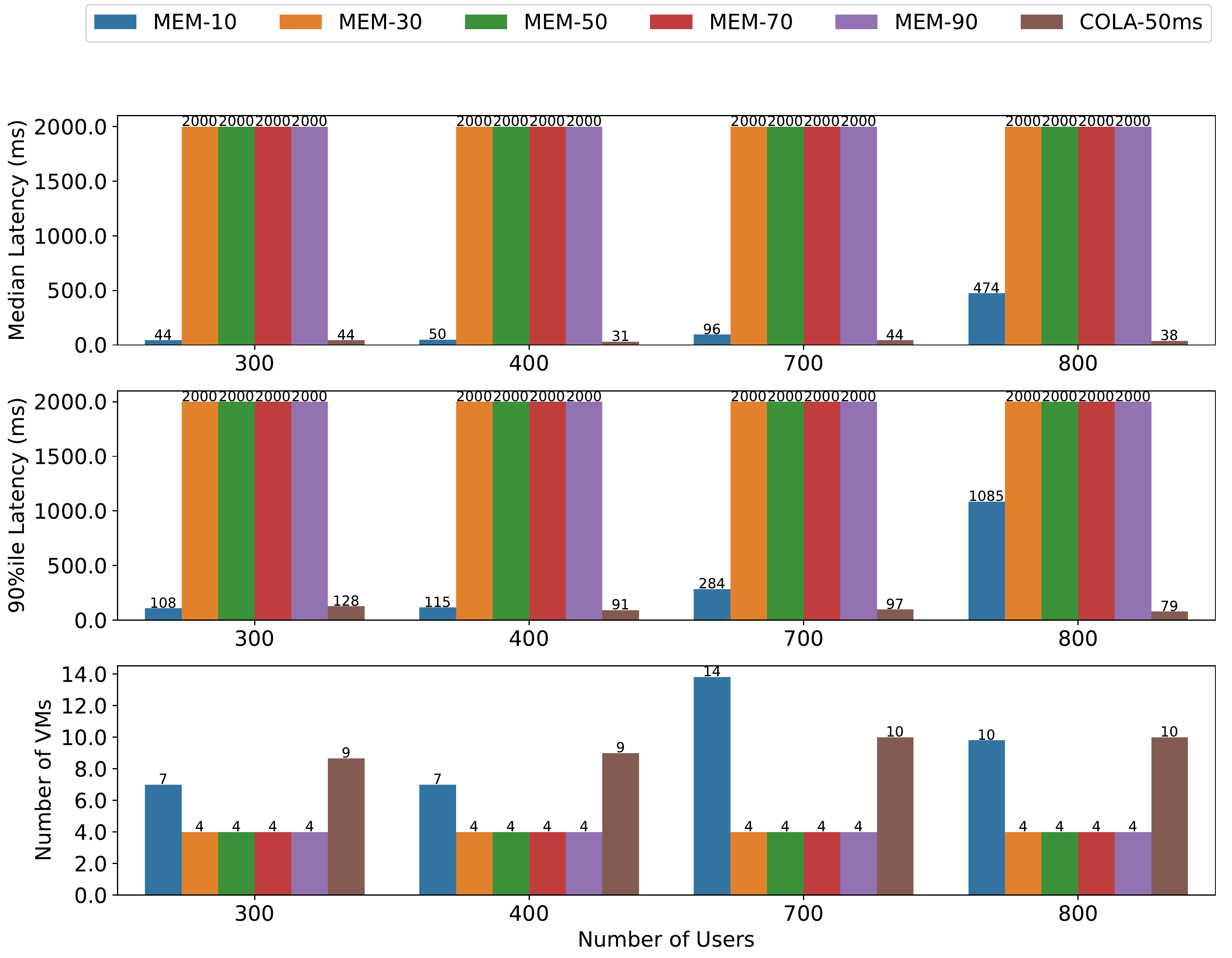}
}%
\captionsetup{justification=centering}
\caption{Constant Rate Evaluation - Memory Autoscaler Comparison for Simple Web Server (Left) and Book Info (Right)}\label{fig:mem-autoscalers}
\end{figure*}

\begin{figure*}
\subfigure{%
	\label{fig:bi-noca}
	\includegraphics[width=8.5cm]{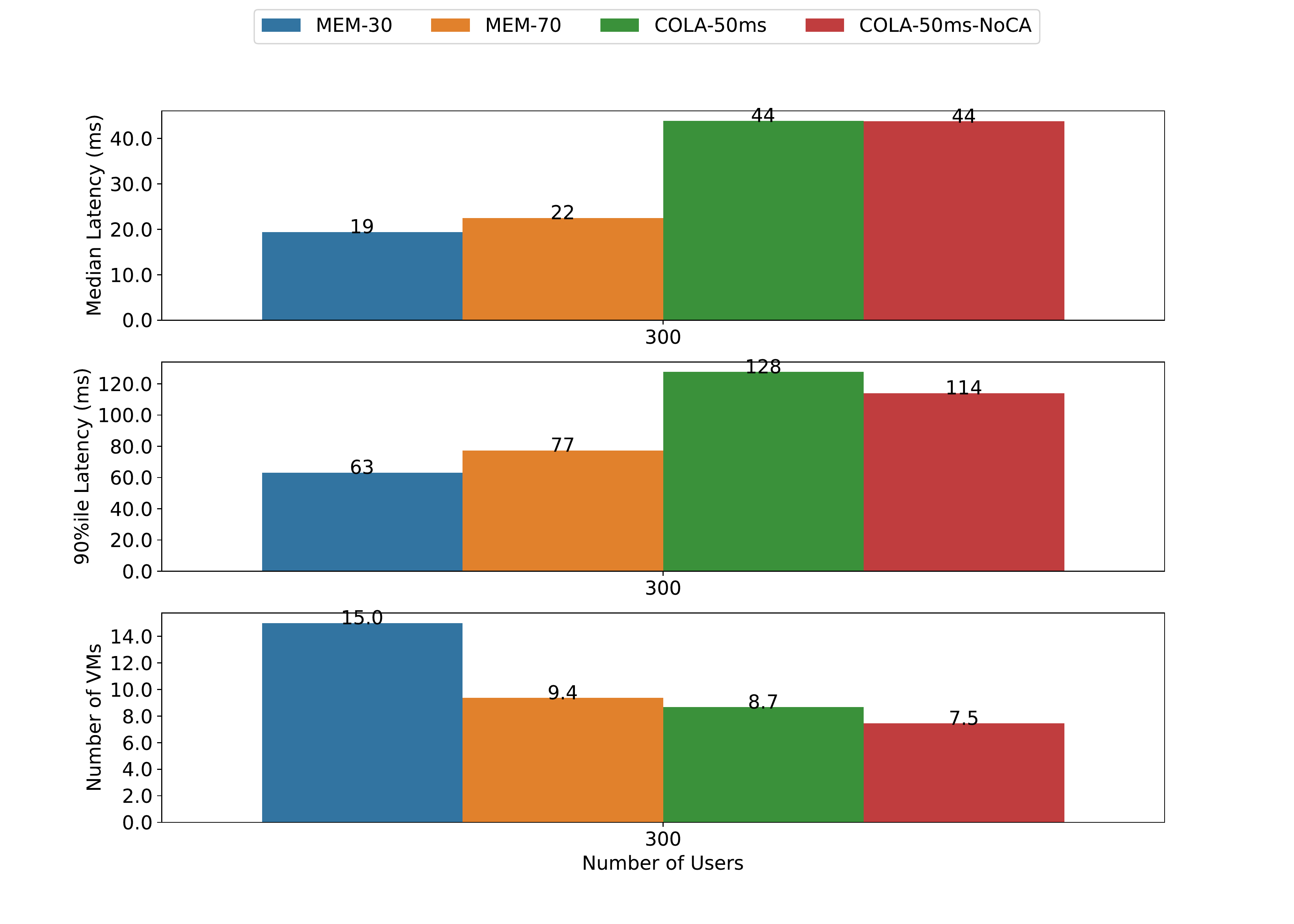}
}%
\hspace*{\fill}
\subfigure{
	\label{fig:ob-noca}
	\includegraphics[width=8.5cm]{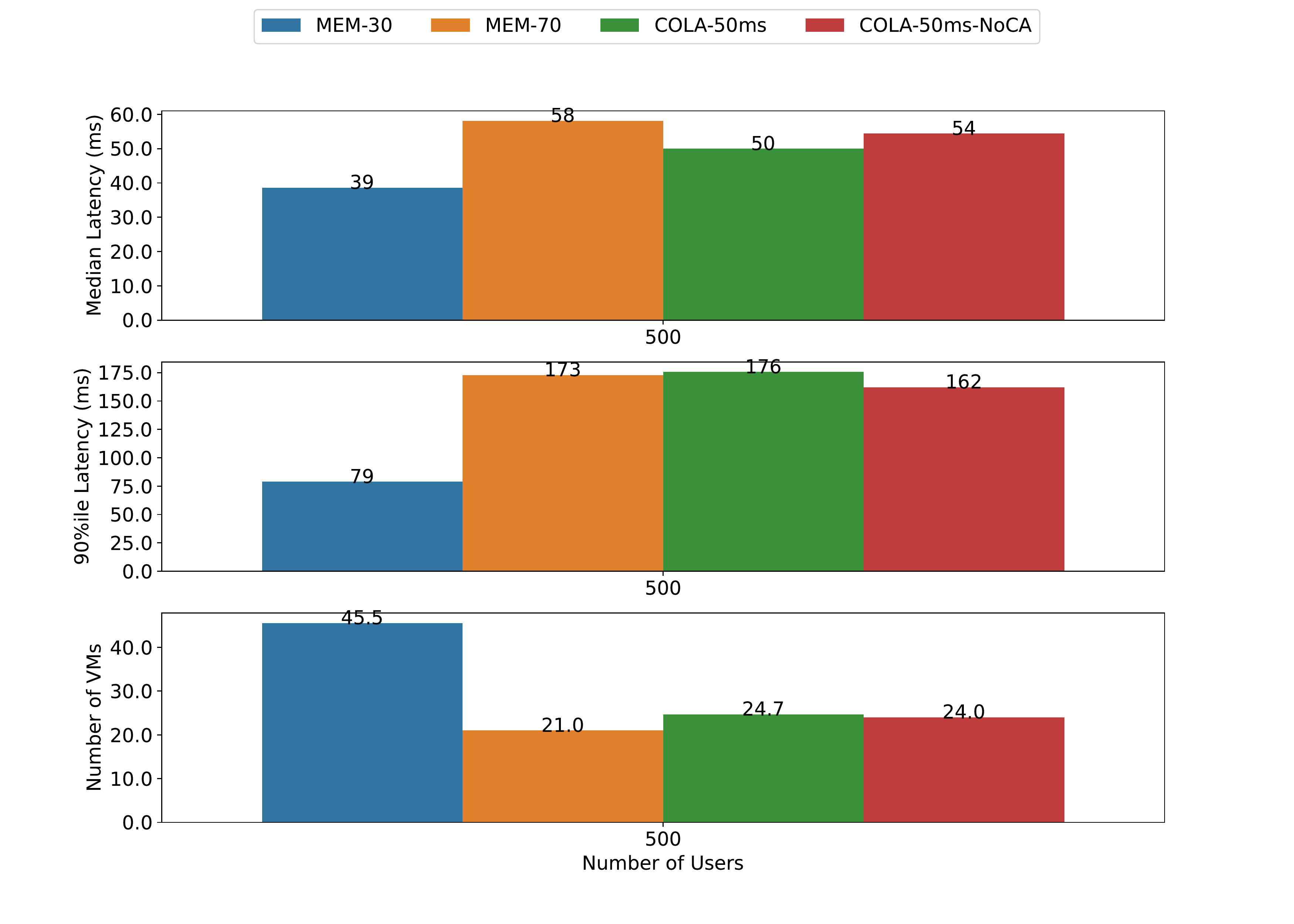}
}%
\captionsetup{justification=centering}
\caption{\sysname{} HPA with GKE Cluster Autoscaler for Book Info (left) and Online Boutique (right).}
\label{fig:fr-noca}
\end{figure*}

\subsection{Model Robustness}
\label{subsec:model-robustness}



\Para{Control Lag and Measurement Noise} All autoscalers we evaluate operate a control loop. Some measurement is input to the controller at which point the controller will take an action to progress towards a set point of interest. In \sysname{} the total requests per second along with a distribution of requests across operations are the input to the controller. The set point our controller guides toward is the cluster state associated with this measurement input. We examine two sources of error in this control loop: controller input lag and sensor noise.

For \sysname{}, there is a lag between when a workload changes (e.g. aggregate RPS increases) and our controller is notified of this change. For our evaluation setting, workload metrics are logged by Google Cloud Monitoring agents. We use the default logging settings provided by Google Cloud for all experiments which means agents that collect and write new logs pertaining to requests do so every 60 seconds.

To better understand the effect of this input lag, we run an experiment where we invoke a load change from no load to a new request per second rate. To isolate this input lag, in this experiment virtual machines have been pre-allocated and autoscalers must simply increase the number of pods in response to an input load. We observe that it takes between 60-90 seconds to react to a load change once a workload has changed enough to warrant action. Although \sysname{} takes longer to start scaling up compared to CPU threshold based autoscaling policies, the cluster state for the new workload and corresponding latency reaches a steady state more quickly.

Secondly, the number of requests we log can be a source of noise. In our settings we empirically find this to not be the case. We evaluate the mean average percent error (MAPE) of five 1-minute intervals of a workload against our measurement taken during training to build our context. The MAPE for Online Boutique, BookInfo and Hello World are .58\%, .34\% and 1.1\% respectively.

\begin{figure*}[!t]
\subfigure{%
	\label{fig:samp-duration-estimation}
	\includegraphics[width=8cm]{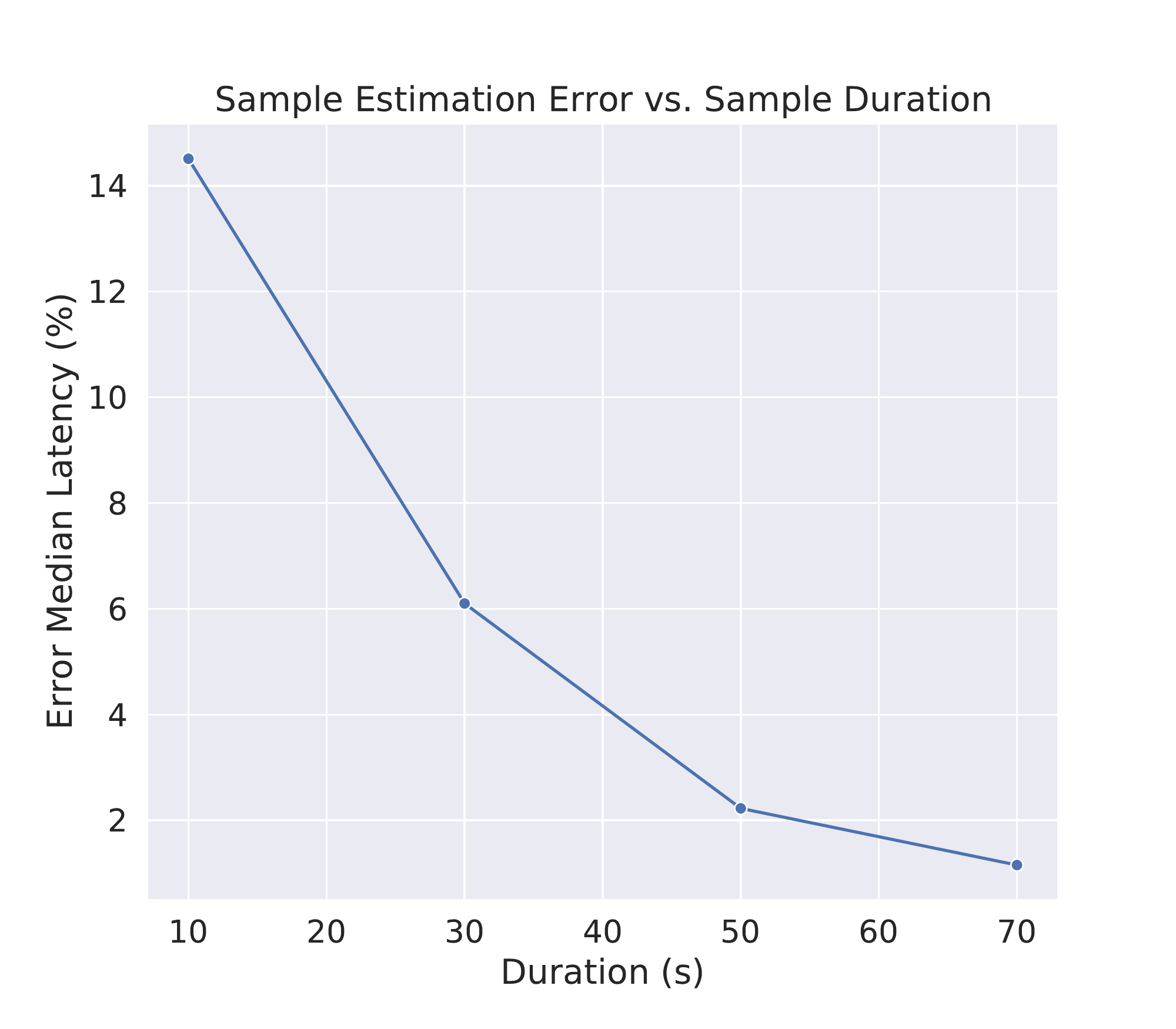}
}%
\hspace*{\fill}
\subfigure{
	\label{fig:samp-duration-estimation-tail}
	\includegraphics[width=8cm]{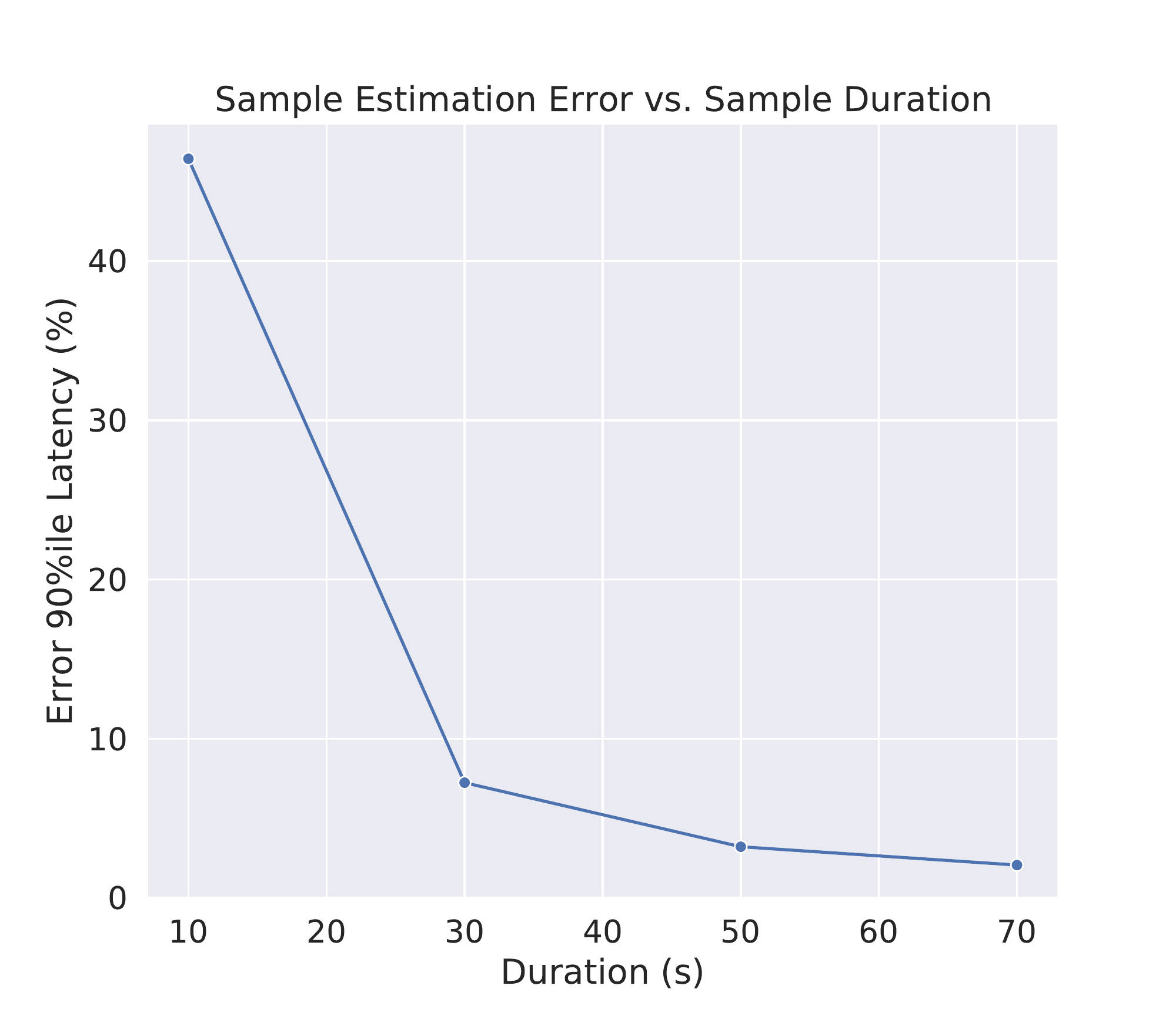}
}%
\captionsetup{justification=centering}
\caption{Latency Sampling Error (\%) for Median (Left) and 90\%ile Latency (Right) vs. Sample Duration (s)}\label{fig:sampling-error}
\end{figure*}

\Para{Sample Duration and Measurement Variation}\label{subsec:sample-duration-variation} We evaluate the selection of a training procedure hyperparameter: sample duration. The duration of our samples can determine the impact of transient noise on our reward estimation during training. Longer training times can help reduce the variance in latency and reward estimation for a fixed cluster configuration and request rate as we average latency over more samples. However, taking longer samples increases the dollar cost of training since we will run a training cluster for a longer period of time. In Figure \ref{fig:samp-duration-estimation}, we show the sample estimation error as it relates to sample duration on one configuration of the Online Boutique application. We calculate the mean percent error of various sample durations from 10 to 80 seconds against a ground truth held out sample of 90 seconds.



\Para{Other robustness experiments} We evaluate a few other settings to further understand failure cases and extensibility of \sysname{}. We evaluate out of range workloads in which we issue a request per second rate larger than the upper bound that \sysname{} is trained for. \sysname{} fails over to a preset Kubernetes CPU autoscaling threshold as shown in Figure \ref{fig:fr-oor-ob}. Additionally, we show the coexistence of \sysname{}'s horizontal pod autoscaler with GKE's cluster autoscaler on the Book Info and Online Boutique applications in Figures \ref{fig:bi-noca} and \ref{fig:ob-noca}.





\subsection{Learned Autoscaler Training Times}

We show the training time for learned autoscalers apart from \sysname{} in Tables \ref{tab:training-cost-lr}-\ref{tab:training-cost-dqn}. The Linear Regression, Bayesian Optimization and Deep Q-Network learned autoscaling policies use more instance hours time during training. Since the space of microservice configurations is explored in ascending size across training samples \sysname{} allocates only a fraction of the full instances needed for the maximum replica range. During training, if the number of replicas grows we add more instances to the training node pool. Both DQN and Bayesian Optimization operate by freely exploring the entire replica range with each subsequent training example requesting an arbitrary number of instances. Consequently we can not perform this optimization for these policies. For the Linear Regression policy we know the full set of training examples before training and consequently a version of this optimization can be performed. The optimization procedure for Linear Regression could sort the full set of training examples by ascending resources requested and incrementally increase the cluster size.


\begin{table}[H]
  \begin{tabular}{llll}
    \toprule
    Application&Time (hrs) & Instance Hrs & Cost (\$)\\
    \midrule
  Simple Web Server&  4.16& 308.33&  \$13.94 \\
  Book Info &  3.89&  346.11&  \$15.79 \\
  Online Boutique&   5.03&  874.00&  \$40.68 \\
  Sock Shop &  6.94&  1000.00&  \$46.33 \\
    Train Ticket&   7.29& 5060.42&  \$239.14 \\
  \bottomrule
\end{tabular}
\caption{Training Cost (Linear Regression Median Latency)\protect\footnotemark}
\label{tab:training-cost-lr}
\end{table}

\begin{table}[H]
  \begin{tabular}{llll}
    \toprule
    Application&Time (hrs) & Instance Hrs & Cost (\$)\\
    \midrule
  Simple Web Server&  4.19& 309.88&  \$14.01 \\
  Book Info &  3.91&  406.47&  \$18.65 \\
  Online Boutique&   5.86&  1020.08&  \$47.46 \\
  Sock Shop &  6.98&  1005.00&  \$46.56 \\
    Train Ticket&   7.32& 5080.66&  \$240.09 \\
  \bottomrule
\end{tabular}
\caption{Training Cost (Bayesian Opt. Median Latency)\protect\footnotemark[\value{footnote}]}
\label{tab:training-cost-bo}
\end{table}
\footnotetext{Time and cost metrics are estimated from number of samples, time per sample, and total number of instances during training. Other tables show metrics obtained from direct measurement rather than estimation.}

\begin{table}[H]
  \begin{tabular}{llll}
    \toprule
    Application&Time (hrs) & Instance Hrs & Cost (\$)\\
    \midrule

  Simple Web Server&  4.22&   312.72&  \$14.14 \\
  Book Info &  3.96&  412.44&  \$18.92 \\
  Online Boutique &  5.96&  1156.84&  \$53.95 \\
  Sock Shop&   7.07& 1018.08&  \$47.17 \\
  Train Ticket&   7.81& 5579.94&  \$263.73 \\
  \bottomrule
\end{tabular}
\caption{Training Cost (DQN Median Latency)}
\label{tab:training-cost-dqn}
\end{table}

\subsection{Analyzing Learned Autoscaler Shortcomings}
\label{subsec:la-shortcomings}

\begin{figure*}[!t]
	\centering
	\includegraphics[width=\textwidth]{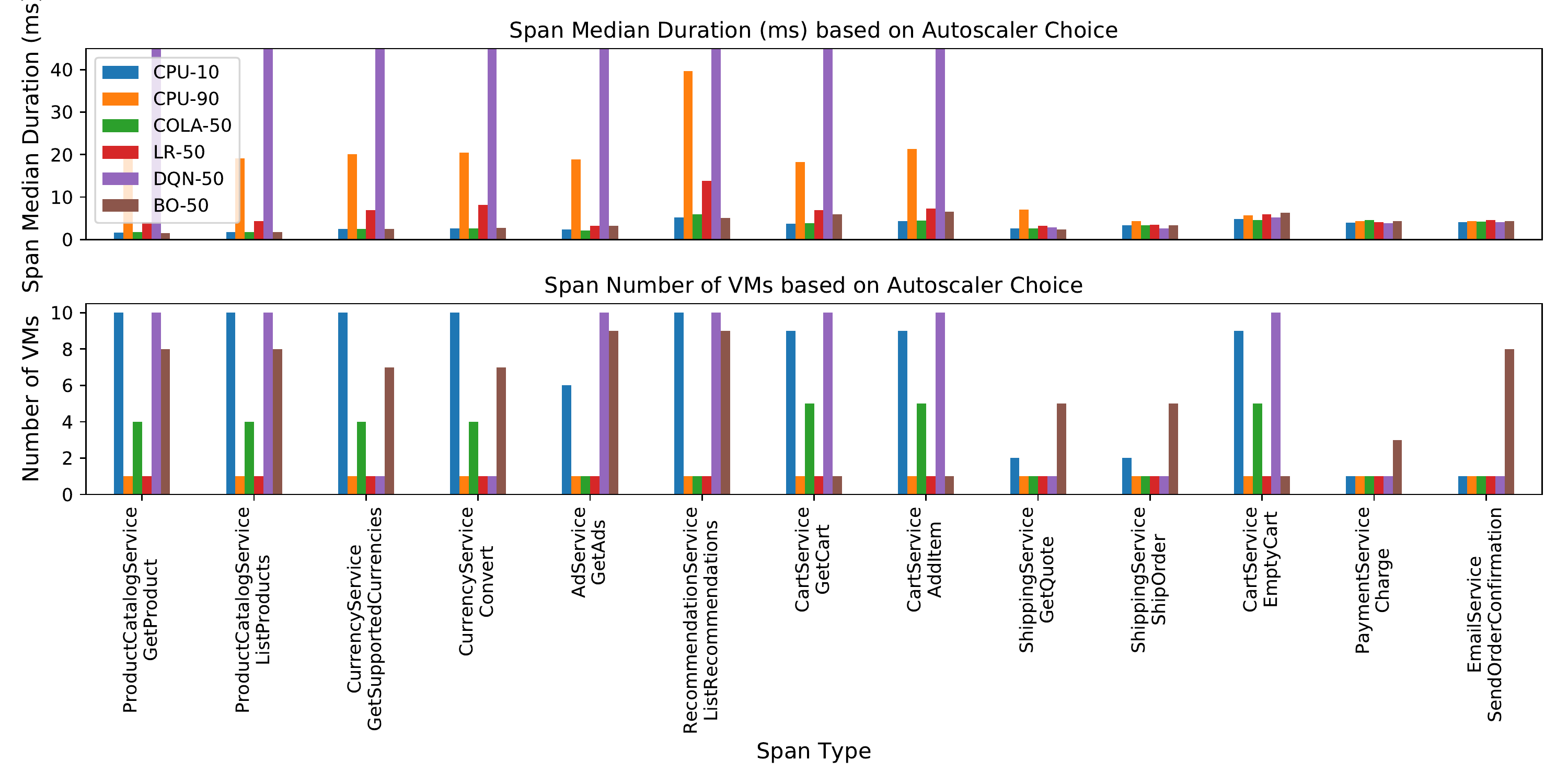}
    \caption{Microservice Span Durations and Number of VMs by Autoscaler Choice.}
	\label{fig:ml-baseline-allocation}
	\vspace{-0.5cm}
\end{figure*}

\begin{figure*}[!t]
	\centering
	\includegraphics[width=.9\textwidth]{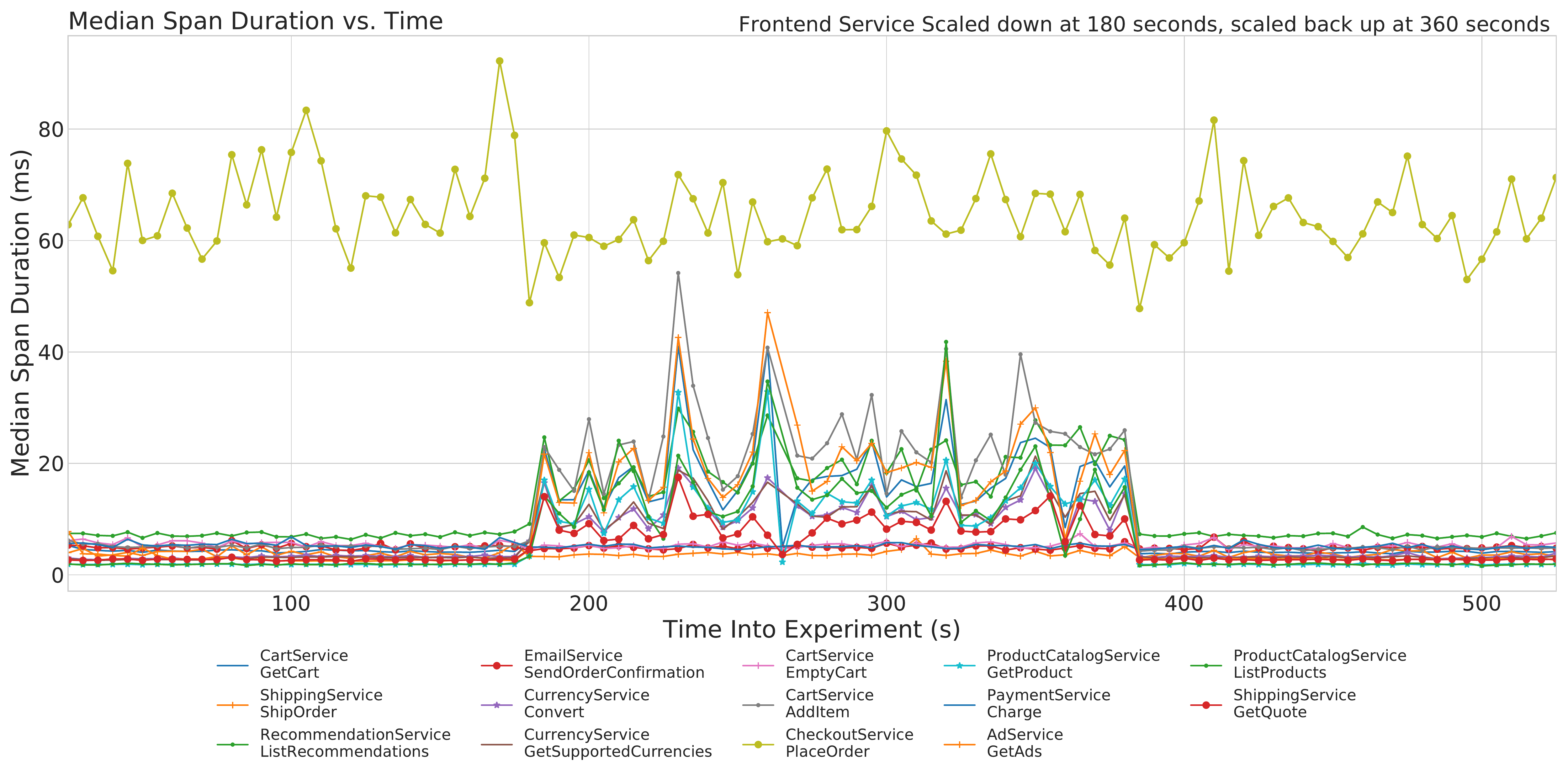}
    \caption{Effect of scaling down \texttt{frontend} service on other microservices in Online Boutique.}
	\label{fig:frontend-time-series}
	\vspace{-0.5cm}
\end{figure*}

\begin{table}
  \begin{tabular}{lrr}
    \toprule
    Autoscaler & Median Span Duration (ms) & Num Instances\\
    \midrule
   CPU-10 &   3.30 &  85 \\
   CPU-90 &   15.81 &  13 \\
   \sysname{}-50 & 3.41 & 26 \\ 
   LR-50 &  5.86  & 15  \\
   BO-50 &  3.84 &  81 \\
   DQN-50 &  117.74 & 65 \\
  \bottomrule
\end{tabular}
\caption{Span Durations and Instances by Autoscaler}
\label{tab:summary-span-duration}
\vspace{-0.25cm}
\end{table}

We analyze the scaling decisions of \sysname{} in relation to other learned autoscaling policies - Linear Regression, Bayesian Optimization and Deep Q-Learning. \sysname{} outperforms these benchmarks on all applications. In Figure \ref{fig:ml-baseline-allocation} we show span durations and number of instances for the Online Boutique application with 400 synthetic users. Table \ref{tab:summary-span-duration} shows a summary of median latency and total number of instances for these policies. 

We find that Linear Regression and Deep Q-Learning result in higher span durations than \sysname{}. In these two cases, the policies allocated 32.6\% and 85.7\% fewer \texttt{frontend} instances than \sysname{}. As detailed in controlled experiments from \S\ref{sec-evaluation} (Figure \ref{fig:span-duration-by-autoscaler}), the frontend instance has downstream latency effects on several other microservices. We perform another controlled experiment to show this effect which can be seen in Figure \ref{fig:frontend-time-series}. In this Figure, we show a timeseries of several spans in our microservice. The first and last 180 seconds in the experiment we apply \sysname{}'s policy. The middle 180 seconds we scale down the frontend to CPU-90's value. From the figure we can observe that while some spans are uneffected in duration from this policy change, the majority of spans rise in latency as a result of scaling down the upstream \texttt{frontend} service. Lastly, the Bayesian Optimization autoscaler results in similar span duration to \sysname{} but expends 3.1$x$ the resources that \sysname{} does. Generally, we observe that our learned autoscaling baselines either allocate too few or too many resources to our microservice cluster. \sysname{} relies on system heuristics in addition to machine learning to prioritize resource allocation to microservices which reduce end to end latency significantly.

\subsection{Training Trajectory Visualizations}

In Figures \ref{fig:sws-training-trajectory}-\ref{fig:ob-training-trajectory} we show the evolution of learned policies during training for our machine learned autoscalers. For \sysname{}, DQN, and Linear Regression we train on the same set of RPS values in our range and compare latency, number of instances and reward as a function of training samples for each of these RPS values. Across the applications tested, we find that \sysname{} is able to converge to a latency which meets our target faster than other machine learned autoscalers. Further, \sysname{} reaches a steady state reward which is better or equal to other autoscalers for all RPS values we examine.


\subsection{Service Selection}

In Figure \ref{fig:bi-training-trajectory-ablation}, we show the effect of altering service selection, a routine described in \S\ref{subsubsec:select-service}, on the Bookinfo application. For all applications \sysname{} chooses services to optimize by prioritizing those with high CPU utilization. We also show in Figure \ref{fig:bi-training-trajectory-ablation} training trajectories where we instead select services by high memory utilization (\sysname{}-mem) and where we select services randomly (\sysname{}-random). We find that in our setting, prioritizing the optimization of services high in CPU utilization works best with autoscalers more quickly finding satisfactory configurations for provided workloads. This may be heavily influenced by our selection of applications which are latency sensitive web applications.

\subsection{Reducing Collateral Damage from Congestion.} From Figure \ref{fig:span-duration-by-autoscaler}, we notice that \sysname{} also reduces the duration of spans on microservices which are not scaled up. We find that this benefit comes from reducing the latency of other spans, which are dependencies of these non-scaled-up spans. As noted in DeathStarBench~\cite{gan2019open}, microservice applications are difficult to optimize as congested upstream microservices can cause ``hotspots'' and introduce queueing delays to downstream microservices as well. We deploy a version of the \sysname{} autoscaling policy, shown in Figure \ref{fig:span-duration-by-autoscaler} as "COLA-50-FE", where we scale down the \texttt{frontend} microservice (which is upstream to all microservices) to the same number of VMs as the overutilized CPU-90 policy. We observe that the span durations increase for almost every span as a result of scaling down only the frontend.

Secondly, DeathStarBench~\cite{gan2019open} documents how backpressure can occur when an upstream microservice span blocks and waits for a reply from a downstream microservice. We find a case of this type of behavior in the Online Boutique microservice code where the \texttt{recommendationservice} code fetches a list of products from the \texttt{productcatalogservice}, blocking until this list is received. We run a second modified version of \sysname{} where we scale down the  \texttt{productcatalog service} to the same value as CPU-90, the overutilized system. As a result the \texttt{listproducts} span duration increases by 85\% (or 1.5ms). Although the resources available to the \texttt{recommendationservice} are unchanged, the \texttt{list \\ recommendations} span duration increases by .85 ms or 13\% of the total span duration. This illustates the benefits of \sysname{}'s global visibility: it observes the consequences on end-to-end latency of each VM allocation choice it makes.

\subsection{\sysname{} Empirical Optimality}
\label{subsec:optimality-exp}

\begin{figure*}
\centering
\subfigure{%
	\includegraphics[width=6cm]{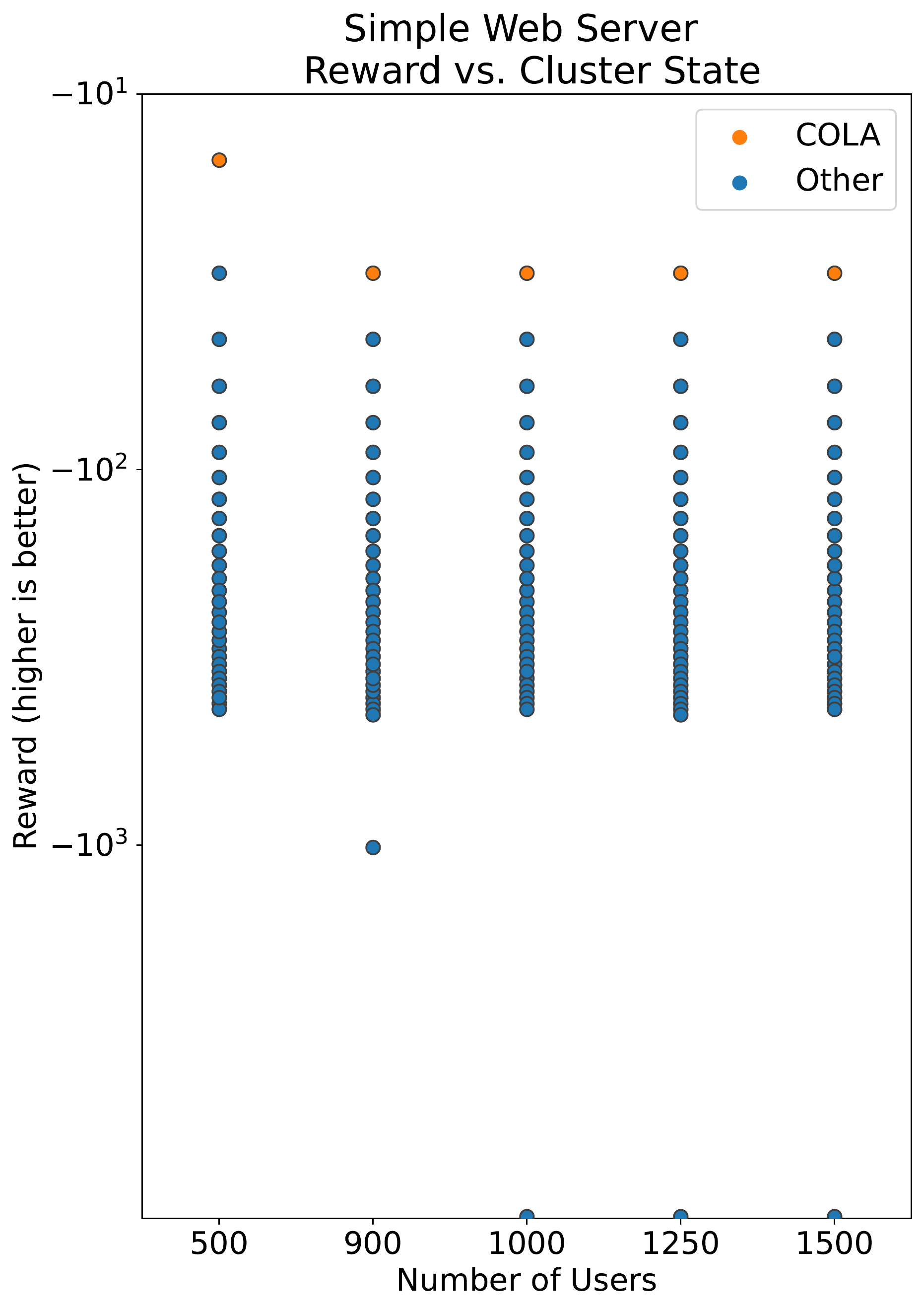}
}
\subfigure{
    \includegraphics[width=6cm]{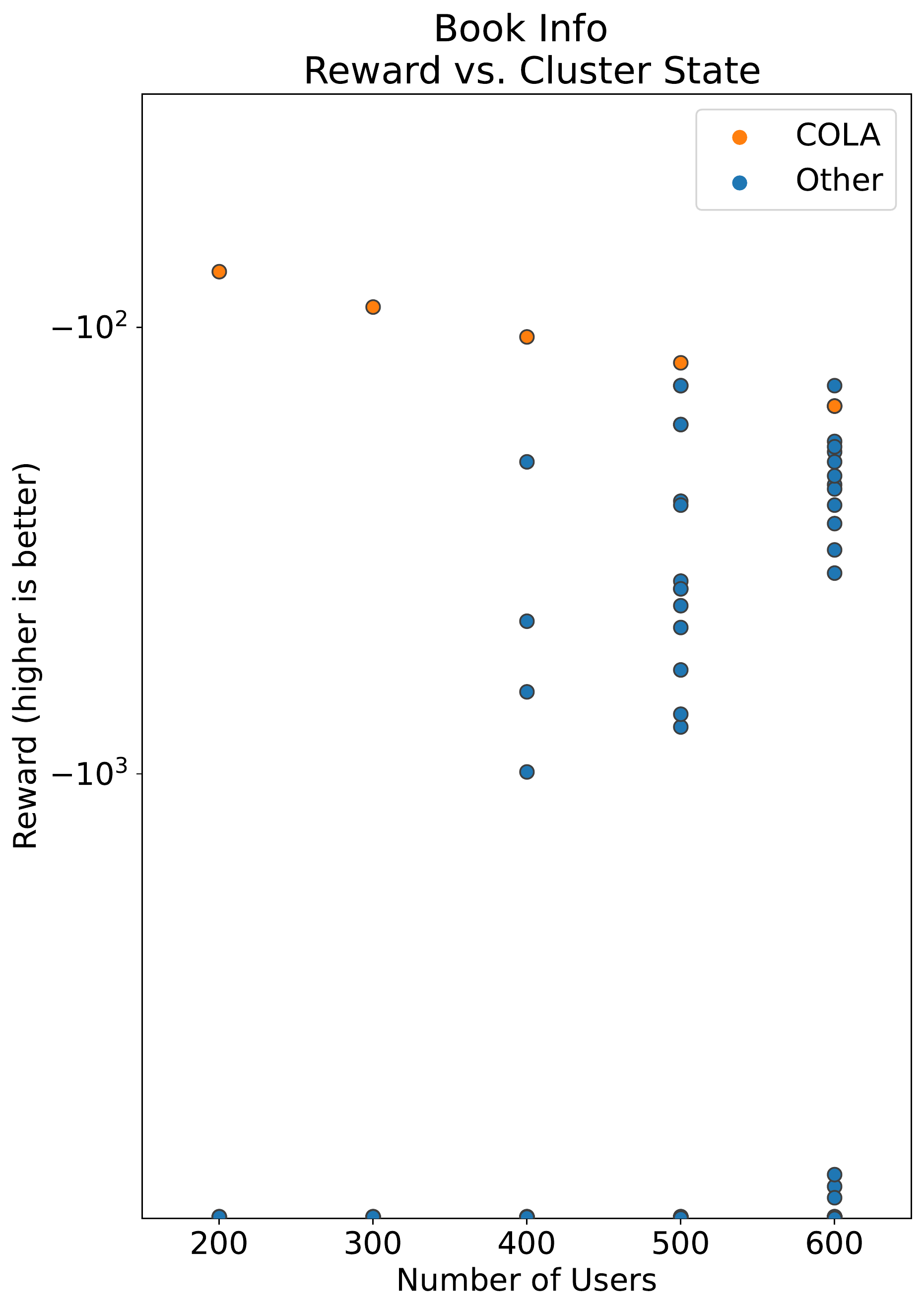}
}%
\captionsetup{justification=centering}
\caption{Reward for each Cluster State for Simple Web Server (left) and Book Info (right). COLA's selected cluster state (orange) is optimal for 9/10 workloads and second best configuration for the other workload.}
\label{fig:optimality-sws-bi}
\end{figure*}

We show evaluation of \sysname{} for empirical optimality -- finding the best possible cluster state for our given reward. Figure \ref{fig:optimality-sws-bi} shows the reward for each cluster state searched with \sysname{}'s solution in orange and all other cluster states in blue. We run each workload for 90 seconds per cluster state and calculate reward for a median latency target of 50ms. \sysname{} is optimal for 9/10 application-workload pairs which we evaluated and is the second best configuration on the last pair. \sysname{} has been trained on 6/10 of these evaluated application-workload pairs and has not seen 4/10. \sysname{} costs .9\% more on average than the optimal cluster state across these experiments.

During training, \sysname{} takes 10 samples per workload for the Simple Web Server application and 13.3 samples on average per workload for Book Info. In total there are 30 possible cluster states for the Simple Web Server application and 320,000 for the Book Info application. We compare the search strategy used by \sysname{} with random search and Bayesian Optimization which we allot 10 samples per workload for the Simple Web Server application (1x \sysname{}'s number of samples) and 40 samples for the Book Info application (3x \sysname{}'s number of samples). The random search procedure takes the alloted samples and chooses the best configuration out of the samples for the workload. Bayesian Optimization is implemented as a Gaussian Process Regressor with a feature vector consisting of number of replicas for each microservice. We train a seperate model per workload, using the implementation from the scikit-learn package \cite{scikit-learn}. Overall, we find that \sysname{} significantly outperforms both random search and Bayesian Optimization and show results in Tables \ref{tab:emp-optimality-sws} and \ref{tab:emp-optimality-bi}.














\subsection{Failed Requests across Workloads}

In Tables \ref{tab:bi-fr-tabular}-\ref{tab:ob-drd-tabular} we include the failed requests per second across all workloads on which we evaluate. For CPU autoscalers and \sysname{} we observe fewer than .1\% of requests being timed out by the client except for in one case for the Train Ticket application. During training and evaluation, requests may respond with a 4xx or 5xx response code and are consequently registered as a failed request. For example, a request may timeout at the client or microservice cluster and be considered a failed request. 

During both training and evaluation we use a timeout of 2000ms on the client side and 30 seconds on the server. This timeout is applied for a few reasons. Firstly, we send requests through a connection pool in which each connection is able to issue a request at a given interval which will decide the aggregate requests per second issued to the cluster. If timeout limits are unbounded we must adjust the size of the connection pool. In order to avoid arbitrarily opening new connections we apply a timeout, $t$, for the connection pool. This value $t$ is also the interval at which we send requests for each connection. In order to issue an aggregate number of requests per second, $r$, to our cluster we adjust the number of connections in our pool once on startup to $r \cdot t$. During evaluation we retain this client side timeout which in this case has a second purpose -- to mimic user requests "bouncing" after some significant waiting time. The amount of time a user waits to "bounce" and reissue a request can depend heavily by application among other factors. However, we find that the timeout we specify is roughly in line with the average bounce time in modern web applications based on an industry benchmark from Google \cite{bounce-rate:online}.

\begin{table}
  \begin{tabular}{lrrrr}
    \toprule
    Autoscaler & Users & Reward & Latency & Replicas\\
    \midrule
   \sysname{}-50 &  500 & -15.0  &  46.0 & 1.0 \\ 
   Random        &  500 & -60.0  &  45.0 & 4.0 \\
   Bayesian Opt. &  500 & -15.0  &  46.0 & 1.0 \\
   \midrule
   \sysname{}-50 &  1000 & -30.0  &  50.0 & 2.0 \\ 
   Random        &  1000 & -60.0  &  46.0 & 4.0\\
   Bayesian Opt. &  1000 & -225.0 &  45.0 &  15.0\\
   \midrule
   \sysname{}-50 &  1500 & -30.0  &  50.0 & 2.0 \\ 
   Random        &  1500 & -30.0  &  50.0 & 2.0\\
   Bayesian Opt. &  1500 & -240.0 &  45.0 & 16.0  \\
  \bottomrule
\end{tabular}
\caption{Simple Web Server Results by Search Procedure}
\label{tab:emp-optimality-sws}
\vspace{-0.25cm}
\end{table}

\begin{table}
  \begin{tabular}{lrrrr}
    \toprule
    Autoscaler & Users & Reward & Latency & Replicas\\
    \midrule
   \sysname{}-50 &  200 & -75.0   &  30.0 & 5.0 \\ 
   Random        &  200 & -210.0  &  33.0 & 14.0\\
   Bayesian Opt. &  200 & -660.0  &  37.0 & 44.0\\
   \midrule
   \sysname{}-50 &  400 & -105.0  &  38.0 & 7.0 \\ 
   Random        &  400 & -210.0  &  36.0 & 14.0\\
   Bayesian Opt. &  400 & -180.0  &  31.0 & 12.0\\
   \midrule
   \sysname{}-50 &  600 & -150.0  &  45.0 & 10.0 \\ 
   Random        &  600 & -255.0  &  56.0 & 15.0\\
   Bayesian Opt. &  600 & -480.0  &  53.0 & 31.0\\
  \bottomrule
\end{tabular}
\caption{Book Info Results by Search Procedure}
\label{tab:emp-optimality-bi}
\vspace{-0.25cm}
\end{table}

Although our latency based timeout is respected in the vast majority of experiments, we find one such experiment where this is not true (in Table \ref{tab:ob-fr-tabular}). Locust \cite{locust:online}, the load generator we use to issue requests and collect aggregate statistics, uses an http client exposed by the gevent \cite{gevent:online} library for all of our experiments. Within the source code of this library, the authors mention that there may be a "fuzz" in the time taken to interrupt the http client for a specific timeout. This fuzz can cause the maximum latency to be larger than the timeout specified and inevitably occurs due to resource contention. For our setting, this fuzz in handling timeout operations effects very few evaluations. Further, it only effects the worst autoscaling policies for a given application and workload.



\subsection{Out of Range Evaluation}

We evaluate the graceful failover in \sysname{} for the Online Boutique application. We train with an upper limit of 200 users and ask \sysname{} to fail over to a 50\% CPU based autoscaling policy when the observed requests are at least 30\% larger than the largest context it has trained on. The results of one evaluation when 600 users request the system are shown below. In Figure \ref{fig:fr-oor-ob} we can see the scaling up of instances and median latency as \sysname{} transitions to a CPU autoscaler.

\begin{figure*}
\subfigure{%
	\includegraphics[width=8cm]{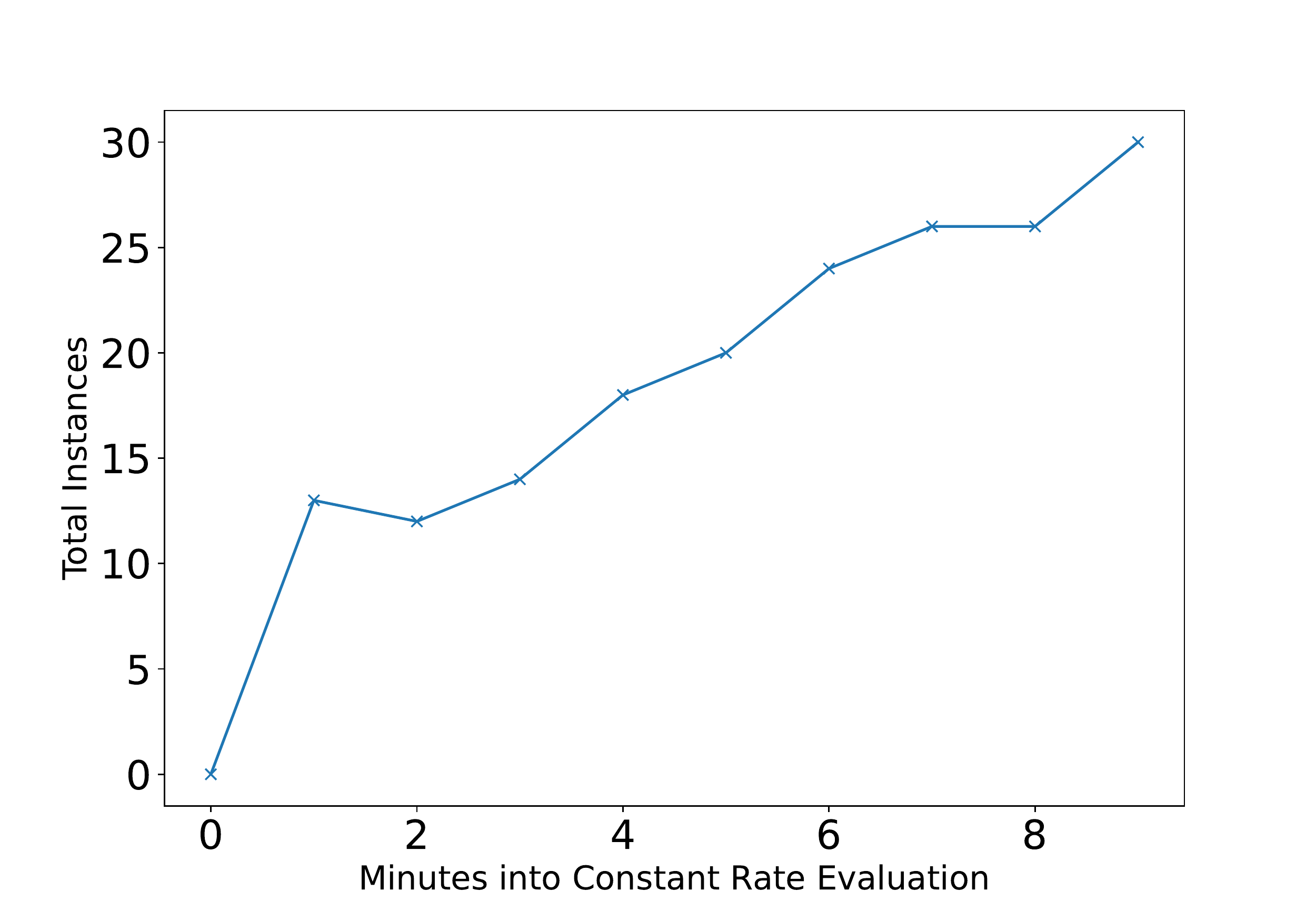}
}%
\hspace*{\fill}
\subfigure{
	\includegraphics[width=8cm]{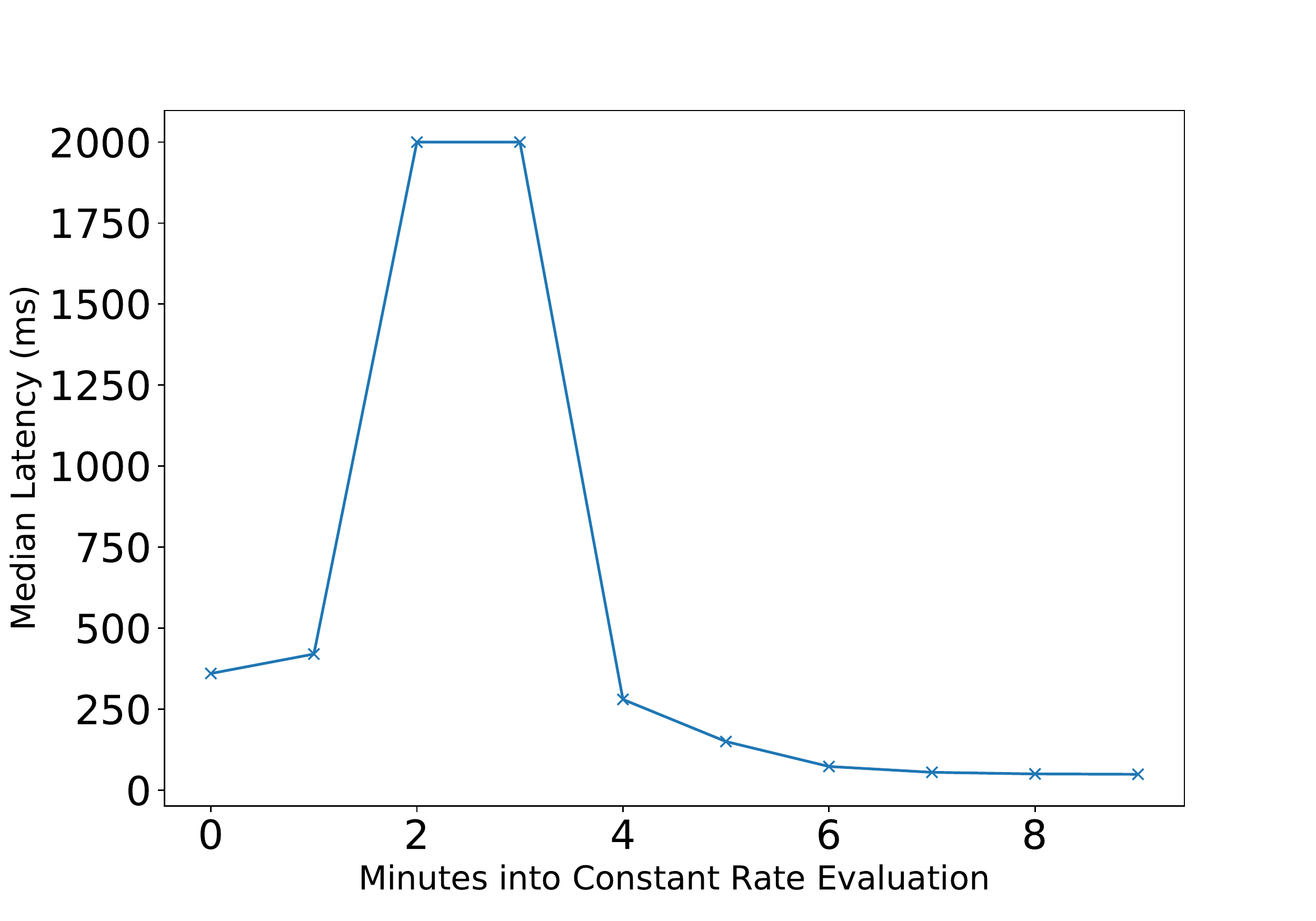}
}%
\captionsetup{justification=centering}
\caption{Online Boutique Out of Range Evaluation, Total Instances (left) and Median Latency (right) over time.}
\label{fig:fr-oor-ob}
\end{figure*}




\subsection{Coexistence with GKE Cluster Autoscaler}
\label{subsec:coexistence-noca}

We evaluate a few Constant Rate workloads with Google Kubernetes Engine handling the cluster autoscaling and \sysname{} handling horizontal pod autoscaling for Book Info in Figure \ref{fig:bi-noca} and Online Boutique in Figure \ref{fig:ob-noca}. Performance is comparable between the cluster autoscaler we develop and the cluster autoscaler offered by GKE.

%

%







\subsection{Bandit Algorithm Selection}
\label{subsec:bandit-algorithm-selection}

\begin{figure*}
\subfigure{%
  \includegraphics[width=0.31\textwidth]{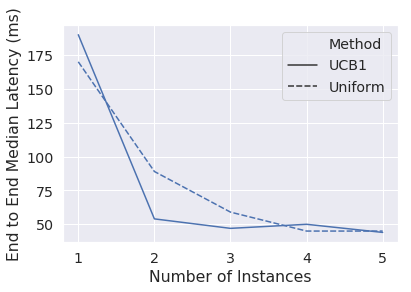}%
}  %
\hspace*{\fill}
\subfigure{
  \includegraphics[width=0.31\textwidth]{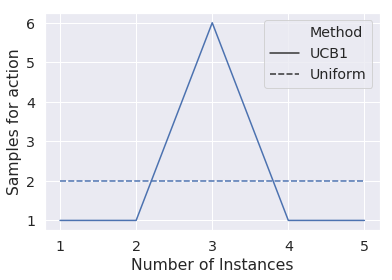}%
}  %
\hspace*{\fill}
\subfigure{
  \includegraphics[width=0.31\textwidth]{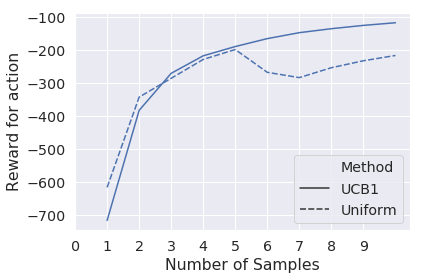}%
}

\captionsetup{justification=centering}
\caption{UCB1 vs. Random Sampling}\label{fig:bandit-search-strat}
\end{figure*}

\begin{figure*}
\subfigure{%
  \includegraphics[width=0.31\textwidth]{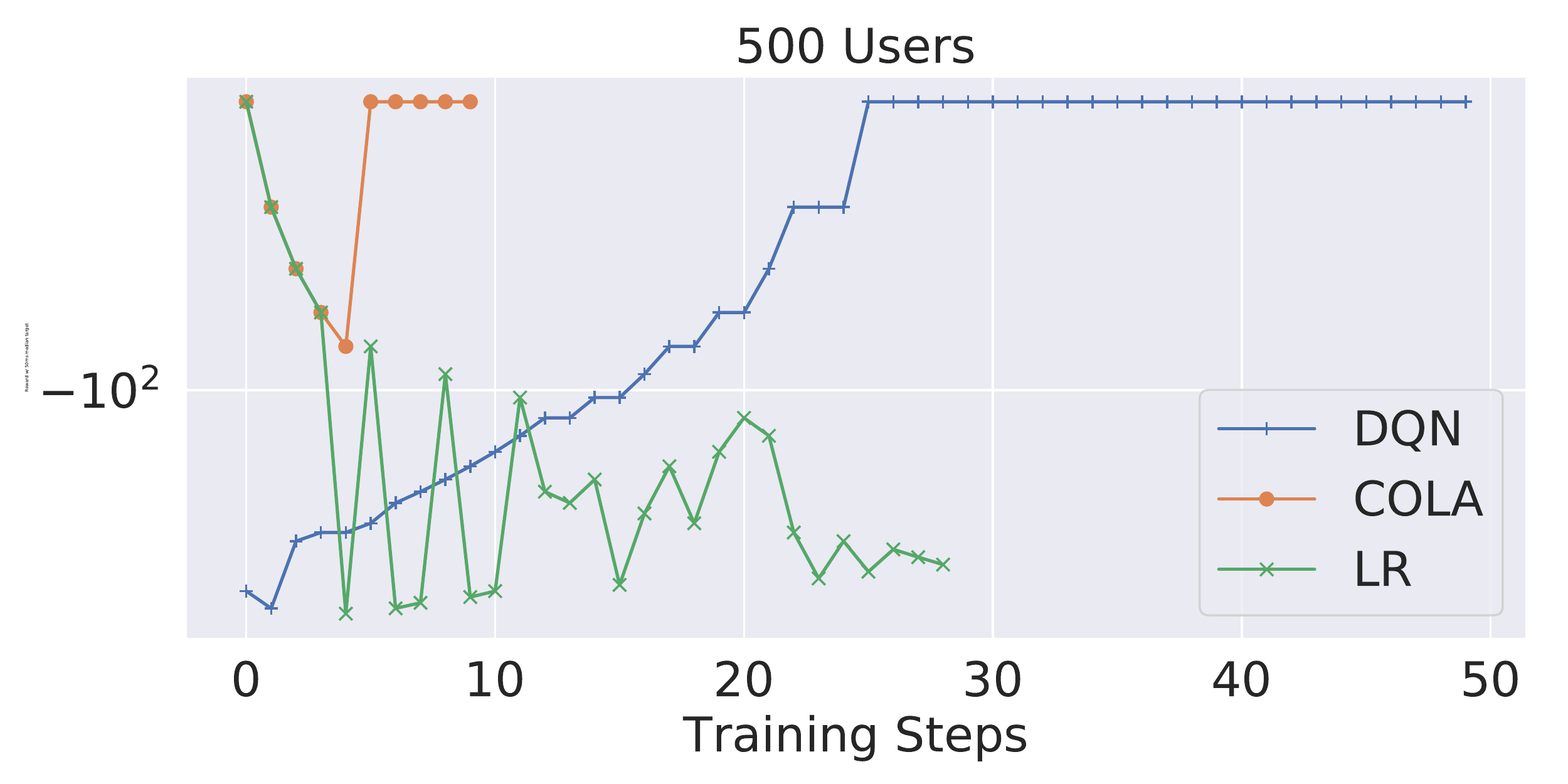}%
}  %
\hspace*{\fill}
\subfigure{
  \includegraphics[width=0.31\textwidth]{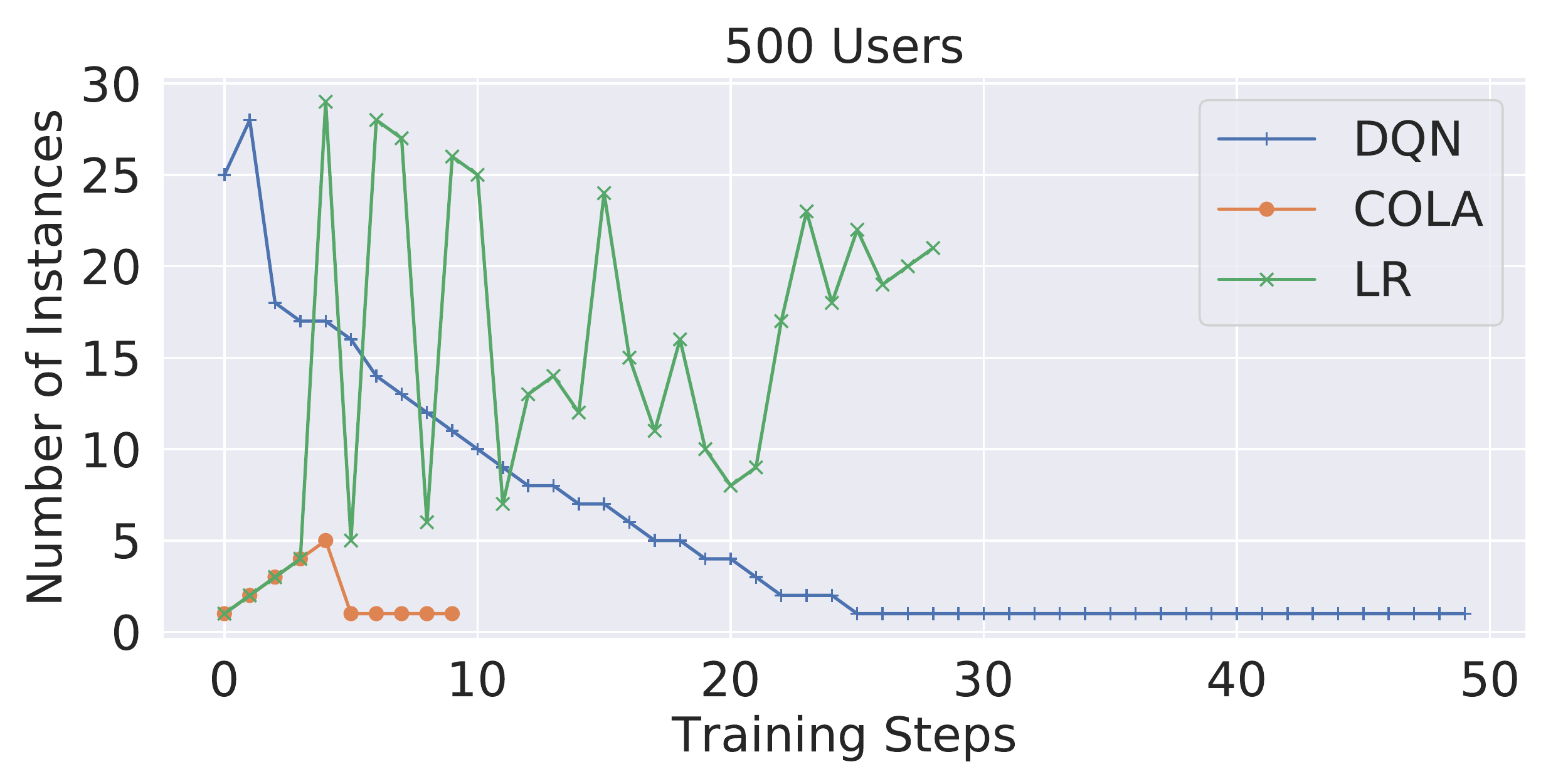}%
}  %
\hspace*{\fill}
\subfigure{
  \includegraphics[width=0.31\textwidth]{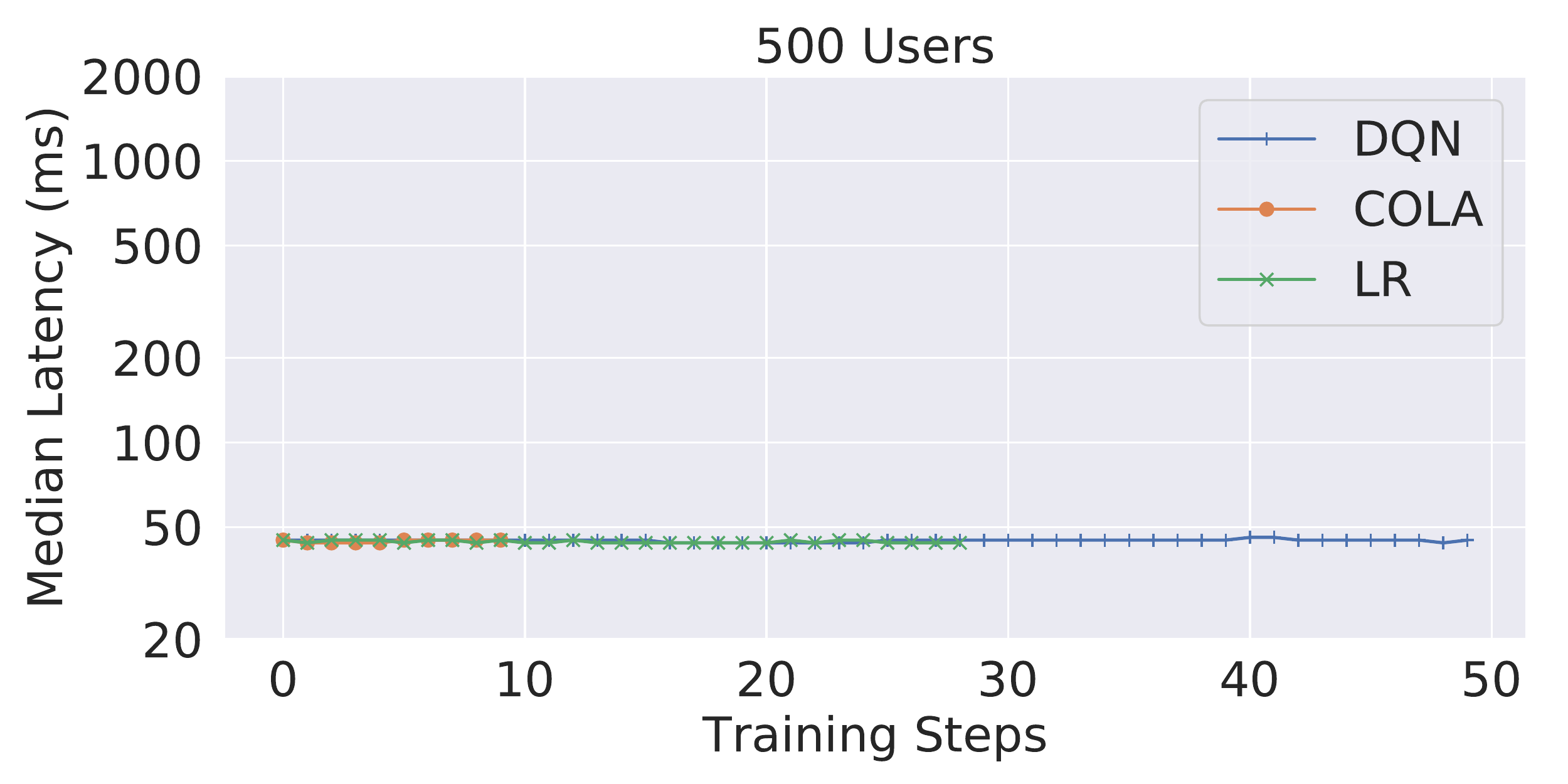}%
}

\subfigure{%
  \includegraphics[width=0.31\textwidth]{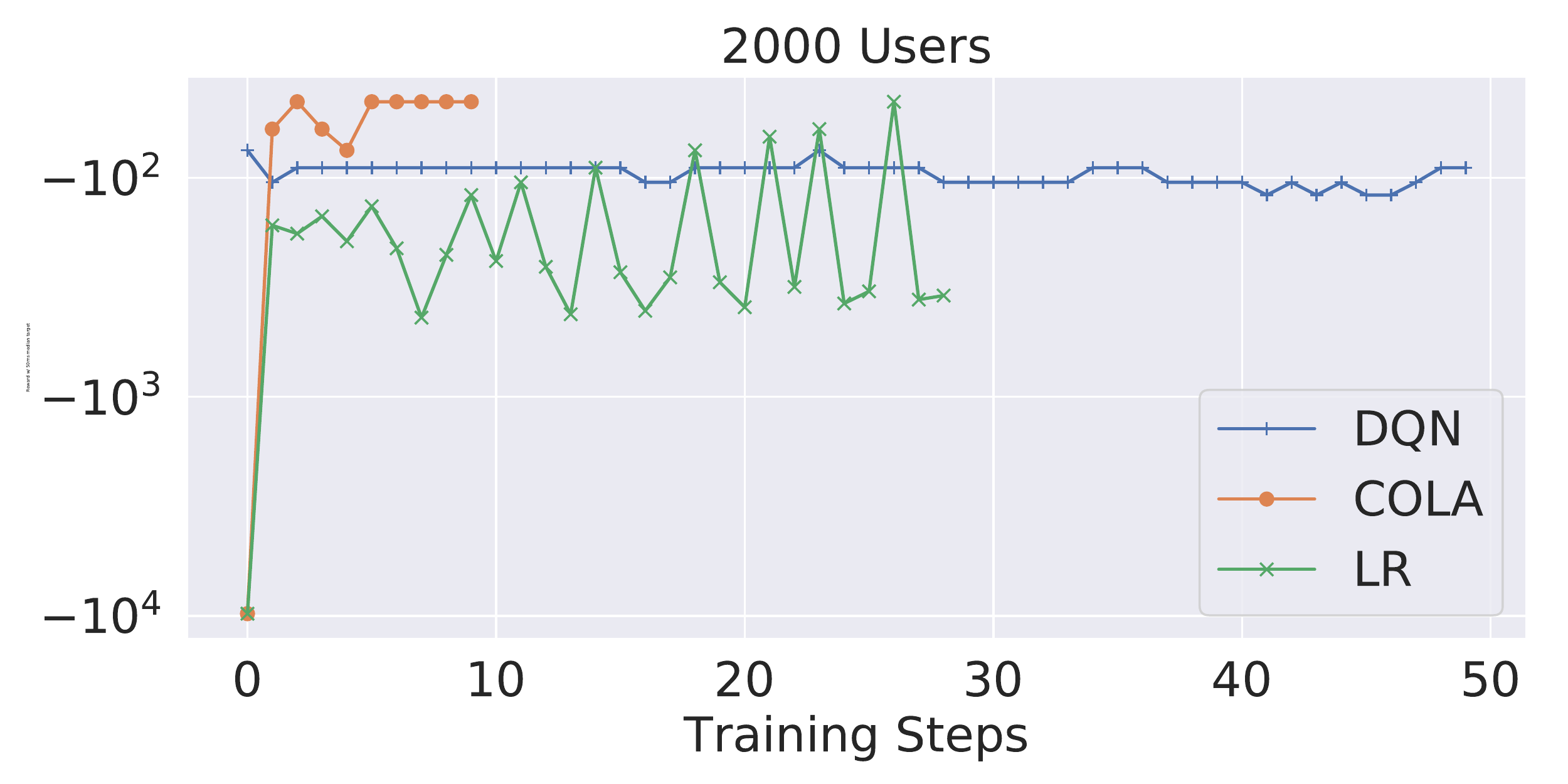}%
}  %
\hspace*{\fill}
\subfigure{
  \includegraphics[width=0.31\textwidth]{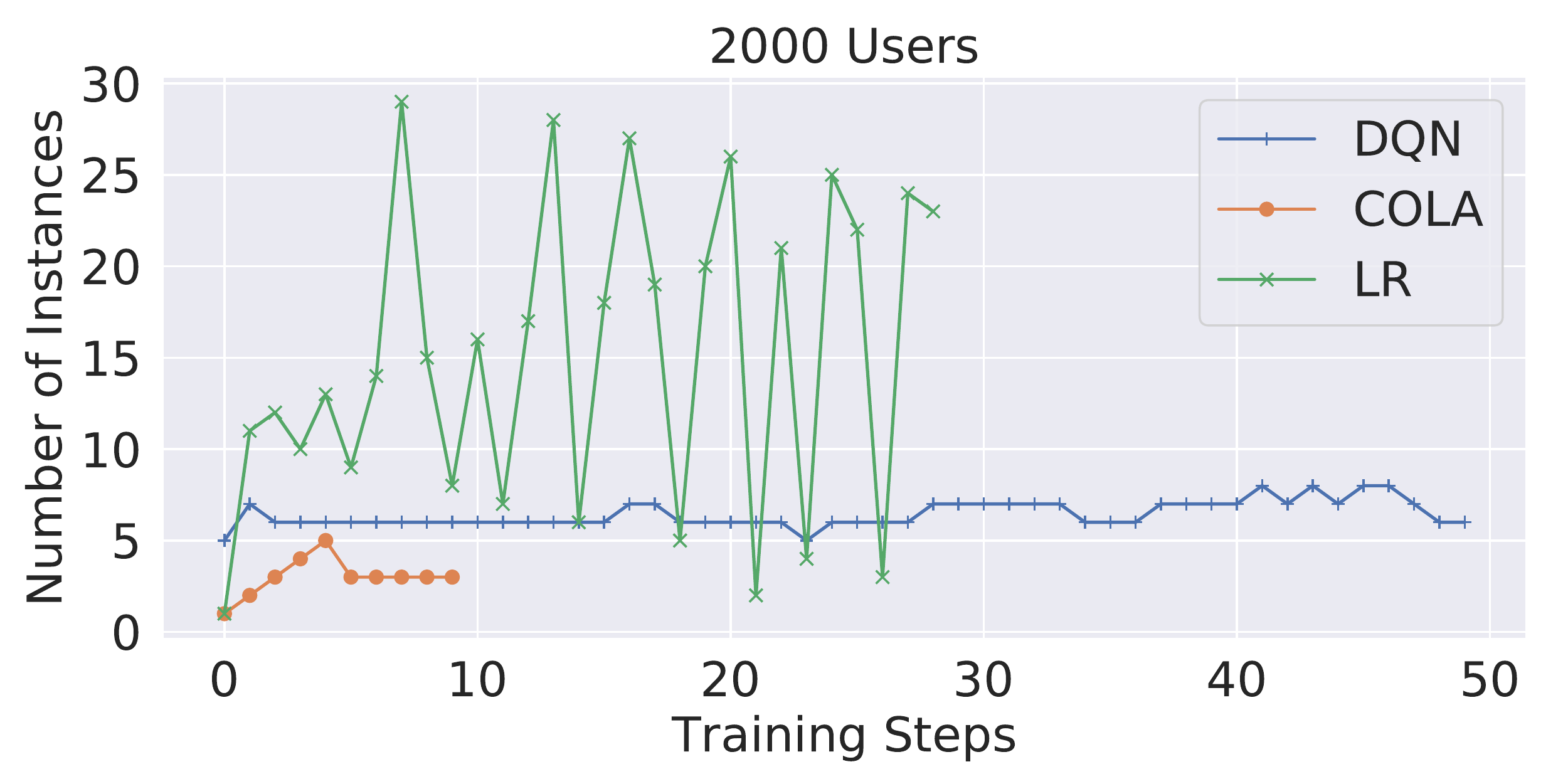}%
}  %
\hspace*{\fill}
\subfigure{
  \includegraphics[width=0.31\textwidth]{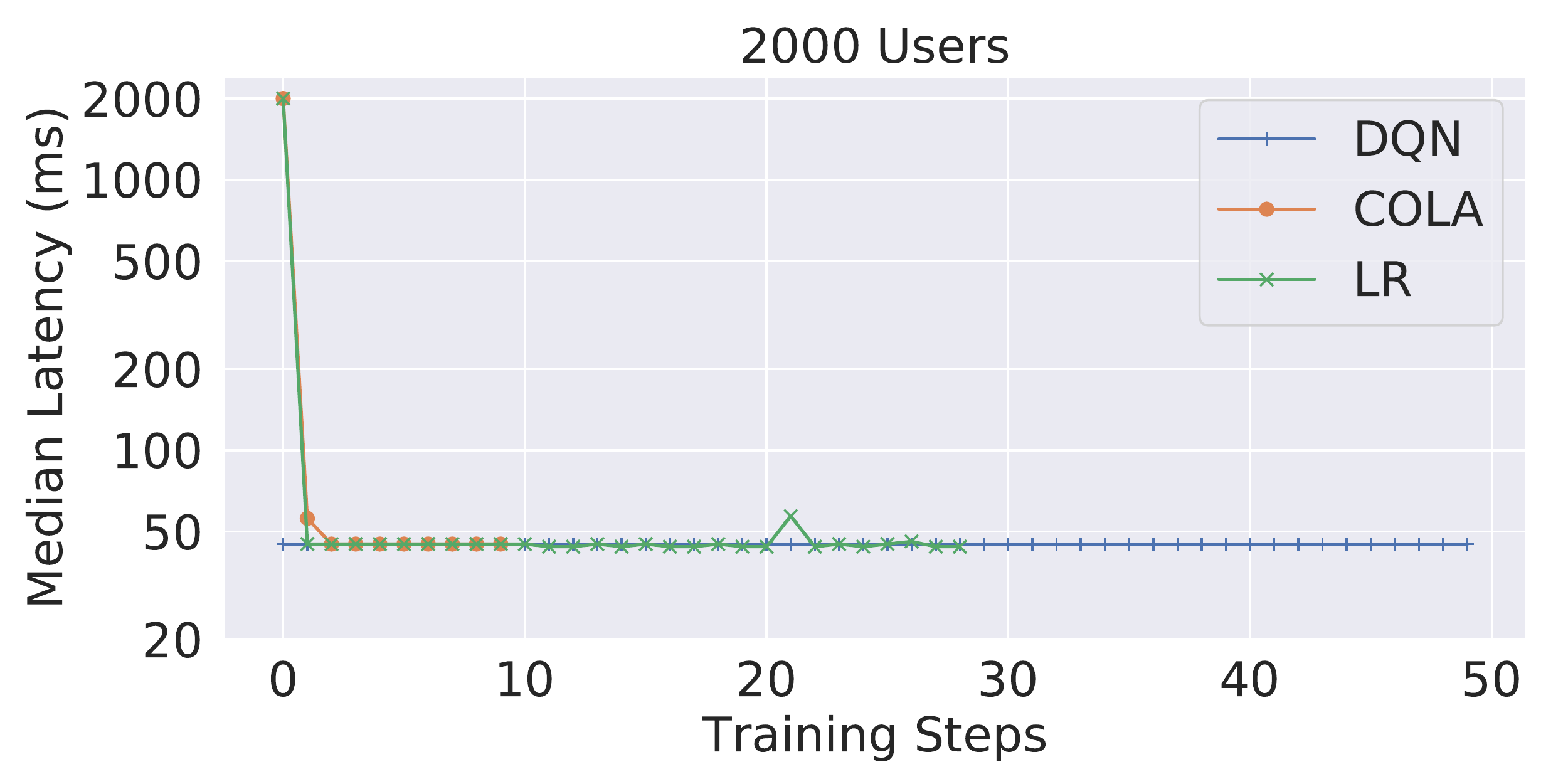}%
}

\captionsetup{justification=centering}
\caption{Simple Web Server Learned Autoscaler Trajectories}\label{fig:sws-training-trajectory}

\subfigure{%
  \includegraphics[width=0.31\textwidth]{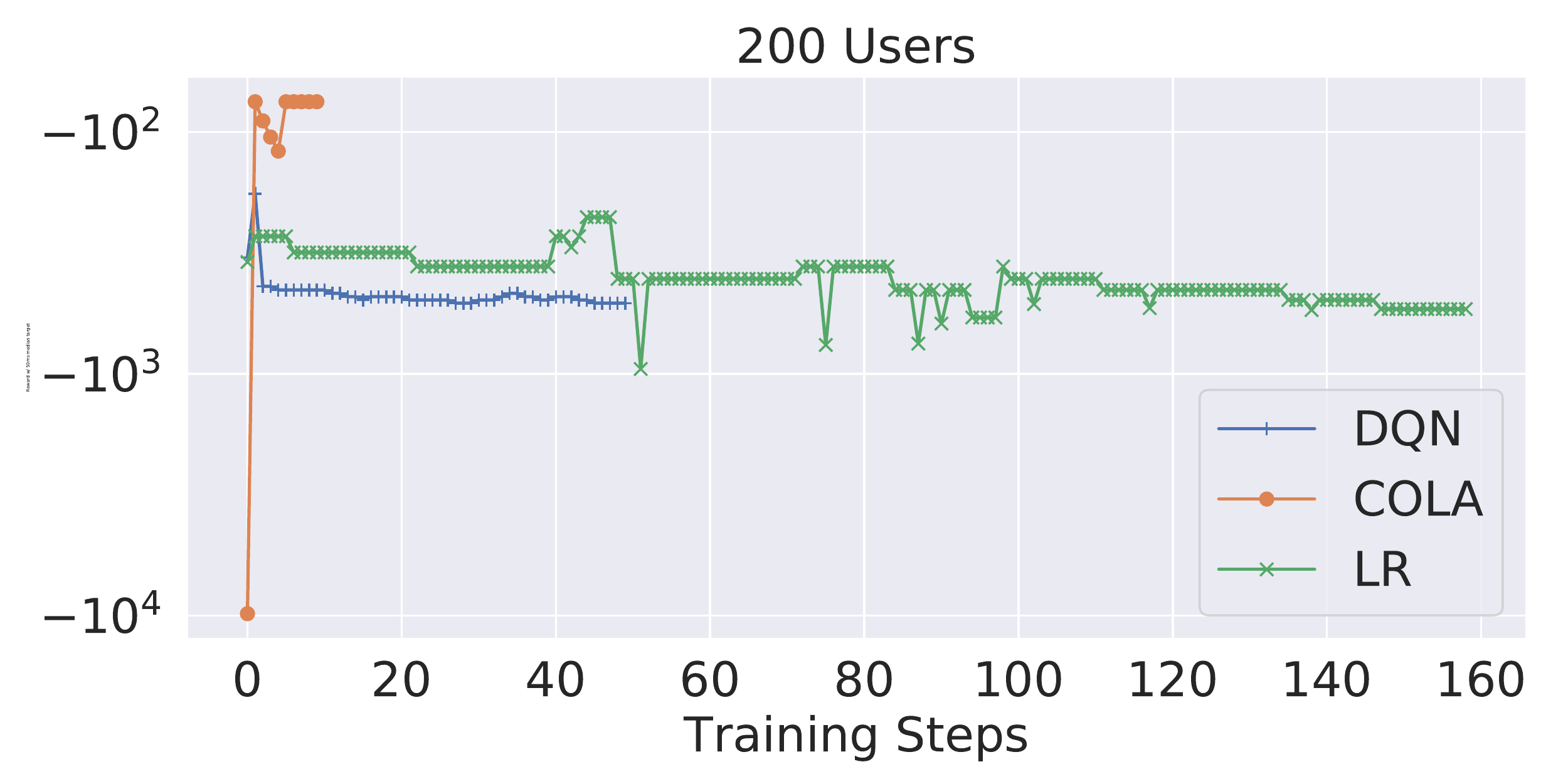}%
}  %
\hspace*{\fill}
\subfigure{
  \includegraphics[width=0.31\textwidth]{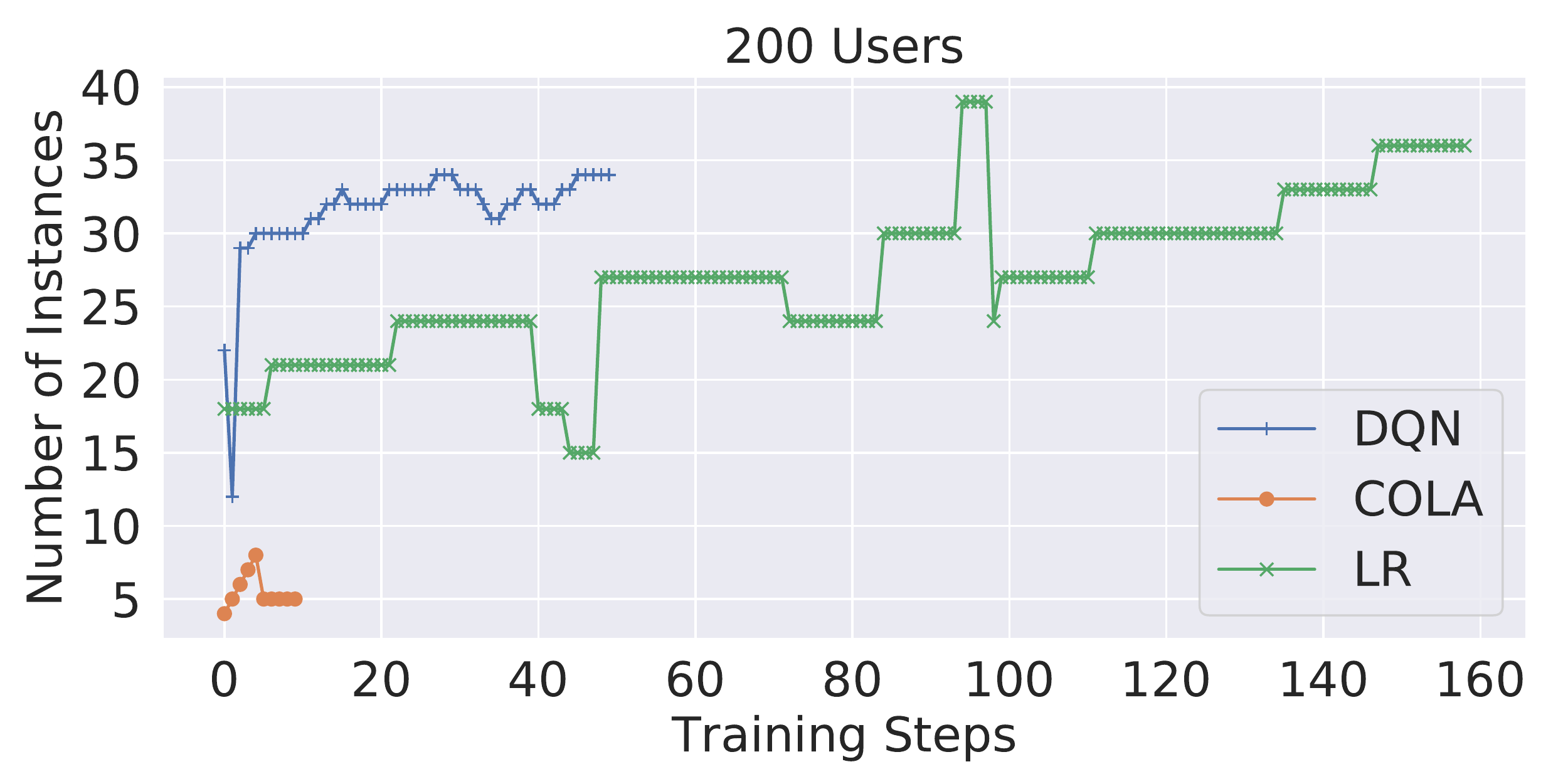}%
}  %
\hspace*{\fill}
\subfigure{
  \includegraphics[width=0.31\textwidth]{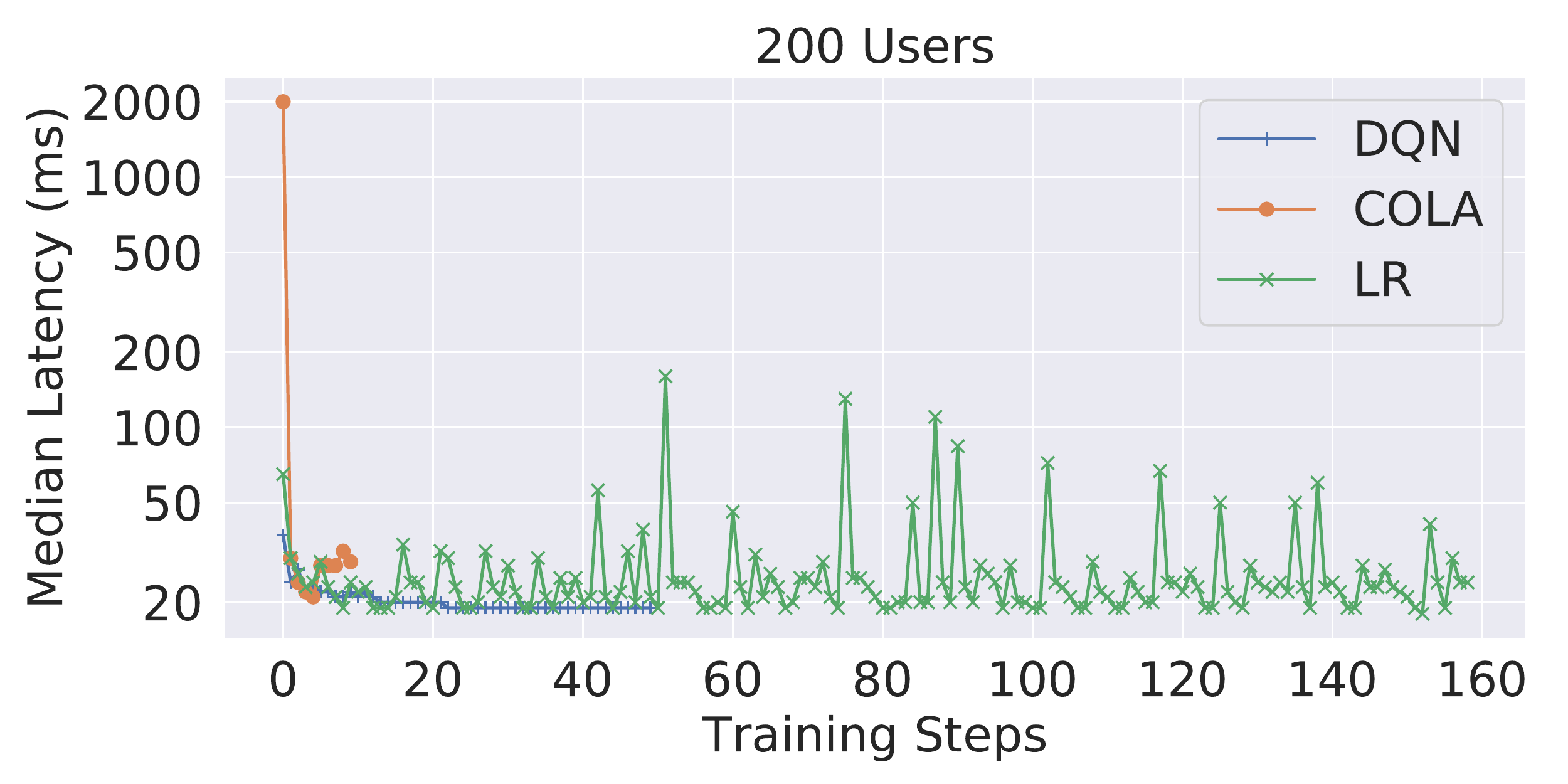}%
}

\subfigure{%
  \includegraphics[width=0.31\textwidth]{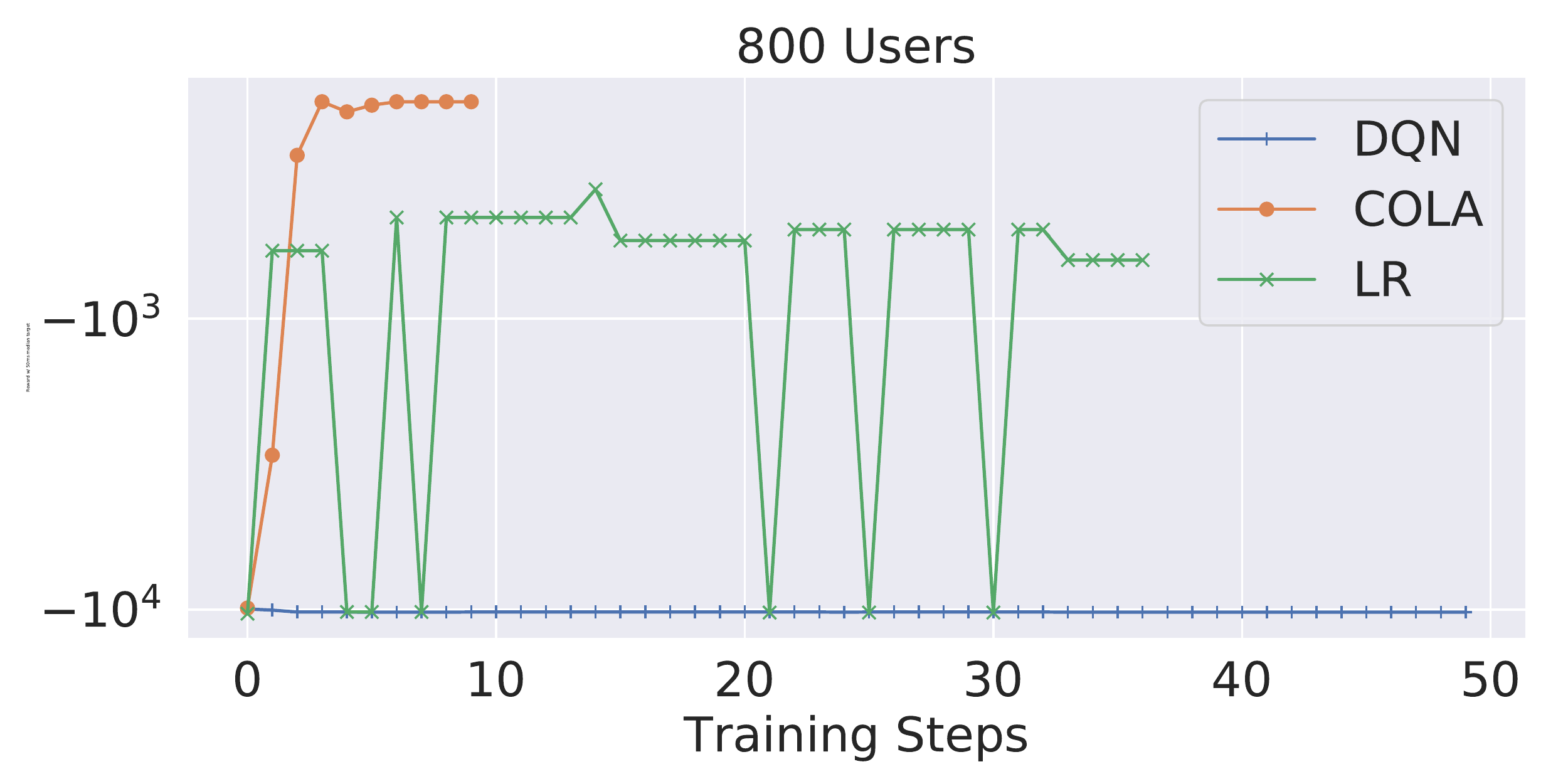}%
}  %
\hspace*{\fill}
\subfigure{
  \includegraphics[width=0.31\textwidth]{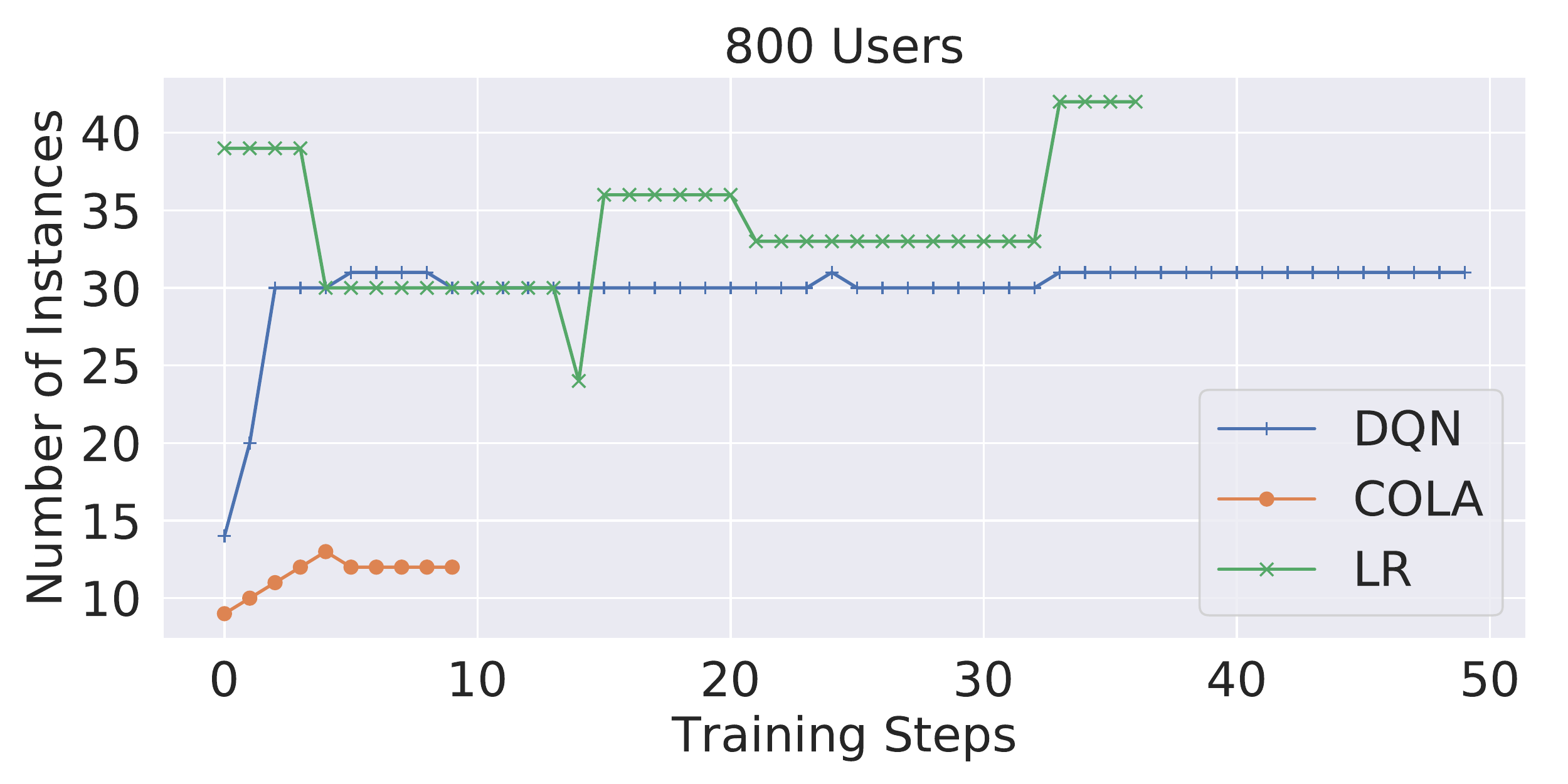}%
}  %
\hspace*{\fill}
\subfigure{
  \includegraphics[width=0.31\textwidth]{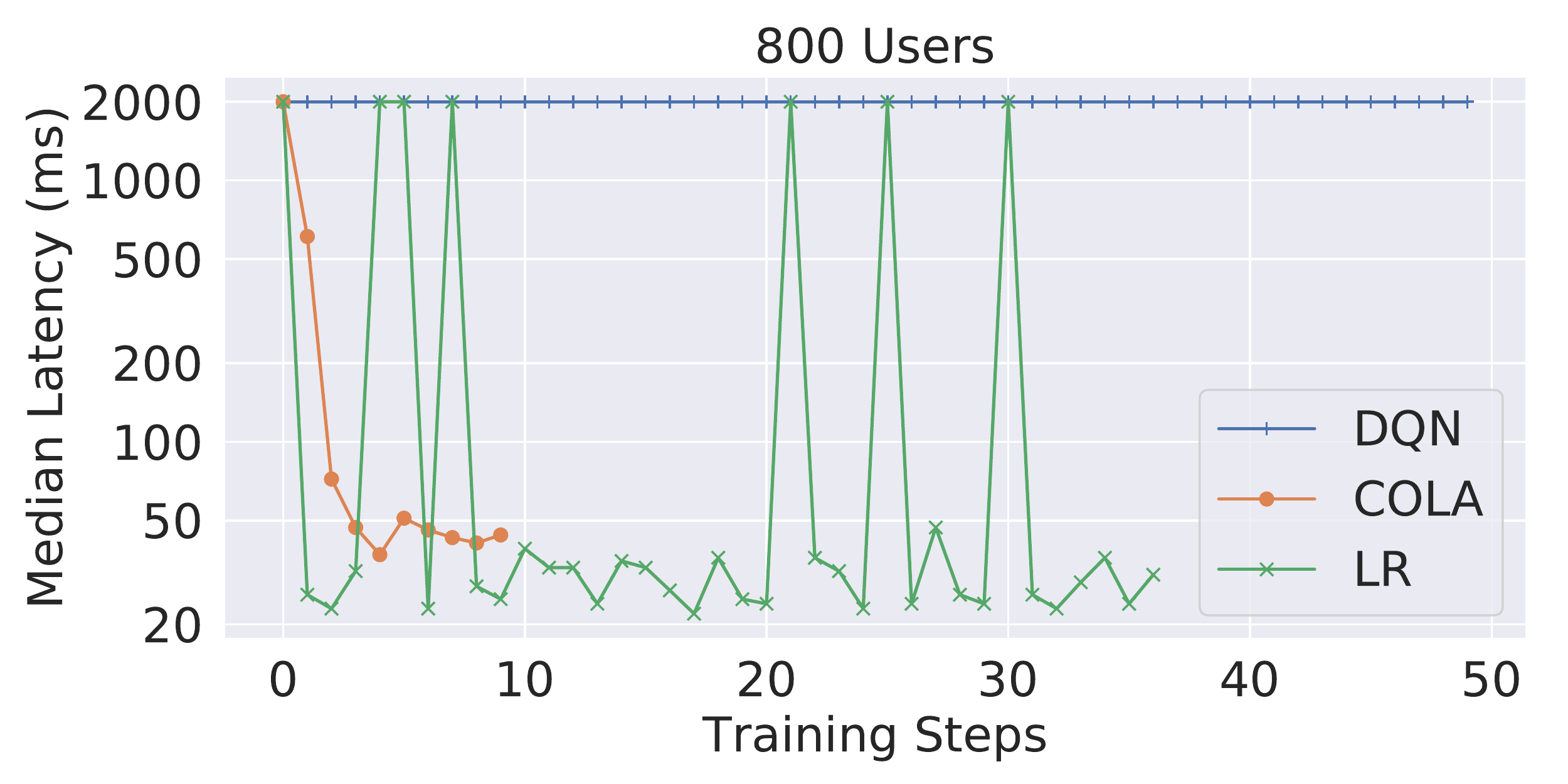}%
}

\captionsetup{justification=centering}
\caption{BookInfo Learned Autoscaler Trajectories}\label{fig:bi-training-trajectory}

\end{figure*}

\begin{figure*}

\subfigure{%
  \includegraphics[width=0.31\textwidth]{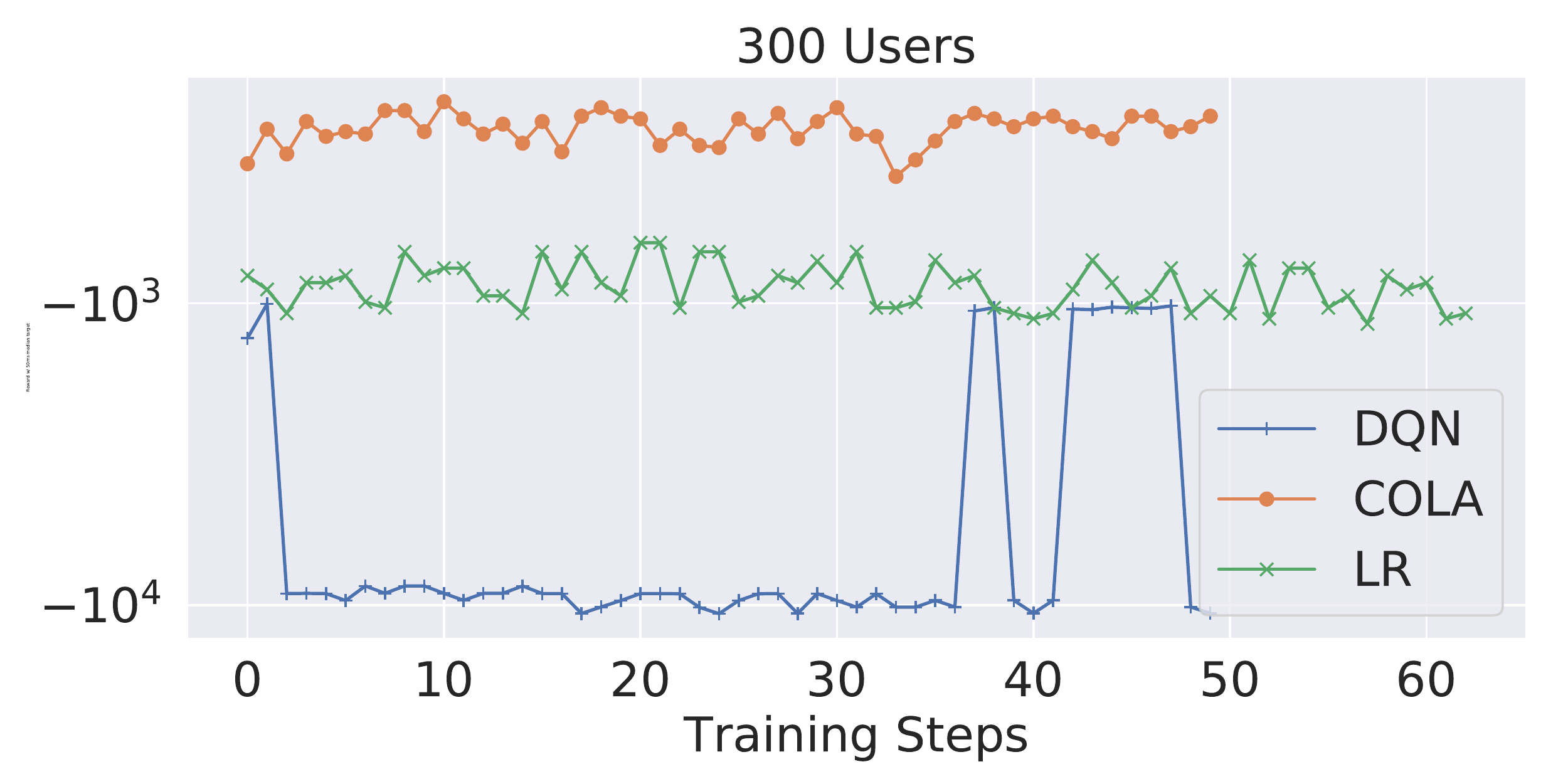}%
}  %
\hspace*{\fill}
\subfigure{
  \includegraphics[width=0.31\textwidth]{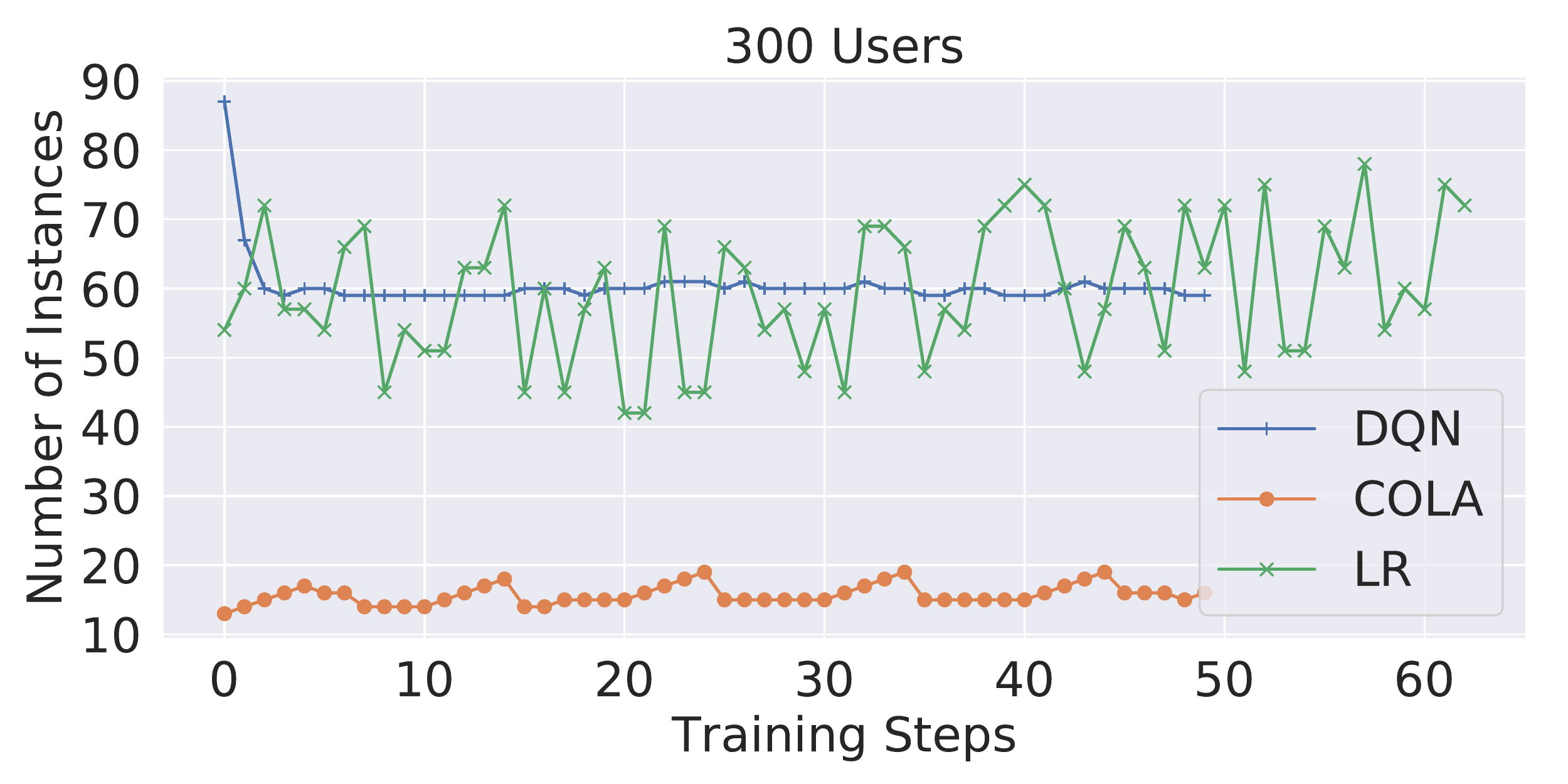}%
}  %
\hspace*{\fill}
\subfigure{
  \includegraphics[width=0.31\textwidth]{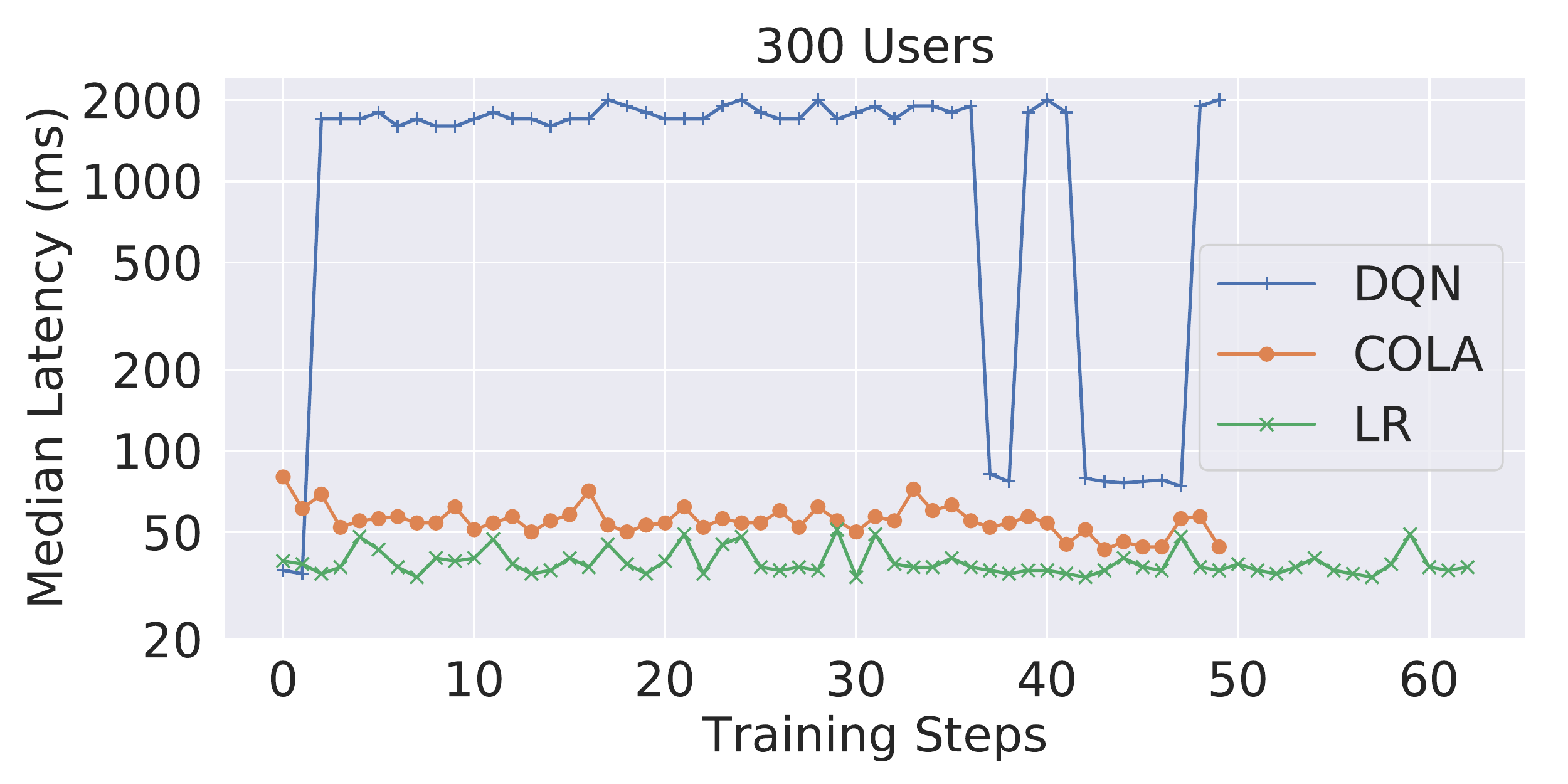}%
}

\subfigure{%
  \includegraphics[width=0.31\textwidth]{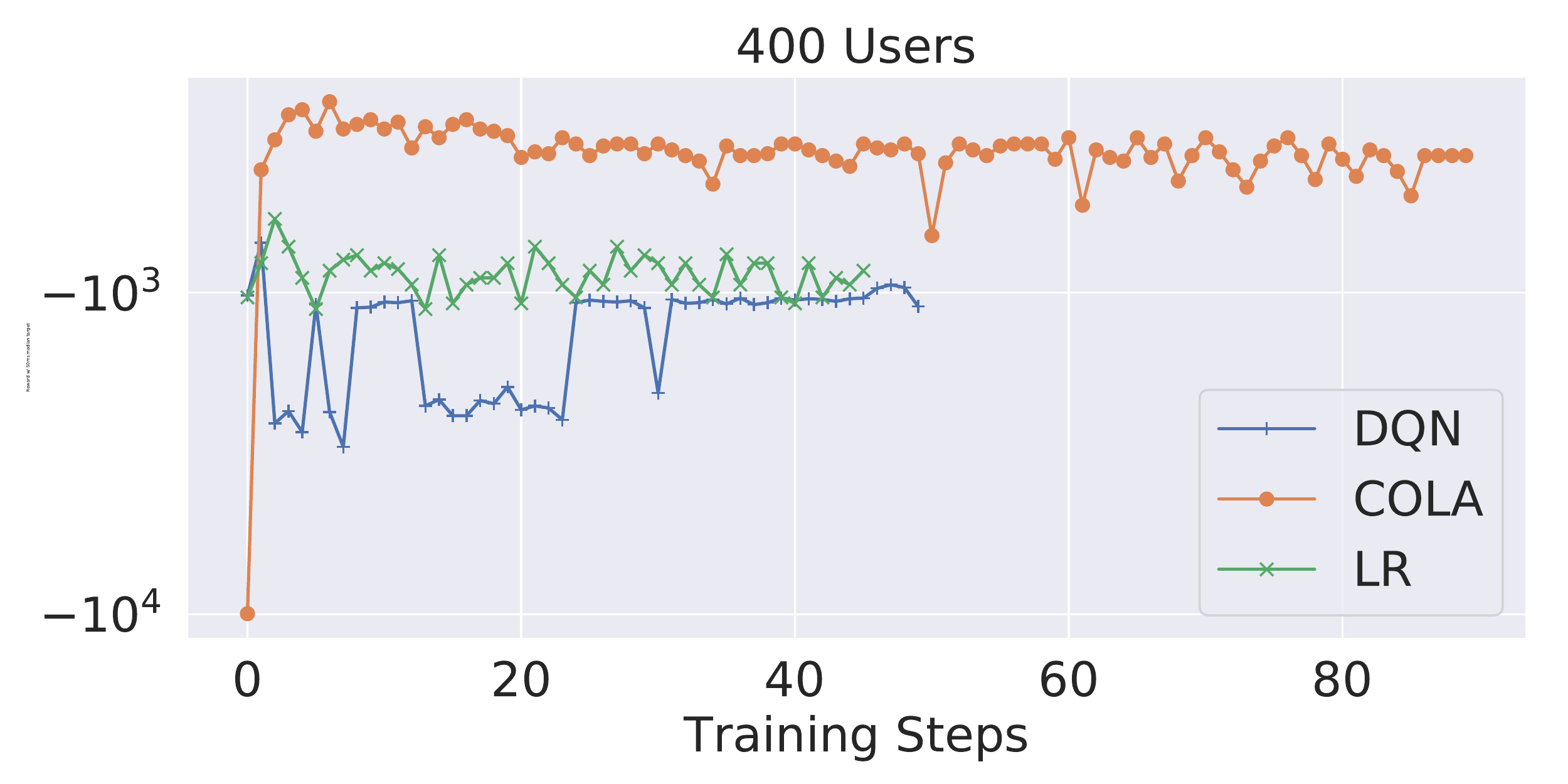}%
}  %
\hspace*{\fill}
\subfigure{
  \includegraphics[width=0.31\textwidth]{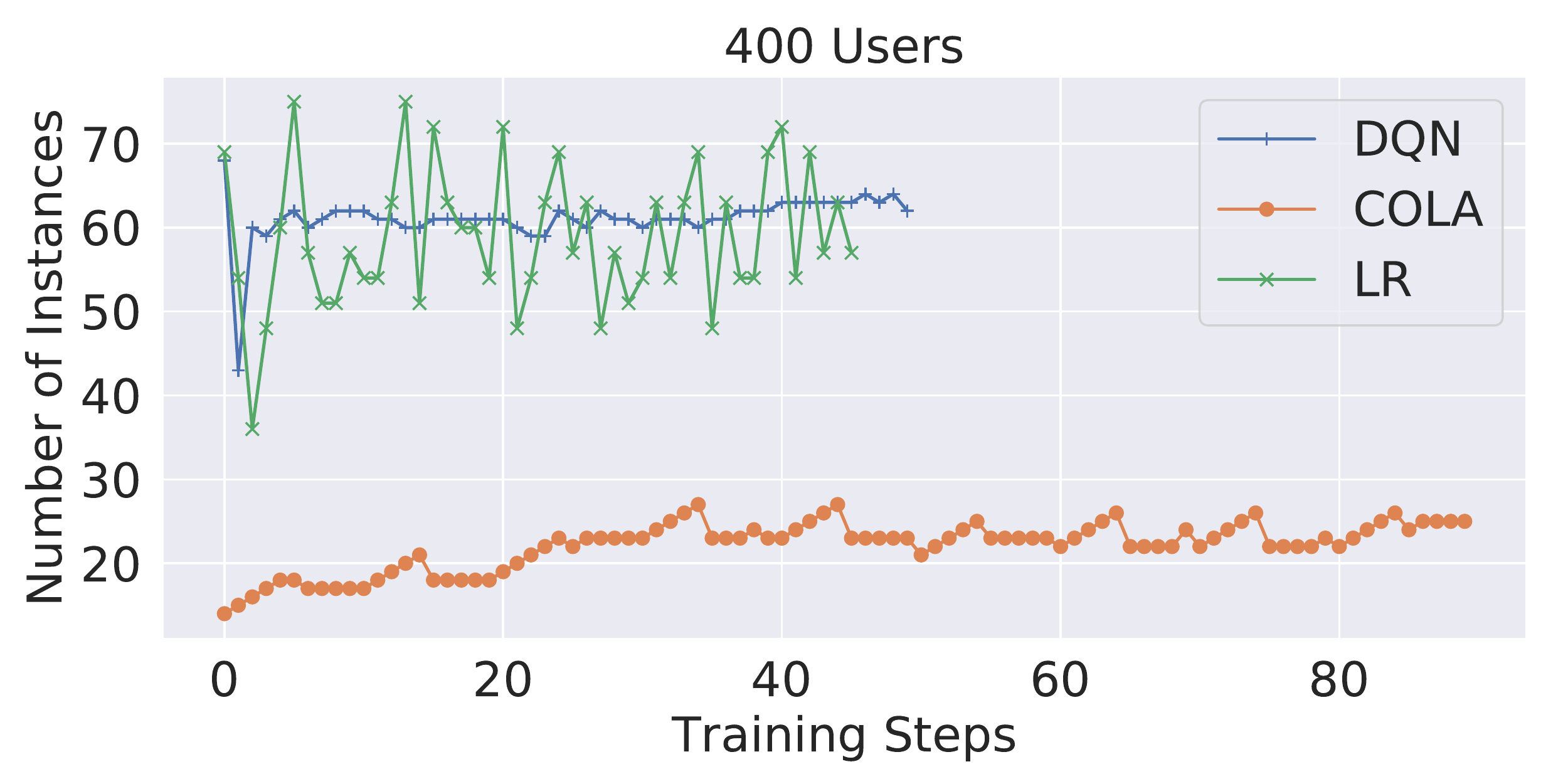}%
}  %
\hspace*{\fill}
\subfigure{
  \includegraphics[width=0.31\textwidth]{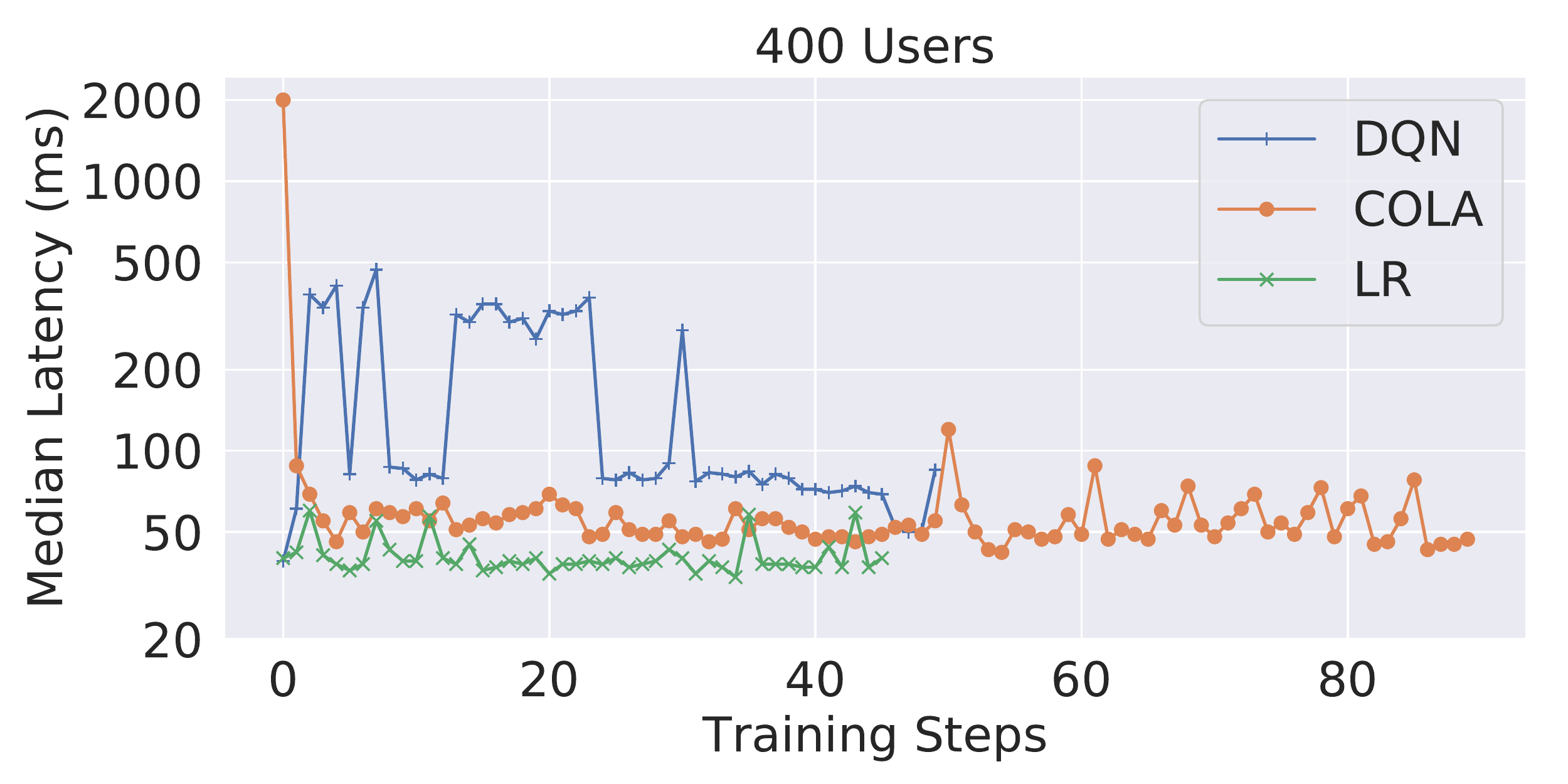}%
}

\captionsetup{justification=centering}
\caption{Online Boutique Learned Autoscaler Trajectories}\label{fig:ob-training-trajectory}

\end{figure*}

\begin{figure*}

\subfigure{%
  \includegraphics[width=0.31\textwidth]{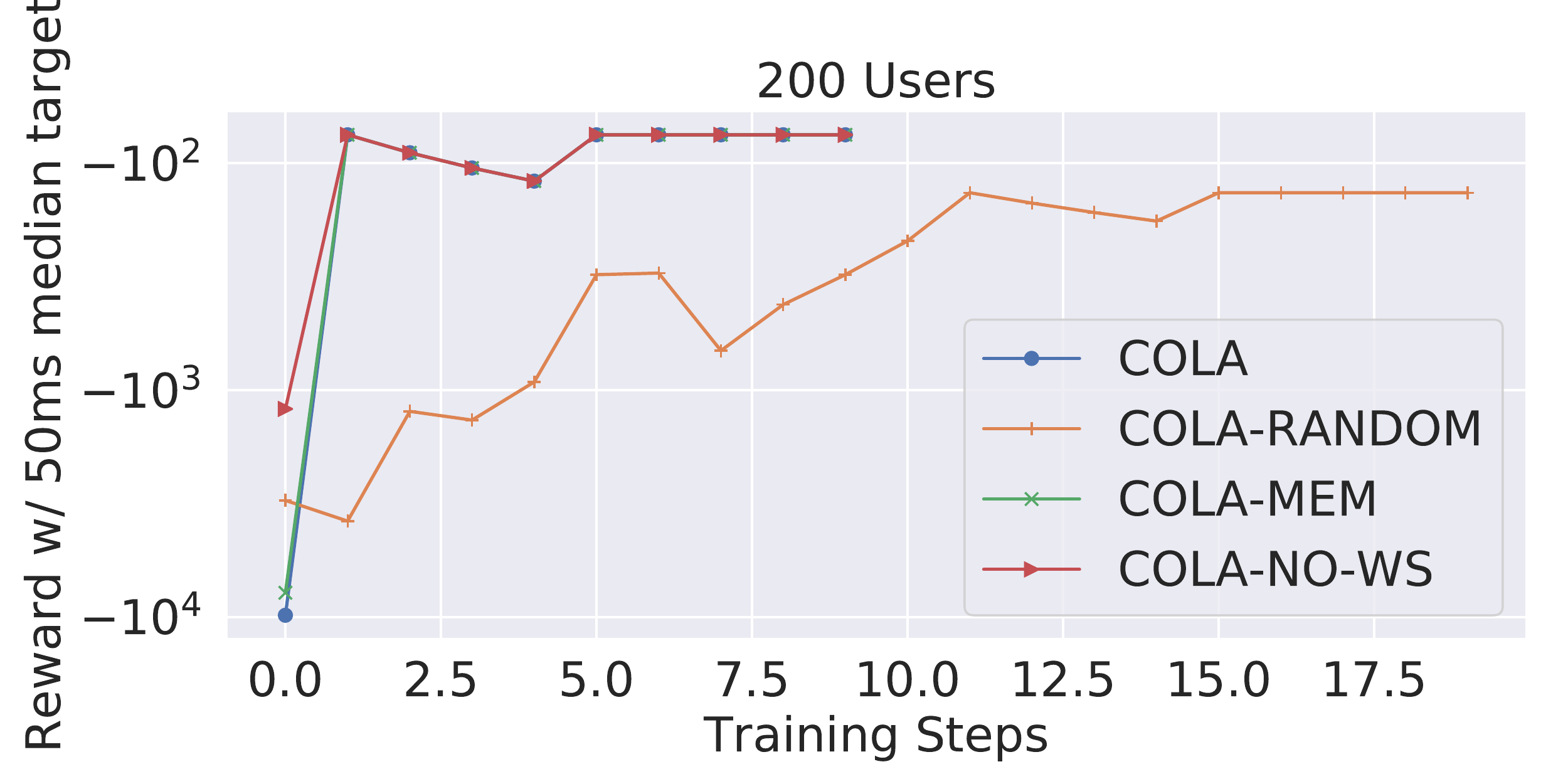}%
}  %
\hspace*{\fill}
\subfigure{
  \includegraphics[width=0.31\textwidth]{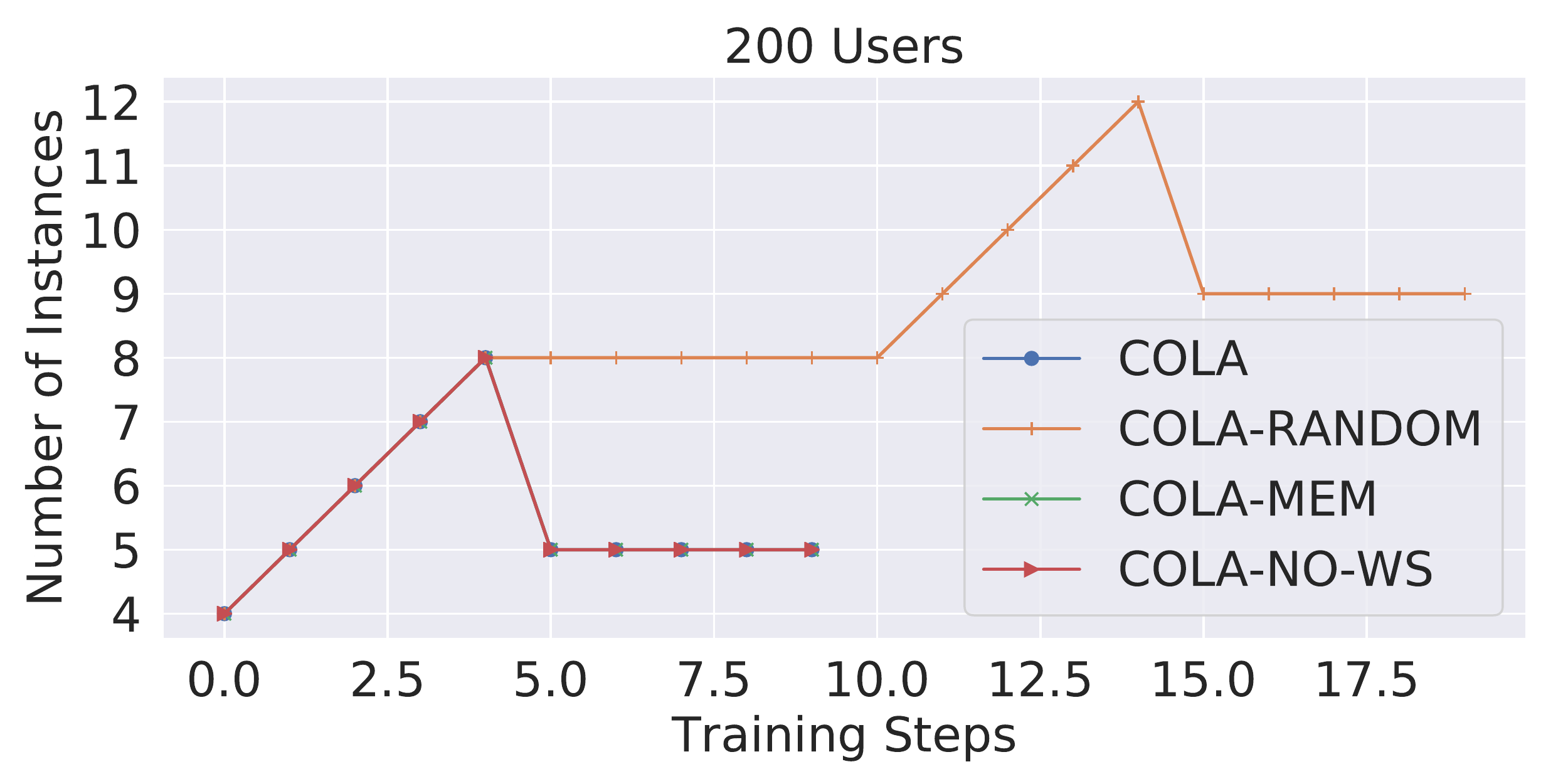}%
}  %
\hspace*{\fill}
\subfigure{
  \includegraphics[width=0.31\textwidth]{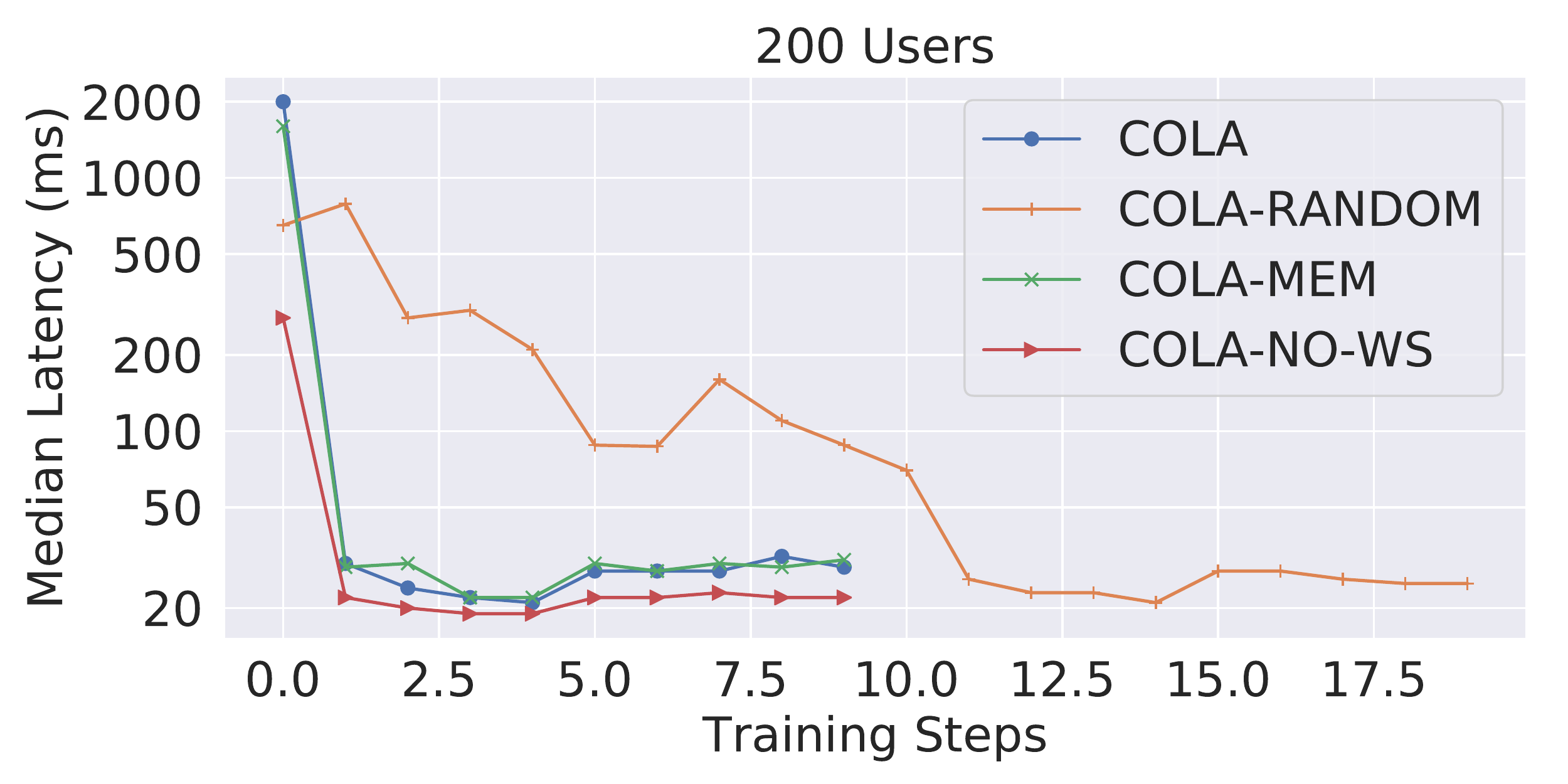}%
}

\subfigure{%
  \includegraphics[width=0.31\textwidth]{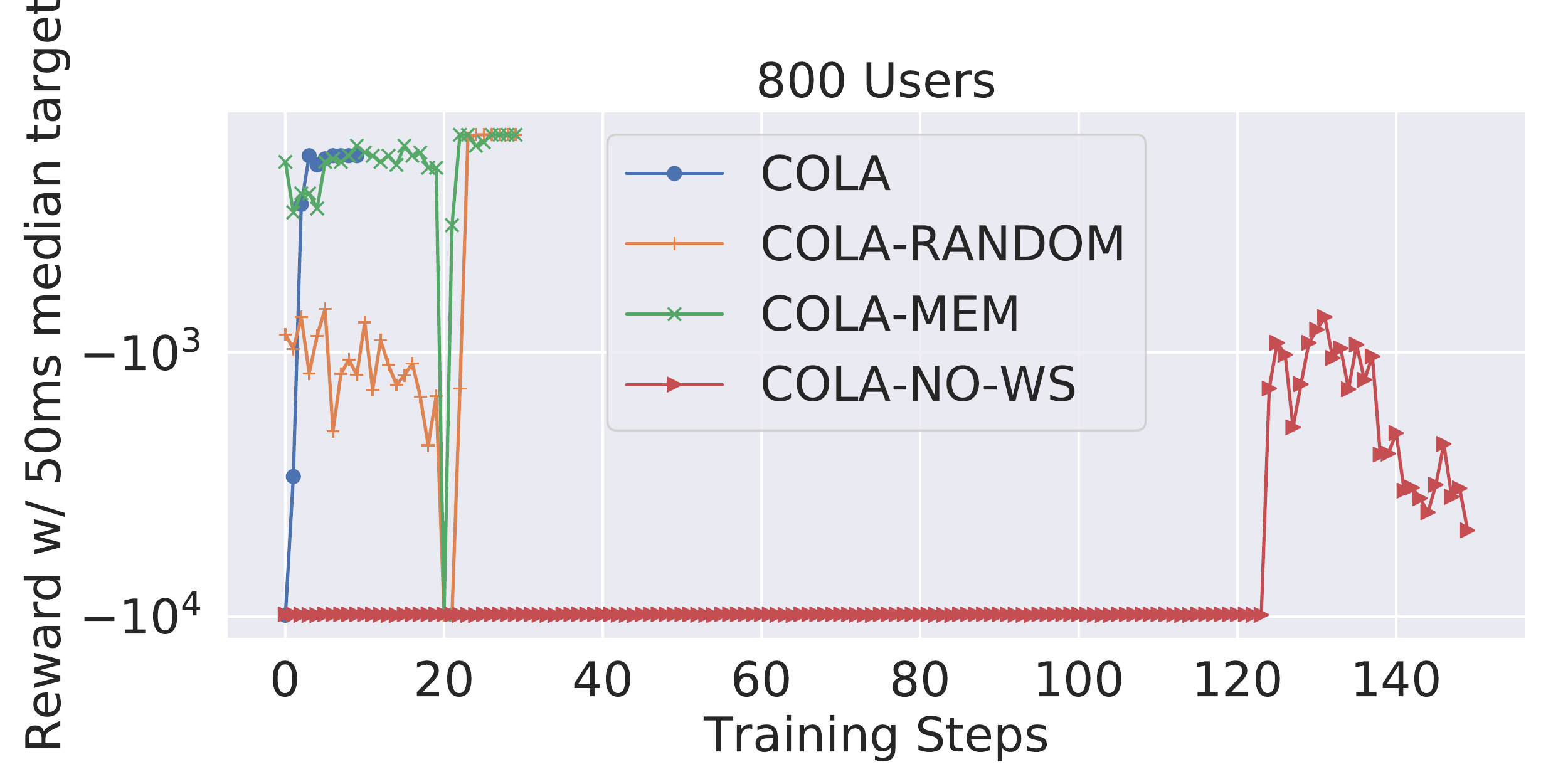}%
}  %
\hspace*{\fill}
\subfigure{
  \includegraphics[width=0.31\textwidth]{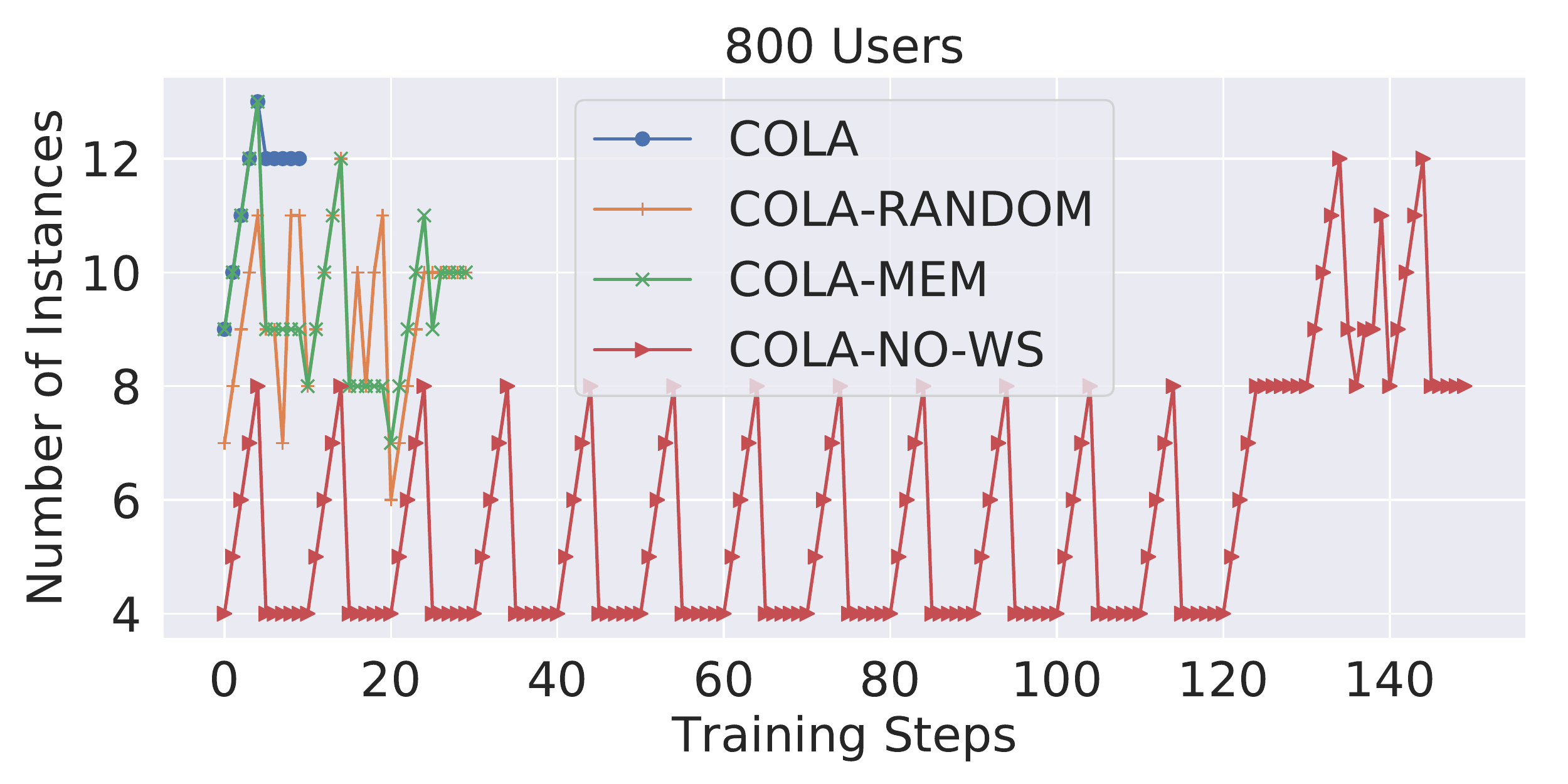}%
}  %
\hspace*{\fill}
\subfigure{
  \includegraphics[width=0.31\textwidth]{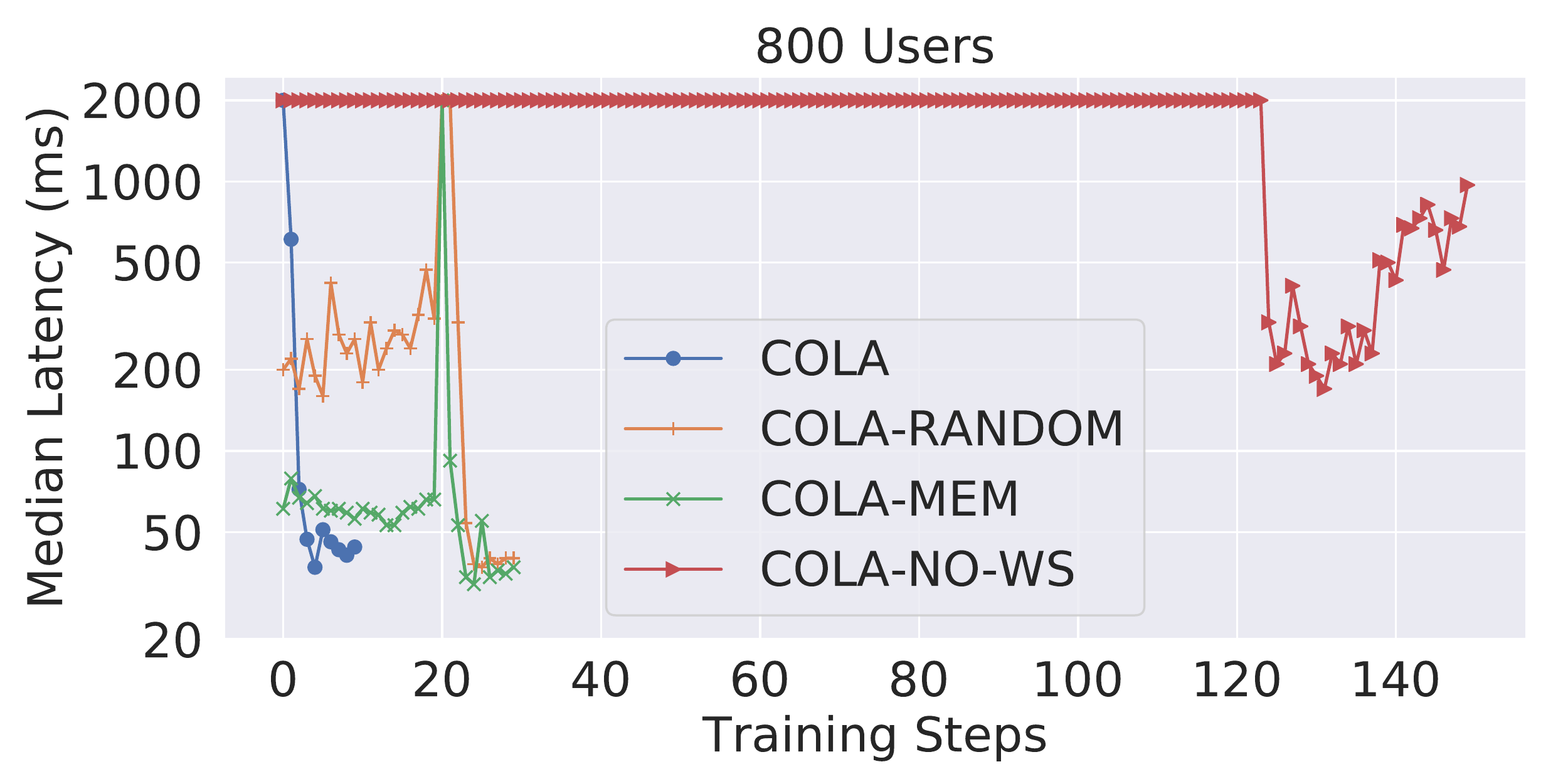}%
}
\captionsetup{justification=centering}
\caption{Book Info Trajectories with different Model Ablations}\label{fig:bi-training-trajectory-ablation}

\end{figure*}

We show a comparison of the UCB1 and random search strategies for a multi armed bandit on the Online Boutique application. In this setting, both bandits are given 10 total trials as their budget and asked to explore across 5 potential instance allocations for a particular service. As mentioned in \S\ref{subsubsec:optimize-service}, the UCB1 strategy more immediately focuses on actions which provide the highest reward whereas the uniform strategy selects across the arms with equal probability regardless of the reward observed in previous trials. The arms explored by these strategies for our sample case is shown in Figure \ref{fig:bandit-search-strat}. We compute error between the Uniform and UCB1 bandits by running a third trial in which we sample the eventually selected arm (4 replicas) 20 times to get a higher fidelity estimate of our expected latency. When compared with this third trial, we find that the UCB1 bandit has an error of 8.9\% whereas the uniform bandit has only evaluated our selected arm twice and obtains an error of 52\%. We expect that increasing the total number of trials should allow both bandits to obtain better estimates of latency for each arm and that the gap in estimation error between the UCB1 and Uniform bandit should shrink.
\subsection{Linear Contextual Bandit vs. Interpolation}
\label{subsec:lcb-v-i}

In our implementation we interpolate microservice configurations to generalize decision making rather than learning a linear regression model to do so (and is standard for linear contextual bandits). By making this choice we assume that microservice configurations are piecewise linear across neighboring context, a weaker assumption than that made by linear regression -- configurations are linear across all context. We include results from an evaluation comparing performance when using interpolation vs. building a linear regression model for the Online Boutique application in Tables \ref{tab:interp-inference} and \ref{tab:lcb-inference}. Both the interpolation and regression bandit autoscalers meet the 50ms target but the linear regression model incurs a higher cost in this case. 

\begin{table}
  \begin{tabular}{lrr}
    \toprule
    Users & Median Latency (ms) & Num Instances\\
    \midrule
   200 &   45 &  14 \\
   300 &   41 &  18 \\
   400 &   50 &  20 \\
  \bottomrule
\end{tabular}
\caption{\sysname{} Interpolated Inference}
\label{tab:interp-inference}
\vspace{-0.25cm}
\end{table}

\begin{table}
  \begin{tabular}{lrr}
    \toprule
    Users & Median Latency (ms) & Num Instances\\
    \midrule
   200 &   44 &  17 \\
   300 &   44 &  22 \\
   400 &   50 &  27 \\
  \bottomrule
\end{tabular}
\caption{\sysname{} Linear Contextual Bandit}
\label{tab:lcb-inference}
\vspace{-0.25cm}
\end{table}
\clearpage

\clearpage

\begin{figure*}


\subfigure{%
  \includegraphics[width=0.31\textwidth]{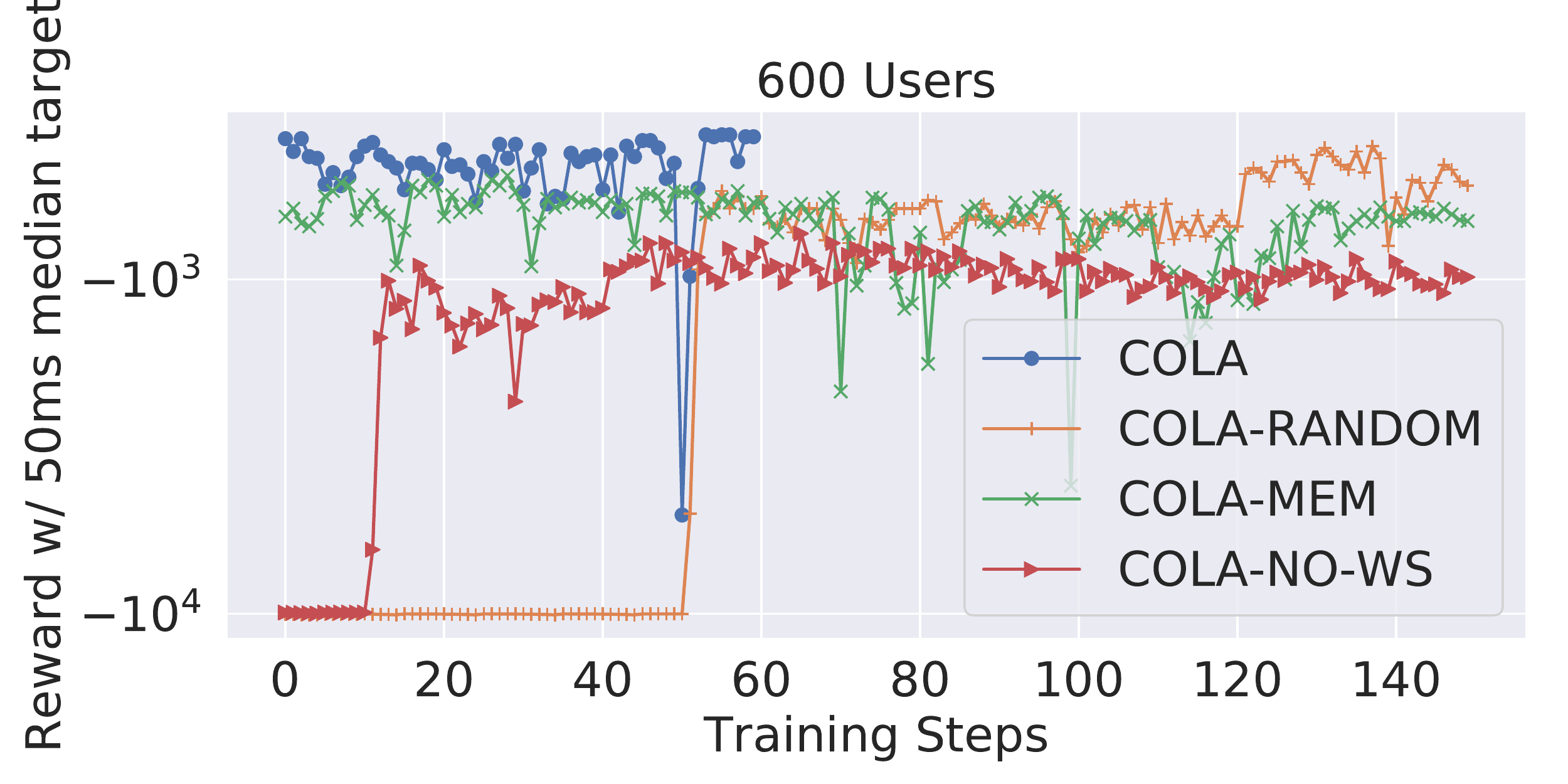}%
}  %
\hspace*{\fill}
\subfigure{
  \includegraphics[width=0.31\textwidth]{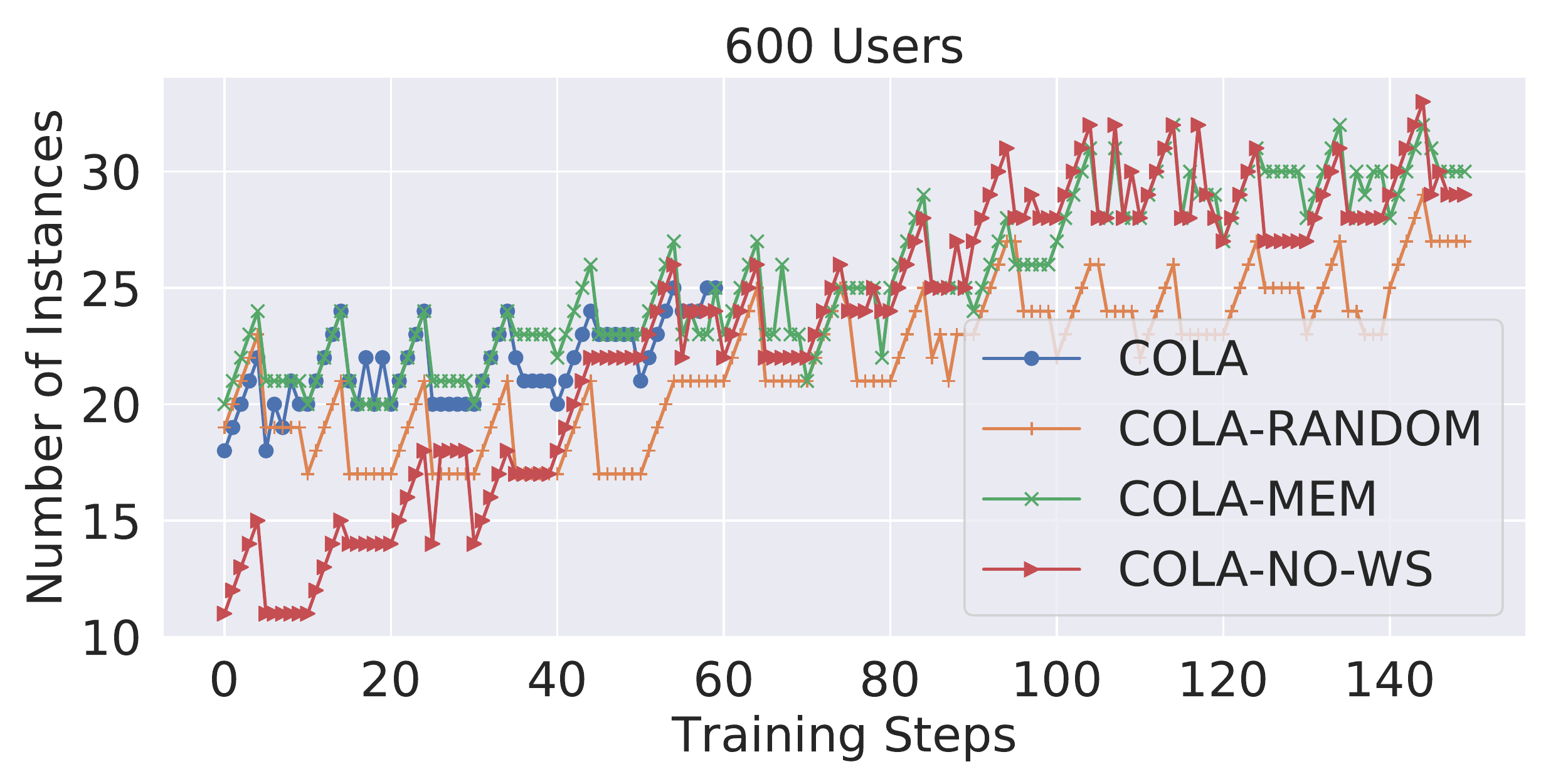}%
}  %
\hspace*{\fill}
\subfigure{
  \includegraphics[width=0.31\textwidth]{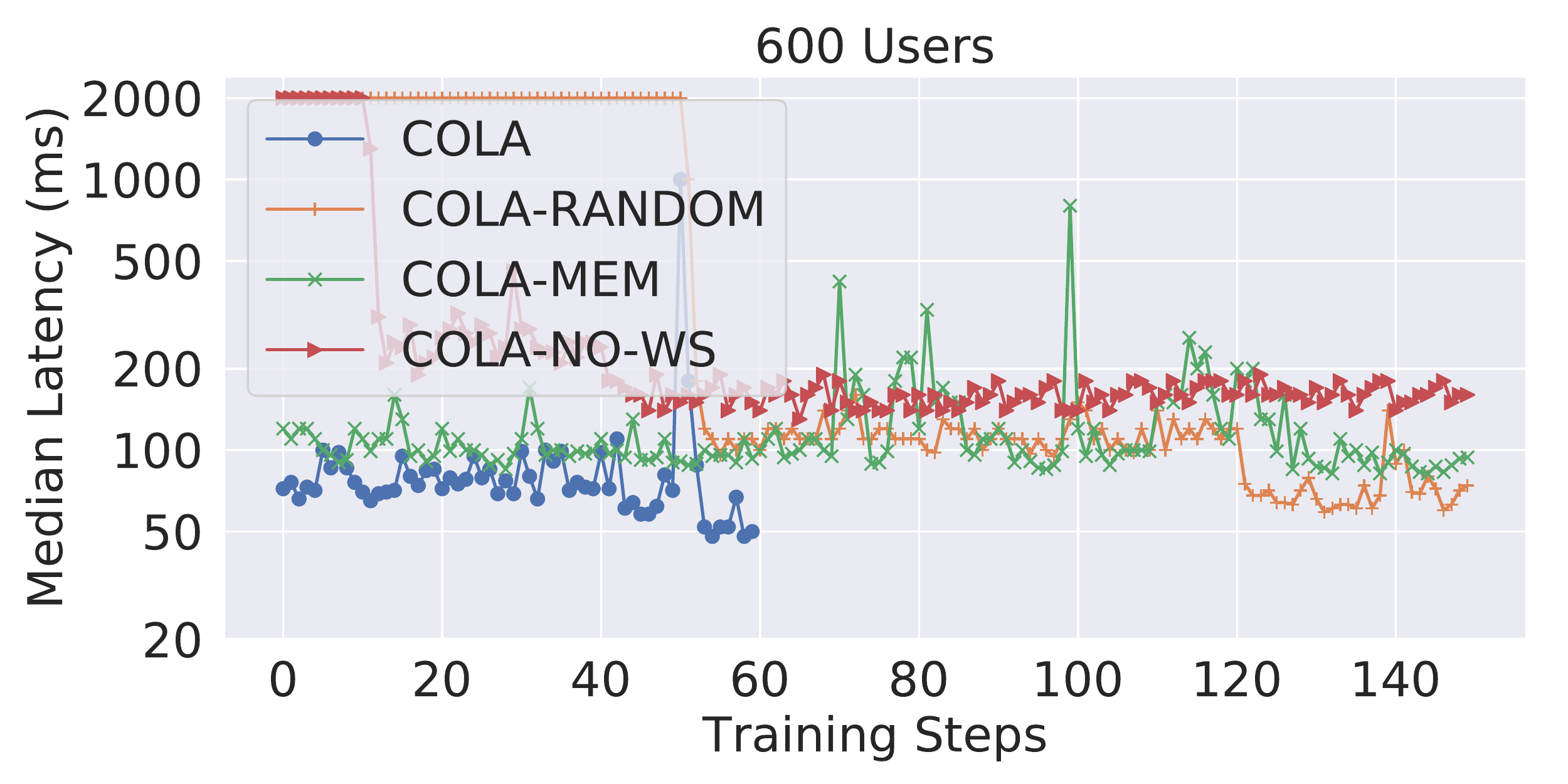}%
}
\captionsetup{justification=centering}
\caption{Online Boutique Trajectories with different Model Ablations}\label{fig:ob-training-trajectory-ablation}

\end{figure*}
\clearpage




\newpage
\pagebreak
\subsection{Open Source Application Modifications}\label{subsec:app-mods}We describe any modifications made to our open source benchmark applications before evaluating on them.

\textbf{Simple Web Server}
We add a 40 ms pause to the response on the server side to simulate delay due to application logic.

\textbf{Online Boutique}
We modify this microservice application by replacing the load generator service provided with our own external load generator based on Locust ~\cite{locust:online}. When deployed as is, the application contains 12 microservices rather than 11.

\textbf{Train Ticket} We populate the application with usernames and credentials for 10,000 synthetic users. Each connection in our load generator samples a synthetic user uniformly at random, logs in with these credentials and then  issues further requests during evaluation. We allow autoscaling on all microservices except the \texttt{ts-auth-service} through which users login.
\subsection{Modeling Details}
We define the model of our microservice autoscaling problem below.

\subsubsection{State} An application consists of $D$ different microservice deployments (e.g., database, web server, cache). Each deployment, $i$, has a defined range of number of replicas which may be launched from 1 to $N_i$. We denote $N_i$ as the replica range for the deployment $i$. We define the state of the cluster, $S$ as a vector of number of current deployments for microservices in our application. If we consider the maximum replicas to be $N$ for all $D$ deployments, the possible states for our microservice application is $S \in \{1, 2, 3, ... ,N\}^D$.

\subsubsection{Actions} Our model has the ability to take actions that modify the number of replicas in our cluster for each microservice to any value in its replica range, consequently altering state $S$. The set of all possible actions $A$ is simply the set of cluster states, $S$. At any given time a model may choose to change the cluster state to any possible configuration.

\subsubsection{Context}
We construct a representation of a workload and denote this representation as a context $C$. This context $C$, contains information on the inbound requests to our application and serves to distinguish the workloads an application may observe. 


In recent years, standards have been introduced by the OpenTracing \cite{opentracing:online} and OpenAPI \cite{openapi-3-1-0:online} organizations to define requests into a distinct set of "operation" names. Roughly speaking, the best practice denotes an operation as a distinct URL without its request parameters. For example, \url{get_account_by_id/id1} and \url{get_account_by_id/id2} both fall under the \url{get_account_by_id} operation name. 

Several popular logging, monitoring and debugging tools can both propogate and utilize these specified operation names. For example microservice tracing tools (e.g. \cite{Jaegerop26:online}, Zipkin \cite{OpenZipk7:online}, and DataDog \cite{datadog:online}) annotate logs of invoked requests with their associated operation name. Monitoring and debugging tools often allow users to filter, process and analyze requests broken down by operation name.

During training and evaluation, we assume that developers have defined $U$ distinct operations for their application. We then construct a context vector $C$ of length $U$ by concatenating the number of requests per operation name. Each element in our context vector $C_i$ is simply the number of requests to our cluster with the operation name $i$.

\subsection{Algorithm Hyperparameter Choices}

We list the hyperparameters chosen for training learned autoscaling policies across all applications in Table \ref{tab:hyperparams}.

\subsection{Tabular Evaluation Results}

In Tables \ref{tab:bi-fr-tabular}-\ref{tab:ob-fc-fr-tabular}, we show full tabular results for all workload evaluations.

\subsection{Microservice Application Diagrams}

Figure \ref{fig:arch-diagrams} shows the architectures for applications on which we evaluate.

\newpage
\subsection{\sysname{} Training Pseudocode}
\label{subsec:pseudocode}
\begin{figure}[!h]
	\centering
	\includegraphics[width=.40\textwidth]{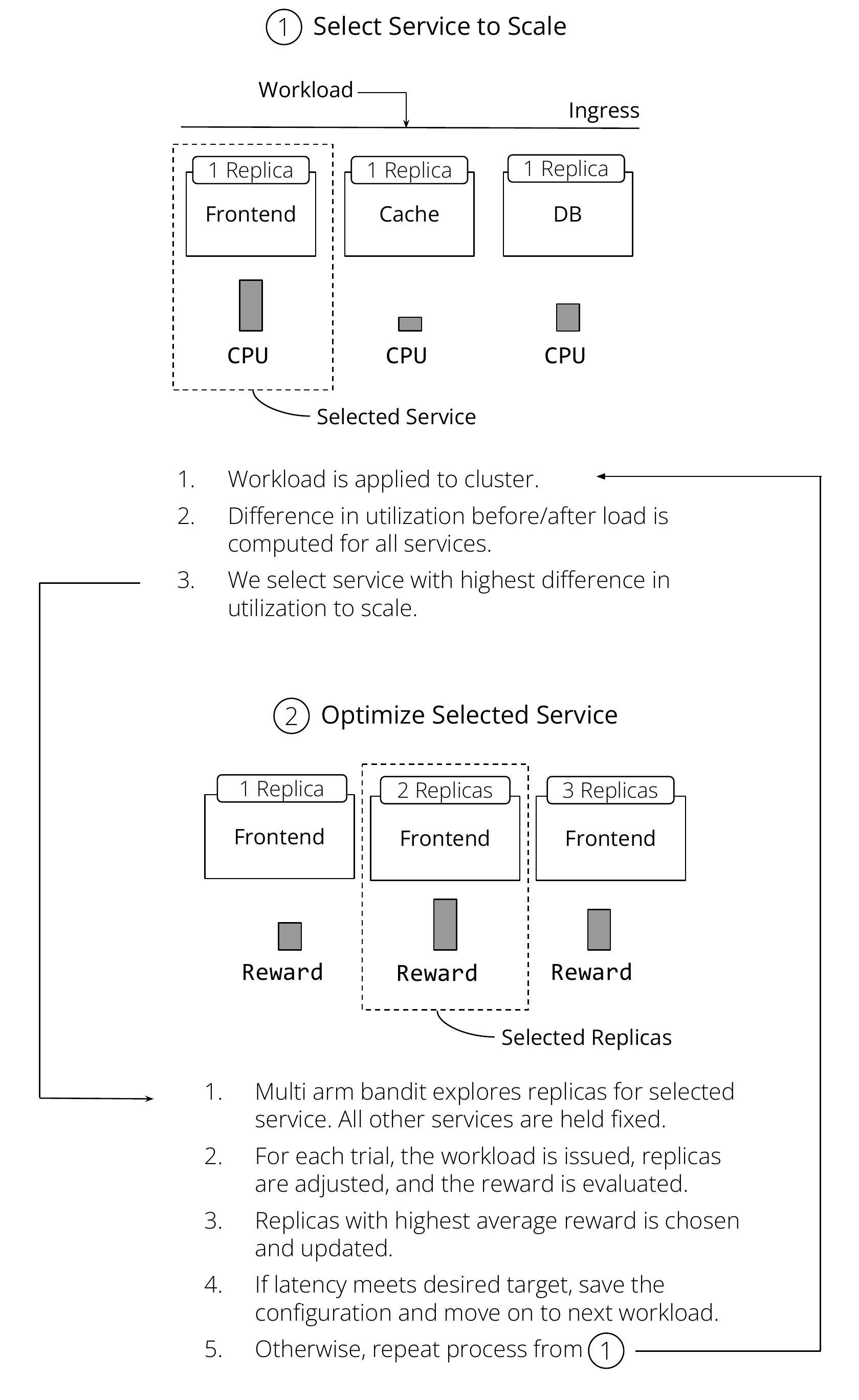}
	\caption{\sysname{} Training Procedure Diagram.}
	\label{fig:training-diagram}
\end{figure}

\begin{table}[!th]
  \begin{tabular}{ll}
    \toprule
    Variable& Description \\
    \midrule
    $A$             &  Set of actions to explore\\
    $l_{target}$    &  Latency target (in ms)\\
    $C$             &  Workload\\
    $\lambda$       &  Current weight hyperparameter\\
    \midrule
    $a_{opt}$      & Best action in $A$\\
    $l_{opt}$      & Latency for best action in $A$\\
      \bottomrule
    \end{tabular}
    \caption{Input and Output variables for \texttt{ucb}}
    \label{tab:vars-ucb}
\end{table}

We provide psuedocode for \sysname{}'s autoscaling training procedure above in Algorithm \ref{alg:hillclimbing}. Also, we document the inputs and outputs to the functions outlined in Algorithm \ref{alg:hillclimbing} in Tables \ref{tab:vars-optcluster} and \ref{tab:vars-ucb}.


\begin{algorithm}[!h]
\caption{\sysname{} Training Procedure}\label{alg:hillclimbing}
  \begin{algorithmic}[1]
    \BlankComment{}

    \Function{optimizeCluster}{$C$,$S^{0}$,$l_{target}$,$T$,$\lambda$,$\lambda_{max}$, $\epsilon$}
    
        \While{$\lambda<\lambda_{max}$}
          \For{$t=1,...,T$}
            \State{$S^t = S^{t-1}$}
            \LineComment{Choose a service to optimize. (Section \ref{subsubsec:select-service})}
            \State Apply workload $C$ to cluster with state $S^{t}.$ \LineComment{Aggregate output of \texttt{kubectl top pod} by service.}
            \State  Get $U$, list of average CPU util for each service. 
            \LineComment{Return highest utilized service.}
            \State $opt = \argmax_{i \in {1,...,D}} U$        
            
            \LineComment{Optimize the chosen service. (Section \ref{subsubsec:optimize-service})}
            \State{$S^t_{opt}, l_{opt}^t$ = ucb($A, l_{target}, C, \lambda$)}
            \If{$l_{opt} <= l_{target}$}         \State{\textbf{break}}
            \EndIf
          \EndFor
          \If{$l_{opt} <= l_{target}$}         \State{\textbf{break}}
          \Else \State{$\lambda = \lambda + \frac{1}{\epsilon}$}
          \EndIf
        \EndWhile
        \State
        \Return{$S^T, l_{opt}^t$}
    \EndFunction

    \BlankComment{}
    \Function{ucb}{$A, l_{target}, C, \lambda$}

    \For{$a \in A$}\textcolor{blue}{\Comment{Iterate over all cluster states.}}
        \State $N_a = \epsilon$ \textcolor{blue}{\Comment{Number of times state $a$ is sampled.}}
        \State $R_a = 0$ \textcolor{blue}{\Comment{Average reward for state $a$.}}
        \State $L_a = 0$ \textcolor{blue}{\Comment{Average latency for state $a$.}}
    \EndFor
    \For{$t=1,2,...,F$}
        \State $a_t = \argmax_{a \in A} R_a + \frac{\sqrt{2log(t)}}{N_a}$ \textcolor{blue}{\Comment{Action for trial $t$.}}
        \State Update cluster replicas based on action $a_t$
        \State Compute total cost $M$ used for action $a_t$
        \State Observe latency $L_t$ for action $a_t$
        \State $R_t = \lambda min((l_{target} - L_t), 0) - M $
        \State $N_{a_t} = N_{a_t} + 1$ 
        \State $R_{a_t} = R_{a_t} + \frac{1}{N_{a_t}} (R_t - R_{a_t})$
        \State $L_{a_t} = L_{a_t} + \frac{1}{N_{a_t}} (L_t - L_{a_t})$
    \EndFor
    \State $a_{opt} = \argmax_{a \in A} R_a$ \textcolor{blue}{\Comment{Get action with highest reward.}}
    \State $l_{opt} = L_{a_{opt}}$ \textcolor{blue}{\Comment{Get corresponding latency.}}
    \State \textbf{return} $a_{opt}, l_{opt}$
    \EndFunction

  \end{algorithmic}
\end{algorithm}

\begin{table}[!th]
  \begin{tabular}{ll}
    \toprule
    Variable& Description \\
    \midrule
    $C$             &  Workload\\
    $S^{0}$         &  Initial cluster state\\
    $l_{target}$    &  Latency target (in ms)\\
    $T$             &  Number of iterations\\
    $\lambda$       &  Current weight hyperparameter\\
    $\lambda_{max}$ &  Maximum weight hyperparameter\\
    $\epsilon$      &  Weight increment\\
    \midrule
    $S^{T}$         &  Final cluster state after optimizing\\
    $l_{opt}^t$      & Final latency after optimizing\\
      \bottomrule
    \end{tabular}
    \caption{Input and Output variables for \texttt{optimizeCluster}}
    \label{tab:vars-optcluster}
\end{table}

\newpage

\begin{table*}
  \begin{tabular}{lrrr}
    \toprule
    Application& $\lambda$ (initial)  & Sample Duration &  Number of Samples  (non \sysname{}) \\
\midrule
Simple Web Server &  1/3&  30&  200\\
Book Info.        &  1/3&  25&  200\\
Online Boutique.  &  1/3&  60&  200\\
Sock Shop.        &  1/3&  80&  200\\
Train Ticket      &  1/3&  80&  250\\
  \bottomrule
\end{tabular}
\caption{Training Hyperparameters}
\label{tab:hyperparams}
\end{table*}

\begin{table*}
  \begin{tabularx}{\textwidth}{XXrrrr}
    \toprule
    Users& Policy & Median Latency &  Failures/s & VM Instances \\
\midrule
300.0&    CPU-30&   19.4&           0.00&         15.00\\
300.0&    CPU-70&   22.5&           0.24&         9.37\\
300.0&   BO-50ms&   20.1&           0.00&        27.58\\
300.0& COLA-50ms&   43.9&           0.27&         8.67\\
300.0 &   DQN-50ms &  2000.0&       138.36&          32.00\\
300.0&   LR-50ms&   53.6&           0.00&          7.00\\
\midrule
400.0&    CPU-30&   18.9&           0.00&         19.00\\
400.0&    CPU-70&   37.5&          0.00&         11.00\\
400.0&   BO-50ms&   20.4&         0.00&         32.00\\
400.0& COLA-50ms&   30.9&         0.11&          9.00\\
400.0&   DQN-50ms&  2000.0&      181.96&          32.00\\
400.0&   LR-50ms&   60.6&          0.00&          7.00\\
\midrule
700.0&    CPU-30&   22.0&        0.41&         26.40\\
700.0&    CPU-70&   47.6&         5.57&         16.10\\
700.0&   BO-50ms&   24.3&         0.00&         18.00\\
700.0& COLA-50ms&   43.7&          0.01&         10.00\\
700.0&   DQN-50ms&  2000.0&      308.87&          32.00\\
700.0&   LR-50ms& 2000.0&        293.54&          7.00\\
\midrule
800.0&    CPU-30&   21.7&          0.31&        27.77\\
800.0&    CPU-70&   30.8&          0.43&        17.37\\
800.0&   BO-50ms&   26.6&          0.00&         18.00\\
800.0& COLA-50ms&   37.7&          0.00&         10.00\\
800.0&   DQN-50ms&  2000.0&        350.43&          32.00\\
800.0&   LR-50ms& 2000.0&          357.96&          7.00\\
  \bottomrule
\end{tabularx}
\caption{Book Info Constant Rate Results}
\label{tab:bi-fr-tabular}
\vspace{-0.25cm}
\end{table*}

\begin{table*}[h]
  \begin{tabularx}{\textwidth}{XXrrrr}
    \toprule
    Users& Policy & Median Latency &  Failures/s & VM Instances \\
\midrule
200.0&    CPU-30&   11.0&           0.00&        25.48\\
200.0&    CPU-70&   18.4&           0.00&        15.65\\
200.0&   BO-50ms&   43.7&           0.13&        34.97\\
200.0& COLA-50ms&   14.3&           0.04&         15.00\\
200.0&   DQN-50ms&     9.0&        0.00&          48.00\\
200.0&   LR-50ms&   43.7&          0.00&         14.00\\
\midrule
300.0&    CPU-30&   11.1&         0.13&        29.57\\
300.0&    CPU-70&   23.9&        0.02&         15.00\\
300.0&   BO-50ms&   37.9&       42.61&        29.27\\
300.0& COLA-50ms&   26.0&        0.18&         15.00\\
300.0&   DQN-50ms&     9.1&             0.00&          48.00\\
300.0&   LR-50ms&   59.5&          0.01&         14.00\\
\midrule
400.0&    CPU-30&   10.5&         10.56&        32.03\\
400.0&    CPU-70&   17.4&          0.00&         16.00\\
400.0&   BO-50ms&  193.2&         25.56&        23.67\\
400.0& COLA-50ms&   33.8&         0.46&         15.00\\
400.0&   DQN-50ms&     9.3&        0.00&          48.00\\
400.0&   LR-50ms&   32.9&           0.00&         15.00\\
\midrule
500.0&    CPU-30&   11.5&           0.10&        41.07\\
500.0&    CPU-70&   23.3&          1.54&        18.65\\
500.0&   BO-50ms&  409.4&        93.34&        23.72\\
500.0& COLA-50ms&   43.0&        0.92&         15.00\\
500.0&   DQN-50ms&     9.5&           0.00&          48.00\\
500.0&   LR-50ms&   44.5&         0.00&         15.00\\
  \bottomrule
\end{tabularx}
\caption{Sock Shop Constant Rate Results}
\label{tab:ss-fr-tabular}
\vspace{-0.25cm}
\end{table*}

\begin{table*}[h]
  \begin{tabularx}{\textwidth}{XXrrrr}
    \toprule
    Users& Policy & Median Latency &  Failures/s & VM Instances \\
\midrule
500.0&    CPU-30&   38.6&        0.10&        45.55\\
500.0&    CPU-70&   58.1&       0.48&         21.00\\
500.0&   BO-50ms&  842.4&      94.62&        44.88\\
500.0& COLA-50ms&   50.0&       0.61&         24.70\\
500.0&   DQN-50ms&  2000.0&   209.09&          55.00\\
500.0&   LR-50ms&  562.0&        0.75&        19.22\\
\midrule
600.0&    CPU-30&   38.7&         0.10&         51.3\\
600.0&    CPU-70&   59.3&        0.17&        22.85\\
600.0&   BO-50ms&  443.7&       59.16&        43.97\\
600.0& COLA-50ms&   47.9&        0.23&         27.00\\
600.0&   DQN-50ms&  2000.0&      244.25&         54.25 \\
600.0&   LR-50ms& 1171.0&        49.26&         19.00\\
\midrule
700.0&    CPU-30&   39.1&         0.14&        58.07\\
700.0&    CPU-70&   61.0&         0.26&         24.80\\
700.0&   BO-50ms& 1837.0&       247.64&        41.28\\
700.0& COLA-50ms&   51.3&        0.52&        26.65\\
700.0&   DQN-50ms&  2000.0&       275.57&          55.00 \\
700.0&   LR-50ms& 1970.0&      156.16&         19.00\\
\midrule
800.0&    CPU-30&   43.6&        0.36&        67.42\\
800.0&    CPU-70&   73.3&        1.07&        27.23\\
800.0&   BO-50ms&  550.7&       92.13&        46.47\\
800.0& COLA-50ms&   54.0&        1.05&        29.38\\
800.0&   DQN-50ms&  2000.0&     324.64&   55.00\\
800.0&   LR-50ms& 2000.0&       299.95&        19.00\\
  \bottomrule
\end{tabularx}
\caption{Online Boutique Constant Rate Results}
\label{tab:ob-fr-tabular}
\vspace{-0.25cm}
\end{table*}

\begin{table*}[h]
  \begin{tabularx}{\textwidth}{XXrrrr}
    \toprule
    Users& Policy & Median Latency &   Failures/s & VM Instances \\
\midrule
250.0&    CPU-30&   45.9&       0.05&       103.07\\
250.0&    CPU-70&   50.4&       0.02&        78.60\\
250.0&   BO-50ms&   60.8&       2.19&        275.80\\
250.0& COLA-50ms&   47.5&       0.00&     77.00\\
250.0&   DQN-50ms&    51.6&     0.00& 352.80\\
250.0&   LR-50ms&  219.6&      20.63&       111.22\\
\midrule
500.0&    CPU-30&   47.0&       0.38&       119.67\\
500.0&    CPU-70&   56.6&       7.03&        84.68\\
500.0&   BO-50ms& 1311.0&     135.49&       279.93\\
500.0& COLA-50ms&   49.8&       0.07&         82.00\\
500.0&   DQN-50ms&    92.7&     0.00&         353.00 \\
500.0&   LR-50ms&   67.9&       1.05&       115.52\\
\midrule
600.0&    CPU-30&   47.3&       0.55&       130.03\\
600.0&    CPU-70&   68.4&      13.92&         84.50\\
600.0&   BO-50ms&  370.3&       60.31&       269.93\\
600.0& COLA-50ms&   56.3&      10.11&        84.43\\
600.0&   DQN-50ms&   922.0&    84.60&         353.00 \\
600.0&   LR-50ms& 1328.0&      184.86&        109.20\\
  \bottomrule
\end{tabularx}
\caption{Train Ticket Constant Rate Results}
\label{tab:tt-fr-tabular}
\vspace{-0.25cm}
\end{table*}

\begin{table*}[h]
  \begin{tabularx}{\textwidth}{XXrrrr}
    \toprule
    Users& Policy &  90\%ile Latency &  Failures/s & VM Instances \\
\midrule

250.0&          CPU-30&        69.8&    0.04&         97.30\\
250.0&          CPU-70&       160.4&    0.03&         76.30\\
250.0& COLA-100ms&        83.4&    0.06&         77.00\\
\midrule
400.0&          CPU-30&        74.6&    0.04&        106.10\\
400.0&          CPU-70&      563.2&     5.60&        78.87\\
400.0& COLA-100ms&       107.0&     0.00&         79.00\\
\midrule
500.0&          CPU-30&         73.8&    0.22&       112.97\\
500.0&          CPU-70&       109.0&     0.10&        83.23\\
500.0&          COLA-100ms&        471.0&    6.62&        87.98\\
\midrule
600.0&          CPU-30&         77.4&    0.26&       115.87\\
600.0&          CPU-70&        180.0&    0.03&        84.27\\
600.0&          COLA-100ms&        91.4&     0.00&         89.00\\
\bottomrule
\end{tabularx}
\caption{Train Ticket Constant Rate Results (Tail Policies)}
\label{tab:tt-tail-fr-tabular}
\vspace{-0.25cm}
\end{table*}

\begin{table*}[h]
  \begin{tabularx}{\textwidth}{XXrrrr}
    \toprule
    Users& Policy &  90\%ile Latency &  Failures/s & VM Instances \\
\midrule

600.0&          CPU-30&        74.2&    0.08&         51.10\\
600.0&          CPU-70&      158.0&    0.22&         23.00\\
600.0& COLA-100ms&         95.8&    0.03&        33.57\\
\midrule
800.0&          CPU-30&         77.6&    0.02&         69.90\\
800.0&          CPU-70&         334.0&    1.07&         28.30\\
800.0& COLA-100ms&              136.0&    0.31&         33.30\\
\bottomrule
\end{tabularx}
\caption{Online Boutique Constant Rate Results (Tail Policies)}
\label{tab:ob-tail-fr-tabular}
\vspace{-0.25cm}
\end{table*}

\begin{table*}[h]
  \begin{tabularx}{\textwidth}{XXrrrr}
    \toprule
    Users& Policy & Median Latency &  Failures/s & VM Instances \\
\midrule

In Sample& CPU-30&  42.3&     0.03&        34.62\\
In Sample& CPU-70&  57.7&     0.41&        15.97\\
In Sample&  BO-50ms& 1050.0&  71.78&        35.28\\
In Sample& COLA-50ms&  52.3&  0.25&        18.38\\
In Sample& DQN-50ms&   1182.0&   59.82&      52.51\\
In Sample& LR-50ms&   65.0&    0.32&        22.63\\

\midrule
Out of Sample&    CPU-30&  38.6&        0.04&        28.77\\
Out of Sample&  CPU-70&  53.3&       0.38&        15.21\\
Out of Sample&   BO-50ms& 715.3&      42.76&        47.73\\
Out of Sample& COLA-50ms&  45.6&       0.15&        17.37\\
Out of Sample& DQN-50ms&   885.3&    34.96&      52.73\\
Out of Sample&  LR-50ms&  58.7&       0.61&        21.57\\

\bottomrule
\end{tabularx}
\caption{Online Boutique Diurnal Results}
\label{tab:ob-tail-diurnal-tabular}
\vspace{-0.25cm}
\end{table*}

\begin{table*}[h]
  \begin{tabularx}{\textwidth}{XXrrrr}
    \toprule
    Users& Policy & Median Latency &  Failures/s & VM Instances \\
\midrule

In Sample&   CPU-30&  21.7&      0.00&        10.47\\
In Sample&   CPU-70&  46.7&     9.66&         6.18\\
In Sample&  BO-50ms&   18.0&     0.00&        25.94\\
In Sample& COLA-50ms&  39.3&    9.62&         6.41\\
In Sample& DQN-50ms&   1430.0&  115.43&      31.94\\
In Sample&  LR-50ms&  21.3&      0.68&        11.88\\

\midrule
Out of Sample&   CPU-30&  21.7&    0.02&        11.19\\
Out of Sample&   CPU-70&  26.3&      0.00&         5.84\\
Out of Sample&  BO-50ms&  19.3&      3.90&         30.00\\
Out of Sample& COLA-50ms&  24.3&     0.00&         6.27\\
Out of Sample& DQN-50ms&   1473.3&   123.81&      31.40\\
Out of Sample&  LR-50ms&  21.3&        0.00&        12.65\\

\bottomrule
\end{tabularx}
\caption{Book Info Diurnal Results}
\label{tab:bi-tail-fr-tabular}
\vspace{-0.25cm}
\end{table*}

\begin{table*}[h]
  \begin{tabularx}{\textwidth}{XXrrrr}
    \toprule
    Users& Policy & Median Latency &  Failures/s & VM Instances \\
\midrule

In Sample&   CPU-30&   45.0&        0.02&         8.44\\
In Sample&   CPU-70&   54.0&        67.29&         3.08\\
In Sample&  BO-50ms&   44.0&          0.00&        14.79\\
In Sample& COLA-50ms&   46.0&         0.00&         1.98\\
In Sample& DQN-50ms&    44.0&     0.00&       6.00\\
In Sample&  LR-50ms&   44.0&        0.00&         6.71\\

\midrule

Out of Sample&   CPU-30&   45.0&         0.00&         8.24\\
Out of Sample&   CPU-70&  45.7&          0.00&          3.30\\
Out of Sample&  BO-50ms&   44.0&         0.10&        18.31\\
Out of Sample& COLA-50ms&  45.3&         0.00&         2.24\\
Out of Sample& DQN-50ms&   44.0&         0.00&       7.63\\
Out of Sample&  LR-50ms&   44.0&         0.00&         6.87\\

\bottomrule
\end{tabularx}
\caption{Simple Web Server Diurnal Results}
\label{tab:sws-fr-tabular}
\vspace{-0.25cm}
\end{table*}

\begin{table*}[h]
  \begin{tabularx}{\textwidth}{XXrrrr}
    \toprule
    Users& Policy & Median Latency &  Failures/s & VM Instances \\
\midrule

In Sample&   CPU-30&    50.0&       0.00&        74.53\\
In Sample&   CPU-70&   49.3&        0.00&        73.16\\
In Sample&  BO-50ms& 1343.3&     184.94&       290.95\\
In Sample& COLA-50ms&   92.3&      62.31&        78.62\\
In Sample& DQN-50ms&   1900.0&    183.13&     350.69\\
In Sample&  LR-50ms&  853.3&      149.59&       110.86\\

\midrule

Out of Sample&    CPU-30&  57.7&       11.11&       131.62\\
Out of Sample&    CPU-70& 166.3&       78.28&        83.07\\
Out of Sample&   BO-50ms& 718.3&       127.76&       270.43\\
Out of Sample& COLA-50ms&   48.0&      0.78&         73.00\\
Out of Sample& DQN-50ms&    48.0&      45.23&     354.20\\
Out of Sample&   LR-50ms&   56.0&      0.84&        102.50\\
\bottomrule
\end{tabularx}
\caption{Train Ticket Diurnal Results}
\label{tab:tt-diurnal-tabular}
\vspace{-0.25cm}
\end{table*}

\begin{table*}[h]
  \begin{tabularx}{\textwidth}{XXrrrr}
    \toprule
    Users& Policy & Median Latency &  Failures/s & VM Instances \\
\midrule

In Sample&    CPU-30&  11.3&          6.61&        28.37\\
In Sample&    CPU-70&  26.3&         9.70&        17.42\\
In Sample&   BO-50ms&   9.7&        50.97&        28.23\\
In Sample&  COLA-50ms&  15.3&       0.00&         15.70\\
In Sample&  DQN-50ms&   10.0&       0.00&      50.00\\
In Sample&   LR-50ms& 306.3&        15.43&         14.00\\

\midrule

Out of Sample&    CPU-30&  10.7&     8.22&        30.06\\
Out of Sample&    CPU-70&   20.0&    0.01&        15.64\\
Out of Sample&   BO-50ms&  23.3&     60.32&        23.38\\
Out of Sample& COLA-50ms&  25.7&     0.00&         15.00\\
Out of Sample& DQN-50ms&     9.7&    0.01&      47.95\\
Out of Sample&   LR-50ms&  148.0&    3.68&        14.28\\

\bottomrule
\end{tabularx}
\caption{Sock Shop Diurnal Results}
\label{tab:ss-diurnal-tabular}
\vspace{-0.25cm}
\end{table*}

\begin{table*}[h]
  \begin{tabularx}{\textwidth}{Xrrrr}
    \toprule
    Policy & Median Latency &   Failures/s & VM Instances \\
\midrule

   CPU-30&   12.5&         4.01&        26.44\\
   CPU-70&  24.3&          8.79&         15.00\\
  BO-50ms&   21.0&        74.85&        23.25\\
COLA-50ms&   41.0&        6.35&         14.10\\
  LR-50ms&  49.3&          7.75&        14.47\\
\bottomrule
\end{tabularx}
\caption{Sock Shop Alternating Constant Rate Results}
\label{tab:ss-acr-tabular}
\vspace{-0.25cm}
\end{table*}

\begin{table*}[h]
  \begin{tabularx}{\textwidth}{XXrrrr}
    \toprule
    Users & Policy & Median Latency &  Failures/s & VM Instances \\
\midrule
300.0&    CPU-30&   39.8&         3.77&         53.00\\
300.0&    CPU-70&   55.4&         6.09&        23.43\\
300.0& COLA-50ms&   49.2&         6.93&         29.00\\
\bottomrule
\end{tabularx}
\caption{Online Boutique Dynamic Request Distribution Results}
\label{tab:ob-drd-tabular}
\vspace{-0.25cm}
\end{table*}

\begin{table*}[h]
  \begin{tabularx}{\textwidth}{XXrrrr}
    \toprule
    Users& Policy & Median Latency &   Failures/s & VM Instances \\
\midrule

100 &     CPU-10 &    19.7 &          0.02 &          20.15 \\
100 &     CPU-30 &    17.1 &          0.00 &      13.07 \\
100 &     CPU-50 &    19.0 &          0.00 &       8.42 \\
100 &     CPU-70 &    20.3 &          0.05 &       8.32 \\
100 &     CPU-90 &    54.7 &          0.00 &          4.00\\
100 &  COLA-50ms &    24.8 &          0.00 &            5.00 \\

\midrule

250 &     CPU-10 &    19.3 &         0.00 &    33.47 \\
250 &     CPU-30 &    19.5 &         0.00 &     8.58 \\
250 &     CPU-50 &    24.6 &         0.00 &          6.10 \\
250 &     CPU-70 &    40.7 &         0.00 &     5.17 \\
250 &     CPU-90 &    39.9 &         0.00 &     5.00 \\
250 &  COLA-50ms &    25.2 &         0.00 &          6.00 \\
\midrule

700 &     CPU-10 &    19.8 &        0.00 &     33.73 \\
700 &     CPU-30 &    23.7 &        0.00 &     17.88 \\
700 &     CPU-50 &    30.8 &        0.00 &     11.83 \\
700 &     CPU-70 &   200.5 &        26.46 &     11.72 \\
700 &     CPU-90 &   205.7 &       111.99 &     12.48 \\
700 &  COLA-50ms &    30.6 &         0.00 &          10.00 \\

\midrule

850 &     CPU-10 &    21.7 &      0.43 &    36.64 \\
850 &     CPU-30 &    28.3 &      8.26 &    22.07 \\
850 &     CPU-50 &    32.6 &      10.77 &    16.62 \\
850 &     CPU-70 &    43.7 &      0.01 &    11.32 \\
850 &    CPU-90 &   43.9 &        3.41 &     10.58\\
850 &  COLA-50ms &    35.3 &      0.00 &         10.00 \\

\midrule
1100 &     CPU-10 &    23.6 &      0.36 &         39.25 \\
1100 &     CPU-30 &    24.7 &      0.47 &         24.65 \\
1100 &     CPU-50 &    37.8 &      0.92 &     20.23 \\
1100 &     CPU-70 &  1055.6 &      247.04 &     12.63 \\
1100 &     CPU-90 &    80.4 &       0.02 &          9.8 \\
1100 &  COLA-50ms &    42.7 &      0.00 &          11.00 \\

\midrule

1250 &     CPU-10 &    25.7 &       0.00 &         41.50 \\
1250 &     CPU-30 &    25.6 &       0.04 &         23.90 \\
1250 &     CPU-50 &    82.1 &       1.77 &         20.10 \\
1250 &     CPU-70 &    75.2 &       0.65 &    15.38 \\
1250 &     CPU-90 &   652.0 &       73.70 &     11.72 \\
1250 &  COLA-50ms &    46.6 &       0.00 &         12.00 \\

\bottomrule
\end{tabularx}
\caption{Book Info Large Dynamic Request Rate}
\label{tab:bi-ldr-tabular}
\vspace{-0.25cm}
\end{table*}

\begin{table*}[h]
  \begin{tabularx}{\textwidth}{XXrrrrr}
    \toprule
    Users& Policy & Median Latency &  Failures/s & Number of Pods & VM Instances \\
\midrule

500&      CPU-30&     45.0&              0.0&       5.266667&   5.833333  \\
500&      CPU-70&     46.2&              0.0&            2.0&        1.0  \\
500&   COLA-50ms&     44.0&              0.0&            2.0&        1.0  \\

\midrule

1000&      CPU-30&     45.0&             0.0&           10.0&        5.0  \\
1000&      CPU-70&     46.0&             0.0&            4.0&        2.0  \\
1000&   COLA-50ms&     45.0&             0.0&            3.0&        2.0  \\
\midrule

2000&      CPU-30&     45.0&             0.0&           18.0&        9.0  \\
2000&      CPU-70&     46.0&             0.0&            7.0&        4.0  \\
2000&   COLA-50ms&     46.0&             0.0&            5.0&        3.5  \\

\bottomrule
\end{tabularx}
\caption{Simple Web Server Two Core Fixed Rate}
\label{tab:sws-tc-fr-tabular}
\vspace{-0.25cm}
\end{table*}

\begin{table*}[h]
  \begin{tabularx}{\textwidth}{XXrrrrr}
    \toprule
    Users& Policy & Median Latency &  Failures/s & Number of Pods & VM Instances \\
\midrule

500&     CPU-30&    44.0&         0.0&           5.0&       2.6 \\
500&     CPU-70&    44.0&         0.0&           2.0&       1.0 \\
500&  COLA-50ms&    44.6&         0.0&           2.0&       1.0 \\
\midrule

1000&     CPU-30&    44.0&        0.0&           9.0&       2.0 \\
1000&     CPU-70&    44.2&        0.0&           4.0&       1.0 \\
1000&  COLA-50ms&    45.0&        0.0&           3.0&       1.0 \\

\midrule

2000&     CPU-30&    44.0&        0.0&          19.0&       4.0 \\
2000&     CPU-70&    45.0&        0.0&           7.0&       2.0 \\
2000&  COLA-50ms&    45.0&        0.0&           6.0&       2.0 \\

\bottomrule
\end{tabularx}
\caption{Simple Web Server Four Core Fixed Rate}
\label{tab:sws-fc-fr-tabular}
\vspace{-0.25cm}
\end{table*}

\begin{table*}[h]
  \begin{tabularx}{\textwidth}{XXrrrrr}
    \toprule
    Users& Policy & Median Latency &  Failures/s & Number of Pods & VM Instances \\
\midrule

200&     CPU-30&    20.6&          0.00&          12.0&      11.0 \\
200&     CPU-70&    26.2&          0.00&           7.0&       4.0 \\
200&  COLA-50ms&    35.6&          0.00&           6.0&       3.5 \\
\midrule

300&     CPU-30&    22.0&          0.0&          18.6&      19.9 \\
300&     CPU-70&    43.0&          0.0&           8.0&       4.0 \\
300&  COLA-50ms&    37.2&          0.0&           8.0&       4.0 \\
\bottomrule
\end{tabularx}
\caption{Book Info Two Core Fixed Rate}
\label{tab:bi-tc-fr-tabular}
\vspace{-0.25cm}
\end{table*}

\begin{table*}[h]
  \begin{tabularx}{\textwidth}{XXrrrrr}
    \toprule
    Users& Policy & Median Latency &  Failures/s & Number of Pods & VM Instances \\
\midrule

200&     CPU-30&    18.6&            0.0&          13.4&       8.0 \\
200&     CPU-70&    25.2&            0.0&           7.0&       2.0 \\
200&  COLA-50ms&    22.8&            0.0&           6.0&       2.0 \\

\midrule

300&     CPU-30&    19.4&            0.0&          16.0&       5.0 \\
300&     CPU-70&    30.4&            0.0&           9.0&       2.0 \\
300&  COLA-50ms&    46.6&            0.0&           7.0&       2.0 \\
\bottomrule
\end{tabularx}
\caption{Book Info Four Core Fixed Rate}
\label{tab:bi-fc-fr-tabular}
\vspace{-0.25cm}
\end{table*}

\begin{table*}[h]
  \begin{tabularx}{\textwidth}{XXrrrrr}
    \toprule
    Users& Policy & Median Latency &  Failures/s & Number of Pods & VM Instances \\
\midrule

100&     CPU-30&    54.6&      0.00&          22.0&      11.0 \\
100&     CPU-70&    69.6&      0.17&          13.0&       7.0 \\
100&  COLA-50ms&    49.4&      0.10&          20.0&       9.0 \\

\midrule

150&     CPU-30&    57.6&     0.28&          28.0&      14.0 \\
150&     CPU-70&    88.8&     0.42&          14.0&       7.0 \\
150&  COLA-50ms&    52.0&     0.02&          22.0&       9.0 \\

\bottomrule
\end{tabularx}
\caption{Online Boutique Two Core Fixed Rate}
\label{tab:ob-tc-fr-tabular}
\vspace{-0.25cm}
\end{table*}

\begin{table*}[h]
  \begin{tabularx}{\textwidth}{XXrrrrr}
    \toprule
    Users& Policy & Median Latency &  Failures/s & Number of Pods & VM Instances \\
\midrule

100&     CPU-30&    47.6&     0.41&     23.06&       5.0 \\
100&     CPU-70&    61.4&     0.63&     13.63&       3.0 \\
100&  COLA-50ms&    48.4&     0.19&     15.00&       3.5 \\

\midrule

150&     CPU-30&    46.4&     0.37&          30.6&       6.57 \\
150&     CPU-70&    53.6&     0.97&          17.0&       4.00 \\
150&  COLA-50ms&    47.0&     0.60&          19.0&       4.50 \\

\bottomrule
\end{tabularx}
\caption{Online Boutique Four Core Fixed Rate}
\label{tab:ob-fc-fr-tabular}
\vspace{-0.25cm}
\end{table*}

\begin{table*}[h]
  \begin{tabularx}{\textwidth}{XXrrrr}
    \toprule
    Users& Policy & Median Latency  &  Failures/s & Number of Pods \\
\midrule
200&   CPU-30 &    26.8  &      0.93 &     10.73  \\
200&   CPU-70  &   2000.0 &    92.76 &         6.95   \\
200&   COLA-50ms&  27.7   &     8.90 &     9.13   \\
\midrule
300&     CPU-30&    26.6&            0.0&          14.0 \\
300&     CPU-70&  2000.0&         137.77&           7.8 \\ 
300&  COLA-50ms&    29.1&         0.0&          10.0 \\
\midrule
400&     CPU-30&    27.1&            0.0&          16.0 \\ 
400&     CPU-70&  2000.0&          181.89&          8.55 \\  
400&  COLA-50ms&    28.3&         0.545&          13.0 \\  
\midrule
500&     CPU-30&    42.0&      20.23&          28.9 \\   
500&     CPU-70&  2000.0&      224.39&             9.45 \\  
500&  COLA-50ms&    39.0&      0.00&           13.0 \\  
\bottomrule
\end{tabularx}
\caption{Book Info Autopilot Results}
\label{tab:bi-ap-tabular}
\vspace{-0.25cm}
\end{table*}

\begin{table*}[h]
  \begin{tabularx}{\textwidth}{XXrrrr}
    \toprule
    Users& Policy & Median Latency &   Failures/s & Number of Pods \\
\midrule
100&     CPU-30&    48.0&        0.12&         18.85 \\   
100&     CPU-70&   180.0&        0.24&          11.0 \\   
100&  COLA-50ms&    46.3&        0.00&     14.18 \\   
\midrule
150&     CPU-30&    47.1&        0.00&     21.52 \\   
150&     CPU-70&   132.4&        0.24&          12.0 \\   
150&  COLA-50ms&    47.3&        0.14&     17.87 \\   
\midrule
200&     CPU-30&    50.1&        0.06&     25.02 \\   
200&     CPU-70&   200.0&        0.38&     12.97 \\   
200&  COLA-50ms&    53.9&        0.56&          18.0 \\   
\bottomrule&
\end{tabularx}
\caption{Online Boutique Autopilot Results}
\label{tab:ob-ap-tabular}
\vspace{-0.25cm}
\end{table*}

\begin{table*}[h]
  \begin{tabularx}{\textwidth}{XXrrrr}
    \toprule
    Users& Policy & Median Latency & Failures/s & Number of Pods \\
\midrule
100&     CPU-30&    39.0&        0.0&          73.0 \\ 
100&     CPU-70&    37.8&       14.97&          73.0 \\ 
100&  COLA-50ms&    39.8&       0.53&     73.23 \\ 
\midrule
200&     CPU-30&    43.8&       0.00&     82.53 \\ 
200&     CPU-70&    38.4&       29.44&          73.0 \\ 
200&  COLA-50ms&    37.0&       0.0&          75.0 \\ 
\midrule
350&     CPU-30&    49.4&       1.42&     92.77 \\ 
350&     CPU-70&    40.2&       50.32&          74.3 \\ 
350&  COLA-50ms&    39.6&       0.0&          76.0 \\ 
\bottomrule&
\end{tabularx}
\caption{Train Ticket Autopilot Results}
\label{tab:tt-ap-tabular}
\vspace{-0.25cm}
\end{table*}

\begin{figure*}[h]
\subfigure[Book Info Diagram - Courtesy of Istio \cite{binfo:online}]{%
  \label{fig:bi-diagram}%
  \includegraphics[width=.50\textwidth]{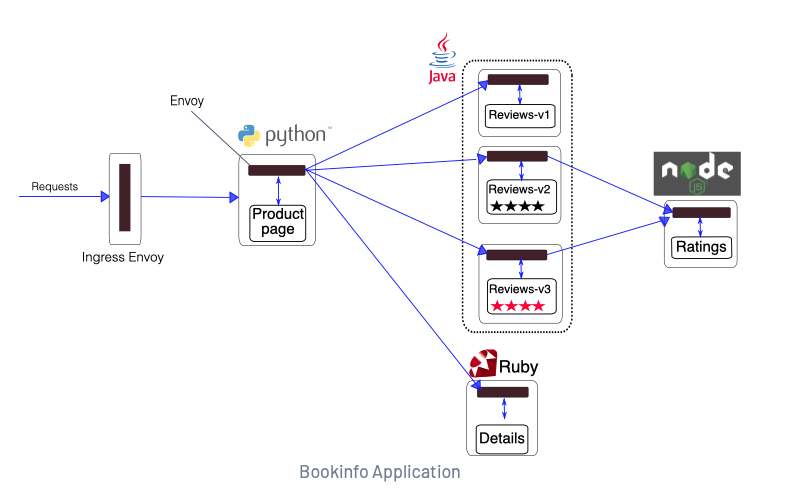}%
}%
\hspace*{\fill}
\subfigure[Online Boutique Diagram - Courtesy of Google \cite{ob:online}]{
  \label{fig:ob-diagram}%
  \includegraphics[width=0.50\textwidth]{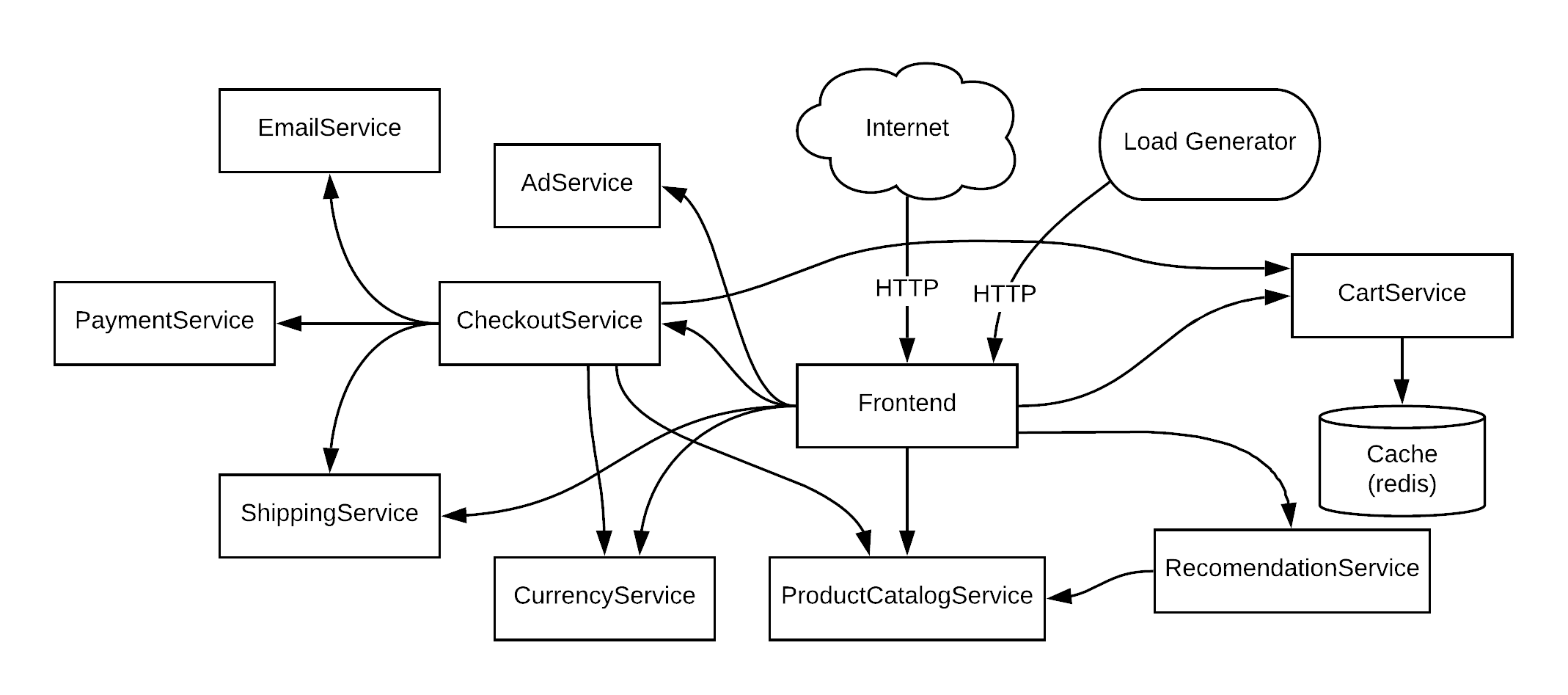}%
}%

\subfigure[Sock Shop Diagram - Courtesy of Weaveworks and Container Solutions \cite{ss:online}]{%
  \label{fig:ss-diagram}%
  \includegraphics[width=0.50\textwidth]{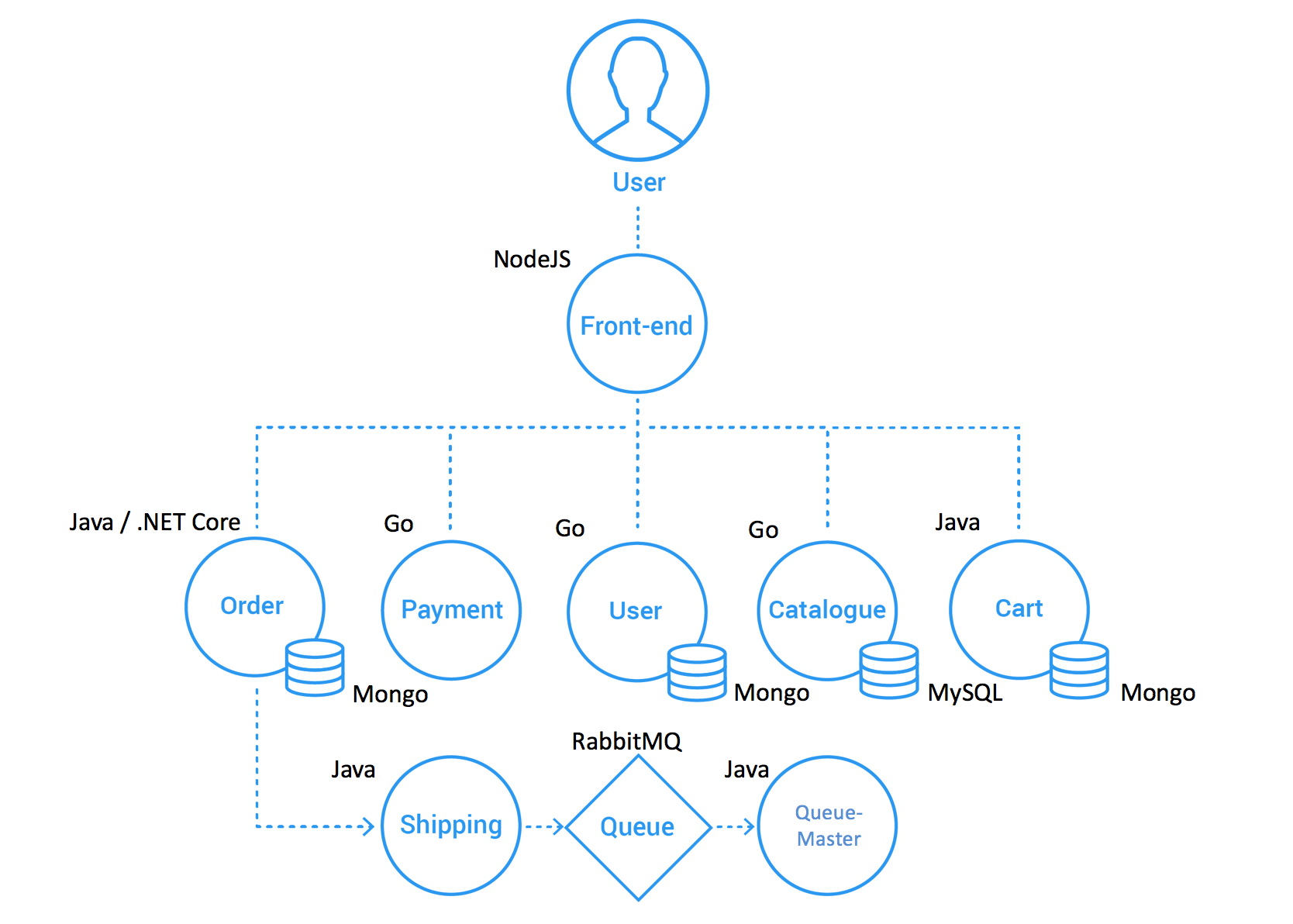}%
}%
\hspace*{\fill}
\subfigure[Trian Ticket Diagram - Courtesy of Fudan SE \cite{zhou2018poster}]{
  \label{fig:tt-diagram}%
  \includegraphics[width=0.50\textwidth]{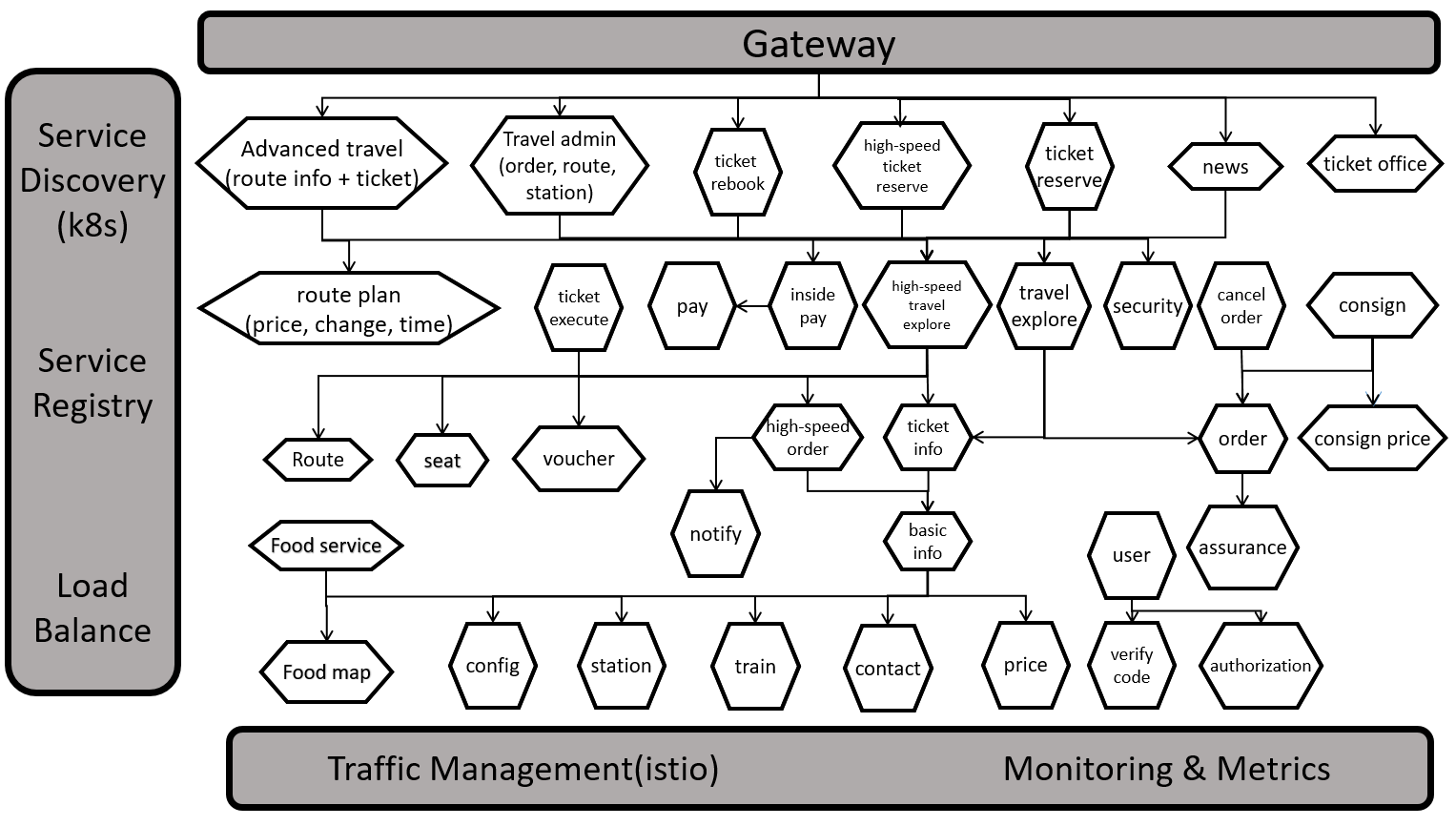}%
}%
\captionsetup{justification=centering}
\caption{Microservice Application Diagrams}\label{fig:arch-diagrams}
\end{figure*}

\end{document}